\documentclass{article}
\usepackage{arxiv}
\setcitestyle{numbers,square,comma}
\usepackage[utf8]{inputenc}
\usepackage[T1]{fontenc}
\usepackage[hidelinks]{hyperref}
\usepackage{url}
\usepackage{booktabs}
\usepackage{amsfonts}
\usepackage{nicefrac}
\usepackage{microtype}
\usepackage{balance}
\usepackage{amsmath}
\usepackage{amsthm}
\usepackage{multirow}
\usepackage{graphicx}
\usepackage{rotating}
\usepackage{graphicx}
\usepackage{adjustbox}
\newcommand{\Z}{\mathbb{Z}}
\newcommand{\Sset}{\mathcal{S}}
\newcommand{\Aset}{\mathcal{A}}
\newcommand{\ind}{\mathbf{1}}
\newcommand{\lev}{\lambda}
\newcommand{\Pup}{P_{\text{up}}}

\title{Spatial Pattern Formation from Multi-Agent Learning\\ in Public Goods Dilemmas}

\author{
  Yefei Zhang \\
  Department of Mathematics\\
  University of Illinois Urbana-Champaign\\
  Urbana, IL 61801, USA \\
  \texttt{yefeiz2@illinois.edu} \\
  \And
  Yuxuan Zhao \\
  Department of Mathematics \\
  University of Illinois Urbana-Champaign\\
  Urbana, IL 61801, USA \\
  \texttt{yuxuan44@illinois.edu} \\
}

\begin{document}

\maketitle

\begin{abstract}
Spatial public goods models show that prescribed movement toward richer locations can generate spatial patterns. We ask how such patterns emerge when agents learn where to move and how learning rates shape their consequences for collective welfare. Fixed populations of cooperators and defectors independently learn movement policies using tabular Q-learning and local observations. Cooperator learning generates clusters around resource peaks, while co-adaptation changes their strength and motion. At a fixed training budget, the largest welfare losses occur when cooperators learn at high rates and defectors at low rates. In part of this regime, learned policies also generate traveling bands supported by a shared directional preference. The conditions supporting travel change with further training, so these patterns reflect training history rather than an established asymptotic outcome. Across the tested learning-rate conditions with cooperator learning, mean collective welfare falls below random movement because increased crowding outweighs gains in resource benefit. Charging agents for the crowding they impose on others during learning recovers much of the welfare loss in the tested conditions. These results connect learning rates to the emergence and welfare costs of spatial organization driven by individual rewards.
\end{abstract}

\section{Introduction}
\label{sec:intro}

Spatial public goods models couple the movement of producers and consumers to a shared resource. Cooperators produce the good, both types consume it, and individuals can move between locations with different resource availability. Prescribed movement rules based on local payoffs or environmental conditions can generate various spatial patterns~\cite{Wakano2009,Funk2019,yao2026pattern,yao2025spatial,Zhao2026pattern}. Such models explain how a specified movement rule organizes a population. Here we ask what happens if movement rules are learned through interaction with a public good field that agents themselves produce and consume. Which parts of the learned policy generate aggregation and collective behavior? How do these policies depend on the partners encountered during training, and what are the consequences for collective welfare?

Learning adds another layer to this feedback. Agents learn where to move from the payoffs they receive, while their movements relocate resource production and consumption and change local crowding. These environmental changes, in turn, shape what other agents learn. If both populations adapt, the resulting spatial organization depends not only on the policies being executed but also on the partners with which those policies were learned.

\begin{figure*}[t]
\centering
\includegraphics[width=\textwidth]{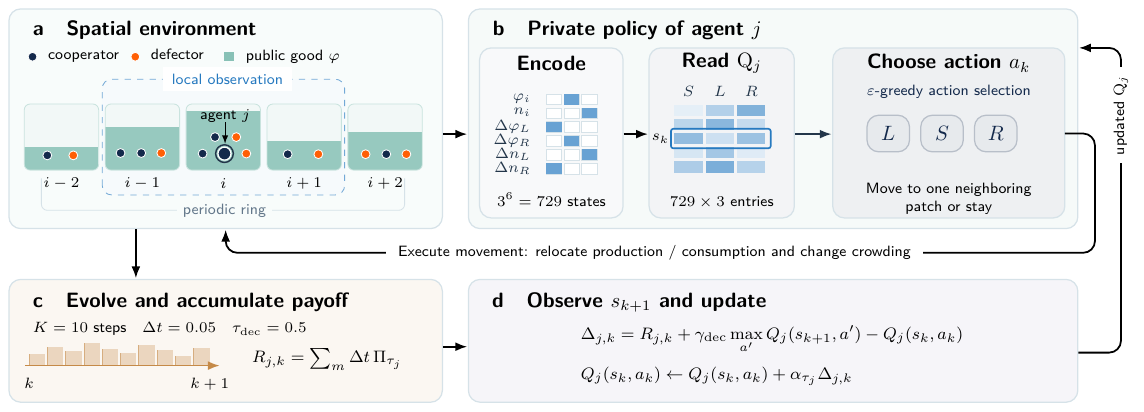}
\caption{The model and the learning loop of one agent. (a)~Agent $j$ on patch $i$ of the ring observes its own and the two adjacent patches. (b)~It bins the observation into one of $729$ states, reads its private table $Q_j$, and chooses $a_k\in\{S,L,R\}$. (c)~The environment evolves over the decision interval while the agent accumulates payoff. (d)~At its next decision it observes $s_{k+1}$ and updates $Q_j$.}
\label{fig:qlearn}
\end{figure*}

We study this feedback on a ring of patches. Each agent has a fixed cooperator or defector type and independently learns whether to stay, move left, or move right using tabular Q-learning (Figure~\ref{fig:qlearn}). It observes only its own patch and its two neighbors and receives its own payoff. There is no prescribed gradient response or preferred direction. Holding types and population sizes fixed isolates learned mobility from changes in the population composition. Private Q-tables also allow us to test the mechanisms through policy interventions. We distill policies into compact movement rules, remove resource-seeking preferences from these rules or directional preferences from the original policies, and recombine policies learned with different training partners.

Previous work on chemotaxis learning and spatial games with learned migration has shown that agents can learn resource-seeking movement and that learned movement can generate spatial clustering (Section~\ref{sec:related}). We build on these findings by examining the mechanisms, training dependence, and welfare consequences of learned spatial organization. First, learning by cooperators is sufficient to generate persistent spatial clusters, and policy interventions show that a preference for richer neighboring patches is necessary for aggregation in the tested distilled rules (Sections~\ref{sec:states} and~\ref{sec:emerge}). Second, co-adaptation changes the learned movement policies: the same defector policies can weaken or intensify clustering depending on the cooperators' training history (Section~\ref{sec:emerge}). When cooperators learn faster than defectors, frozen policies can produce traveling bands supported by a spontaneously learned preference for a common direction. The learning-rate conditions under which these bands appear and whether travel persists with further training depend on the training budget (Sections~\ref{sec:rates} and~\ref{sec:training}). Third, the resulting spatial organization lowers welfare primarily by increasing crowding. Total production is independent of location, and total resource uptake remains nearly unchanged across the evaluated conditions, so aggregation increases crowding costs without a comparable gain in aggregate resource benefit. Charging agents for the crowding they impose on others during learning recovers much of the welfare loss (Section~\ref{sec:welfare}).

\section{Related Work}
\label{sec:related}

Random dispersal can favor cooperation~\cite{Vainstein2007,Meloni2009,Cardillo2012}, while migration in response to success, aspiration, or local resource conditions allows individuals to adapt their movement to their surroundings~\cite{Helbing2009,Yang2010,Wu2012,Inaba2025}. Migration costs and speeds can further shape cooperative outcomes~\cite{Lee2022,Ibrahimli2024,Bara2024,Young2018}. Models with an explicit public good field connect movement to Turing patterns and instabilities induced by directed motion~\cite{Wakano2009,Funk2019,Zhao2026pattern}. Related evolutionary game models show that movement in response to payoff or environmental conditions can similarly destabilize homogeneous states and generate spatial patterns in both agent-based and continuum descriptions~\cite{yao2026pattern,yao2025spatial}. Similar links between gradient-following responses and population patterns also appear in chemotaxis, including classical analyses of aggregation and traveling bands~\cite{Keller1970,KellerSegel1971Bands,Hillen2009,Stevens2000}, as well as in ideal-free distributions and mean-field games~\cite{Cantrell2010,Cantrell2025}.

Reinforcement learning has been used to study contribution decisions in spatial and networked public goods games~\cite{Zheng2024,Mintz2025,Meylahn2025,Xie2026review}, including models with adaptive incentives~\cite{Kong2026}. In these settings, agents do not learn where to move. Movement itself has been learned in a spatial prisoner's dilemma~\cite{Mangold2025} and in a cyclic-dominance model of species coexistence~\cite{Jiang2025}. Related work also shows that agents can learn chemotactic behavior in prescribed fields~\cite{Colabrese2017,Tovey2024,Pramanik2025,Tovazzi2025}, while sequential social dilemmas study multiple learners interacting through shared resources~\cite{Leibo2017,Perolat2017,Hughes2018,Leibo2021}. These works study learned mobility, cooperation, coexistence, or navigation in a given environmental signal. In our setting, the resource field is instead produced and depleted by the mobile agents themselves. We keep social types fixed and ask how this feedback shapes the learned movement policies, the spatial patterns they generate, and their consequences for welfare.

Independent learning makes each agent's environment nonstationary~\cite{Claus1998,Matignon2012,HernandezLeal2019}. The relative learning rates of interacting agents can further change the resulting dynamics~\cite{Bowling2002,Foerster2018,Leonardos2022,Bielawski2025,Barfuss2023}. Repeated interaction can also give rise to shared conventions~\cite{Young1993,Shoham1997,Lerer2019}. In our model, this takes the form of a common directional preference. Left and right are equivalent in the underlying environment, but once a population begins to move, the two directions are no longer equivalent to the agents themselves. We use policy interventions to separate this directional preference from the mechanism that generates aggregation. Finally, we connect the resulting collective behavior to incentive design~\cite{Yang2020lio,Kong2026} by asking how agents change their movement when their reward includes the crowding cost they impose on others, while continuing to evaluate outcomes using the original payoff.

\section{The Model}
\label{sec:model}

The model couples a public good field to agents that learn where to move. Its resource dynamics follow~\cite{Young2018,Zhao2026pattern}, while the populations remain fixed so that learning changes locations but not population sizes.

\subsection{Resource production, consumption, and crowding}
\label{sec:env}

The environment is a ring of $M$ patches, $\mathcal G=\Z_M$, with spacing $\ell=L/M$ and patch volume $\Omega$. Patch indices $i\pm1$ are interpreted modulo $M$. There are $N_C$ cooperators and $N_D$ defectors. Agent $j\in\{1,\ldots,N\}$ has a fixed type $\tau_j\in\{C,D\}$ and a position $x_j(t)\in\mathcal G$, where $N=N_C+N_D$. The local populations are
\begin{equation}
    U_i=\sum_{j:\tau_j=C}\ind\{x_j=i\},\quad V_i=\sum_{j:\tau_j=D}\ind\{x_j=i\},\quad n_i=U_i+V_i.
\end{equation}
There are no births, deaths, or type changes. In particular, a cluster of cooperators represents their redistribution, not an increase in their number.

The public good concentration $\varphi_i\ge0$ obeys
\begin{equation}
    \label{eq:phi}
    \frac{d\varphi_i}{dt}=c\frac{U_i}{\Omega}-\left(\kappa\frac{n_i}{\Omega}+\delta\right)\varphi_i+\frac{D_\varphi}{\ell^2}(\varphi_{i+1}+\varphi_{i-1}-2\varphi_i).
\end{equation}

Cooperators produce the public good, which is consumed by all agents, decays naturally, and diffuses between neighboring patches. Moving a cooperator therefore relocates both production and consumption, while moving a defector relocates consumption alone.

An agent on patch $i$ receives the payoff rate
\begin{equation}
    \label{eq:payoff}
    \begin{aligned}
        \Pi_C(\varphi_i,n_i)&=r_u\varphi_i-c-\gamma n_i/\Omega,\\
        \Pi_D(\varphi_i,n_i)&=r_v\varphi_i-\gamma n_i/\Omega.
    \end{aligned}
\end{equation}
Resource access is valuable to both types, crowding is costly to both, and production costs fall on cooperators. We take $r_u>r_v$ to allow producers to value the resource more strongly. All parameters appear in Table~\ref{tab:params}.

At uniform densities $u_0=N_C/(M\Omega)$ and $n_0=N/(M\Omega)$, the equilibrium resource is $\varphi^*=cu_0/(\kappa n_0+\delta)\approx0.699$, giving $\Pi_D^*\approx3.69$ and $\Pi_C^*\approx3.39$. More generally, defectors have a free-rider advantage on the same patch because $\Pi_D-\Pi_C=1-\varphi_i\ge0$ under our parameters, with $\varphi_i\le1$ preserved by the dynamics from the initial condition $\varphi_i=\varphi^*$ (see Appendix~\ref{sec:numerics}). Agents learn mobility while their contribution types remain fixed.

\begin{table}[t]
\caption{Model and simulation parameters.}
\label{tab:params}
\small
\setlength{\tabcolsep}{3pt}
\centering
\resizebox{0.75\columnwidth}{!}{%
\begin{tabular}{llr}
\toprule
symbol & meaning & value\\
\midrule
$M$, $L$, $\Omega$ & patches, ring length, patch volume & $64$, $8$, $20$\\
$N_C$, $N_D$ & agents per type & $448$, $192$\\
$c$, $\kappa$, $\delta$, $D_\varphi$ & production, uptake, decay, diffusion & $1$, $1$, $10^{-3}$, $0.03$\\
$r_u$, $r_v$, $\gamma$ & benefit (C, D), crowding cost & $7$, $6$, $1$\\
\midrule
$\Delta t$, $K$, $\tau_{\mathrm{dec}}$ & time step, steps per decision, interval & $0.05$, $10$, $0.5$\\
$\beta$, $\gamma_{\mathrm{dec}}$ & discount rate, per-decision discount & $0.5$, $0.779$\\
$\alpha_C$, $\alpha_D$ & learning rates (default) & $0.05$, $0.05$\\
$\varepsilon$, $\varepsilon_{\mathrm{eval}}$ & exploration (training, evaluation) & $0.3\to0.01$, $0.01$\\
$E$, $T_{\mathrm{ep}}$, $T_{\mathrm{eval}}$ & episodes, episode and evaluation length & $200$, $200$, $400$\\
\bottomrule
\end{tabular}
}
\end{table}

\subsection{Agent observations, movement, and rewards}
\label{sec:agents}

An agent chooses $a\in\Aset=\{S,L,R\}$ to stay, move one patch left, or move one patch right. The environment advances in steps $\Delta t=0.05$, while each agent decides once every $K=10$ steps, giving a decision interval $\tau_{\mathrm{dec}}=0.5$. Agent $j$ draws an independent phase $\theta_j\sim\mathrm{Unif}\{0,\ldots,K-1\}$ at the start of each episode and decides at steps $m_{j,k}=\theta_j+kK$. Agents that decide at the same step observe the same pre-movement configuration and execute their moves together. Between decisions, an agent remains at its current patch while the resource and other agents continue to change.

The observation describes the current patch and its differences from the two neighbors. For two thresholds $E=(e_1,e_2)$, define $\lev_E(z)=\sum_{q=1}^2\ind\{z>e_q\}\in\{0,1,2\}$. An agent on patch $i$ observes
\begin{equation}
    \label{eq:state}
    \begin{aligned}
        s=\big(&\lev_{E_\varphi}(\varphi_i),\lev_{E_n}(n_i),\lev_{E_{\Delta\varphi}}(\varphi_{i-1}-\varphi_i),
        \lev_{E_{\Delta\varphi}}(\varphi_{i+1}-\varphi_i),\\
        &\lev_{E_{\Delta n}}(n_{i-1}-n_i),
        \lev_{E_{\Delta n}}(n_{i+1}-n_i)\big).
    \end{aligned}
\end{equation}
The thresholds are $E_\varphi=(0.65,0.75)$, $E_n=(7.5,12.5)$, $E_{\Delta\varphi}=(-0.01,0.01)$, and $E_{\Delta n}=(-2.5,2.5)$. The six components distinguish three levels each, producing $729$ states. Agents do not observe the types of their neighbors or anything beyond the adjacent patches.

Each agent has a private table $Q_j:\Sset\times\Aset\to\mathbb R$ with $2187$ entries. It learns from the payoff earned over a complete decision interval, rather than the payoff immediately after moving (Figure~\ref{fig:qlearn}):
\begin{equation}
    \label{eq:reward}
    R_{j,k}=\sum_{m=m_{j,k}}^{m_{j,k}+K-1}\Delta t\,\Pi_{\tau_j}\!\left(\bar\varphi_{x_j^{(m)}}^{(m)},n_{x_j^{(m)}}^{(m)}\right).
\end{equation}
Positions and occupancies here are measured after the step's moves, and $\bar\varphi^{(m)}$ is the step-averaged resource. This reward includes the consequences of waiting in the selected patch as other agents arrive or leave and the resource evolves.

At the next decision, agent $j$ updates its previous state--action pair by Q-learning~\cite{Watkins1992}:
\begin{equation}
    \label{eq:q}
    \begin{aligned}
        Q_j(s_k,a_k)\leftarrow{}&Q_j(s_k,a_k)+\alpha_{\tau_j}\Delta_{j,k},\\
        \Delta_{j,k}={}&R_{j,k}+\gamma_{\mathrm{dec}}\max_{a'}Q_j(s_{k+1},a')-Q_j(s_k,a_k).
    \end{aligned}
\end{equation}
The rates $\alpha_C$ and $\alpha_D$ control value updates for the two types. We use $\gamma_{\mathrm{dec}}=e^{-\beta\tau_{\mathrm{dec}}}\approx0.779$, with $\beta=0.5$. Each agent maintains a private Q-table and learns from its own reward. All entries are initialized to $\tau_{\mathrm{dec}}\Pi_{\tau_j}^*/(1-\gamma_{\mathrm{dec}})$, and actions are selected $\varepsilon$-greedily with uniform tie-breaking. A zero learning rate thus keeps all entries equal and actions uniformly random. By contrast, freezing a trained table stops updates while preserving learned action preferences.

Each step executes decisions and moves, integrates~\eqref{eq:phi} at the new occupancies, and accumulates payoffs. Integration uses two Euler substeps of length $\Delta t/2$, with negative concentrations clipped to zero after each substep. The substep diffusion number is $0.048$, and Appendix~\ref{sec:numerics} details the update order. Because the local observation need not be Markov and other agents may be learning, we treat Q-learning as the adaptation mechanism being studied rather than assume optimality or convergence~\cite{Claus1998,Matignon2012}.

\subsection{Separating training from spatial dynamics}
\label{sec:protocol}

The main conditions are cooperator-only learning $(0.05,0)$, defector-only learning $(0,0.05)$, and equal-rate learning $(0.05,0.05)$. Further experiments vary both rates. Each condition is trained for $200$ episodes of duration $200$, approximately $8\times10^4$ decisions per agent. Each episode starts with an independent uniform placement, resources reset to $\varphi^*$, and newly drawn decision phases, while Q-tables persist. Exploration decreases linearly from $0.3$ to $0.01$ over the first $160$ episodes and then remains constant.

After training, we freeze the tables and evaluate for $400$ time units from a held-out placement, using $\varepsilon=0.01$. This separates spatial dynamics generated by acquired policies from further adaptation. Payoff statistics use the second half, $t\in[200,400]$. Main comparisons use ten independent training seeds. Random-movement and no-movement baselines use the same evaluation placements and decision clocks. The training seed is the unit of replication, with rollouts from each seed averaged before comparison. Reported intervals are $95\%$ percentile bootstrap intervals over the ten seeds, for a mean or for a per-seed paired difference. Runs with the same seed share the placements of all training episodes, the decision phases, and the evaluation placement, whatever the learning rates or rewards, but action draws are not shared.

Two interventions then test the mechanism. Policy recombination holds one population's tables fixed while changing its partners, separating acquired training effects from movement during evaluation. Policy replay replaces private tables with compact stochastic rules estimated from recorded action frequencies, in which selected preferences can be altered while the others are retained. Full protocols are given in Appendix~\ref{sec:protocols}.

\subsection{Measurements}
\label{sec:measures}

We measure aggregation by $I_U=\mathrm{Var}_i(U_i)/\overline U$, sampled at $t=200$, $300$, and $400$. Independent uniform placement gives $I_U\approx1$, while larger values indicate concentration into fewer patches. The spatial correlation $\rho_{U\varphi}$ measures how closely cooperator concentrations align with resource peaks.

To detect travel, once per time unit we find the cyclic shift $k^*(t)$ best aligning $U(t)$ with $U(t+5)$, with velocity $v=k^*/5$ patches per time unit. We classify travel as coherent when the mean correlation after alignment exceeds $0.7$ and more than $90\%$ of windows have velocities of the same nonzero sign. Appendix~\ref{sec:measure} gives the complete algorithm. The directional action bias of type $\tau$ is $\chi_\tau=(\#R-\#L)/\#\text{decisions}$. The same expression computed for a single agent gives its individual bias.

Welfare is the sum of agents' original payoffs:
\begin{equation}
\label{eq:welfare}
    W=\sum_j\Pi_{\tau_j}=B-N_Cc-\mathcal K,\quad B=\sum_i(r_uU_i+r_vV_i)\varphi_i,\quad\mathcal K=\frac{\gamma}{\Omega}\sum_i n_i^2.
\end{equation}
Here $B$ is resource benefit and $\mathcal K$ is crowding cost. Production costs are fixed, so this decomposition identifies which of the other two terms changes when spatial organization changes.

\section{Results}
\label{sec:results}

\begin{figure*}[t]
\centering
\includegraphics[width=\textwidth]{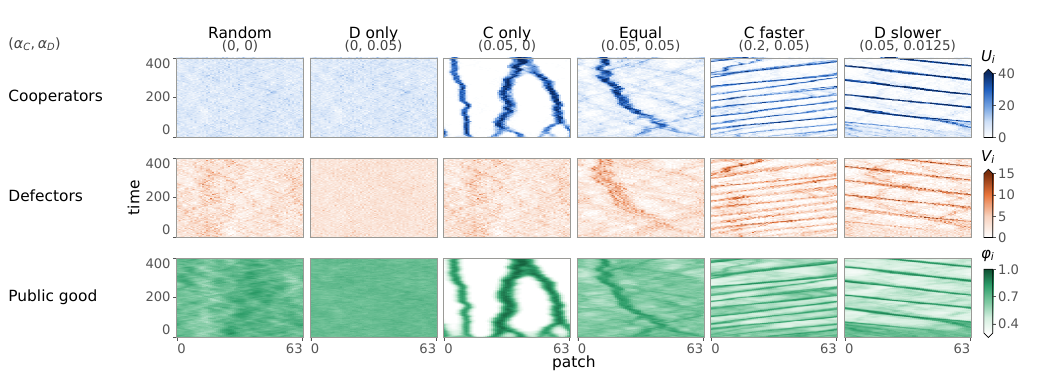}
\caption{Spatial dynamics under frozen policies. Columns show the six learning conditions and rows show cooperators, defectors, and the public good. Cooperator learning produces aggregation, while faster cooperator learning can generate traveling bands. All panels start from the same held-out placement, and learned policies are frozen after $200$ training episodes.}
\label{fig:states}
\end{figure*}

\begin{table}[t]
\caption{Frozen-policy evaluation of the main conditions (means over ten training seeds, matched placements): clustering of cooperators ($I_U$) and defectors ($I_V$), number of coherently traveling seeds, and welfare $W=B-N_Cc-\mathcal K$ with between-seed SD. Rate pairs are $(\alpha_C,\alpha_D)$. $^\dagger$Equal-rate co-trained cooperators with random defectors.}
\label{tab:main}
\small
\setlength{\tabcolsep}{3pt}
\centering
\begin{tabular}{lrrrrrr}
\toprule
Condition & $I_U$ & $I_V$ & Travel & $W$ & $B$ & $\mathcal K$\\
\midrule
No movement & 0.97 & 0.96 & 0/10 & $2199\pm4$ & 2999 & 352\\
Random movement & 0.94 & 0.97 & 0/10 & $2199\pm1$ & 2998 & 352\\
Only D learns & 0.94 & 0.45 & 0/10 & $2205\pm1$ & 2997 & 344\\
Only C learns & 24.7 & 0.97 & 0/10 & $1759\pm49$ & 3066 & 858\\
Both learn $(0.05,0.05)$ & 7.9 & 1.46 & 0/10 & $1977\pm94$ & 3001 & 576\\
C co-trained, D random$^\dagger$ & 11.0 & 0.97 & 0/10 & $2022\pm25$ & 3039 & 570\\
Both learn $(0.2,0.05)$ & 14.7 & 2.61 & 10/10 & $1783\pm129$ & 3008 & 777\\
Both learn $(0.05,0.0125)$ & 22.6 & 3.21 & 10/10 & $1599\pm169$ & 3014 & 967\\
\bottomrule
\end{tabular}
\end{table}

\subsection{Spatial organization under frozen policies}
\label{sec:states}

Figure~\ref{fig:states} shows one evaluation trajectory per condition, and Table~\ref{tab:main} summarizes the statistics over ten training seeds. Under random movement, cooperator clustering is $I_U=0.94$, close to the independent-placement baseline. With defector-only learning, cooperators retain random movement, while defector clustering falls from $I_V=0.97$ to $0.45$ and the spatial correlation between cooperator occupancy and resource availability rises from $\rho_{U\varphi}=0.32$ to $0.61$. Cooperator-only learning instead generates pronounced clusters that coincide with resource peaks, with $I_U=24.7$ and $\rho_{U\varphi}=0.83$. Equal-rate co-learning produces weaker clusters with $I_U=7.9$ and no coherent travel in any seed. When cooperators learn four times faster, the clusters form bands that travel coherently around the ring in all ten seeds of both tested rate pairs. Defector occupancy and the resource field show the same bands in the examples of Figure~\ref{fig:states}.

\subsection{Learned resource seeking and the role of training partners}
\label{sec:emerge}

\begin{figure*}[t]
\centering
\includegraphics[width=\textwidth]{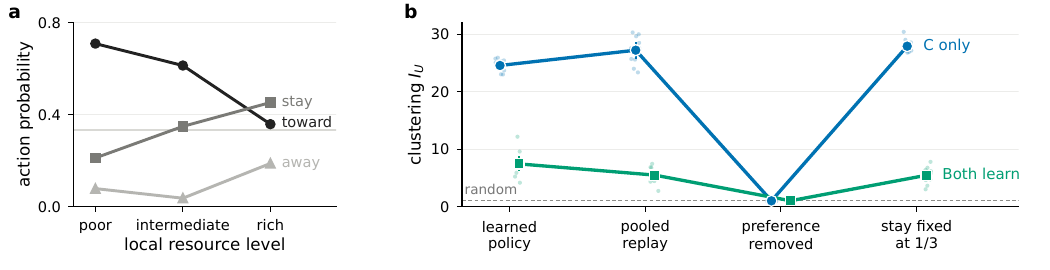}
\caption{Resource seeking drives aggregation. (a)~Under cooperator-only learning, the pooled policy favors movement toward richer patches. (b)~Replaying this preference preserves clustering, while removing it collapses clustering to the random-movement level. Fixing the stay probability at one third does not remove the pattern. Blue shows cooperator-only learning at $\alpha_C=0.05$ and green shows equal-rate co-learning at $\alpha_C=\alpha_D=0.05$.}
\label{fig:mechanism}
\end{figure*}

What have the cooperators learned under cooperator-only training? To summarize their private $729$-state tables, we pool recorded actions across cooperators and mirror-reflected observations. The resulting rule depends on the current patch's resource level and whether exactly one neighbor is classified as richer. In these contexts, we relabel the three actions as moving toward the richer neighbor, staying, or moving away from it. Otherwise, the rule retains the observed stay probability and splits the remaining probability equally between left and right. This construction removes absolute directional preferences but reveals a clear preference for richer neighbors (Figure~\ref{fig:mechanism}a). On poor patches, cooperators move toward the richer neighbor with probability $0.71$ and away with probability $0.08$. These probabilities are $0.61$ and $0.04$ on intermediate patches. On rich patches, staying is the most common action with probability $0.45$, while moving toward the richer neighbor remains more likely than moving away, with probabilities $0.36$ and $0.19$, respectively.

To test whether resource seeking supports aggregation, we replay the pooled rule and two targeted modifications of it. Each rule is estimated from the evaluation rollout of its training seed and run on five fresh rollouts with new placements and decision phases. The learned tables are run on the same rollouts, and comparisons are paired by training seed. Despite discarding occupancy observations and individual differences between agents, the pooled rule reproduces strong aggregation, with $I_U=27.3$ compared with $24.6$ under the learned tables, and also produces a welfare loss relative to random movement.

Equalizing the probabilities of moving toward and away from the richer neighbor while keeping the stay probability unchanged eliminates pronounced aggregation in all $50$ rollouts. The clustering measure $I_U$ falls to $0.8$--$1.2$, and welfare returns to approximately the random-movement level. Conversely, setting the stay probability to $1/3$ while preserving the relative probabilities of moving toward and away from the richer neighbor retains strong aggregation, with $I_U=28.0$ (Figure~\ref{fig:mechanism}b). Equal-rate co-learning gives the same qualitative result, with weaker clusters under the pooled rules that disappear when the resource-seeking preference is removed (Figure~\ref{fig:mechanism}b). Aggregation also persists under greedy evaluation of the learned tables (see Appendix~\ref{sec:eps}).

These interventions identify the preference for richer neighboring patches as necessary for aggregation in the tested distilled policies, whereas the learned dependence of staying on resource availability is not required. The resource dynamics provide a consistent feedback. A cooperator entering a patch contributes $(c-\kappa\varphi_i)/\Omega\ge0$ to the local rate of resource change (here $c=\kappa$ and $\varphi_i\leq1$), whereas a defector contributes $-\kappa\varphi_i/\Omega$. Resource-seeking cooperators can therefore reinforce the concentrations that attract further arrivals, as in prescribed-taxis models, but here the preference is acquired through individual learning.

For comparison, we evaluate three hand-designed resource-seeking rules for cooperators while defectors move randomly. On the same fresh rollouts, these rules produce stronger aggregation, with $I_U$ ranging from $39.9$ to $51.3$ compared with $24.6$ under the learned policies, and lower welfare, with $W$ ranging from $1105$ to $1378$ compared with $1742$. None of these mirror-symmetric rules produces coherent travel (see Appendix~\ref{sec:hand}).

Training partners change what cooperators learn. Equal-rate co-learning yields weaker clustering with $I_U=7.9$ compared with $24.7$ under cooperator-only learning. This difference could arise from the cooperator policies acquired during training or the learned defectors' behavior during evaluation, or both. To distinguish these effects, we pair co-trained cooperators with random defectors. Clustering rises to $I_U=11.0$ but remains well below $24.7$, showing that co-training changes the cooperator policies themselves. Reintroducing learned defectors reduces mean clustering from $11.0$ to $7.9$, with a reduction in every seed, without a detectable welfare improvement. Welfare is $1977$ with learned defectors versus $2022$ with random defectors, giving a paired difference of $-45$ with a $95\%$ interval of $[-96,4]$.

The same learned defector policies have the opposite effect when paired with cooperators trained with random defectors. Clustering increases from $24.7$ to $45.6$, and welfare falls from $1759$ to $883$. Thus, the effect of learned defectors depends on the cooperator policies they encounter, which in turn depend on the partners present during training (see Appendix~\ref{sec:env-app}).

\subsection{Learning-rate dependence and directional conventions}
\label{sec:rates}

\begin{figure*}[t]
\centering
\includegraphics[width=\textwidth]{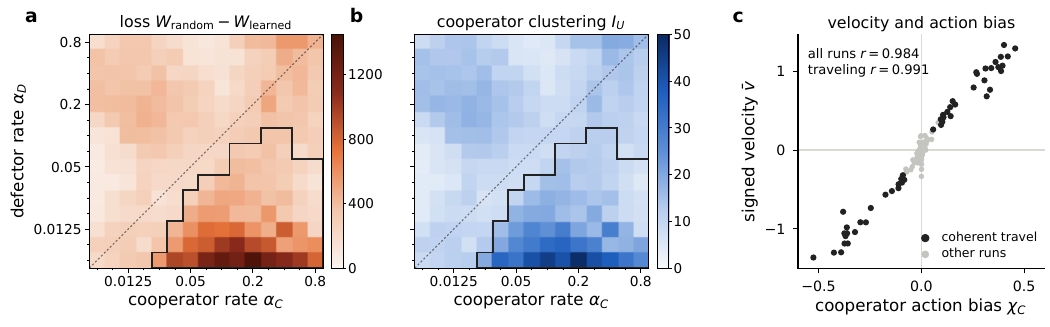}
\caption{Learning-rate dependence and directional bias after $200$ training episodes. 
Panels (a) and (b) show welfare loss relative to random movement, $W_{\mathrm{random}}-W_{\mathrm{learned}}$, and cooperator clustering $I_U$, averaged over ten training seeds and twenty seeds when $\alpha_D=0.05$. 
The black outline marks cells in which at least half of the seeds travel coherently, and the dashed line marks $\alpha_C=\alpha_D$. 
Panel (c) shows signed velocity against cooperator left--right action bias for the $172$ runs from the main conditions and the $5\times5$ rate grid. 
These maps summarize finite-budget outcomes rather than an asymptotic phase diagram.}
\label{fig:rates}
\end{figure*}

We next vary both learning rates over fifteen values between $0.00625$ and $0.8$, with ten training seeds per rate pair, twenty at $\alpha_D=0.05$, and the same $200$-episode budget (Figure~\ref{fig:rates}a, b). Across cells with learning cooperators, mean clustering ranges from $I_U=3.5$ to $49$, while mean welfare losses relative to random movement range from 74 to 1417. Cell means of $W$ and $I_U$ are strongly negatively correlated with $r=-0.99$. Losses are largest at high cooperator and low defector learning rates, but coherent travel occupies only part of this region. All $69$ cells where at least half the seeds travel satisfy $\alpha_C\ge2\alpha_D$ and $\alpha_D\le0.1$. No seed travels when $\alpha_C\le\alpha_D$, and at most one in ten travels when $\alpha_D\ge0.14$. A finer sweep along $\alpha_D=0.05$ shows traveling counts rising from $0$ to $19$ of $20$ as $\alpha_C$ increases from $0.079$ to $0.126$ (see Appendix~\ref{sec:sweep}). These observed boundaries depend on the training budget and shift with further training (Section~\ref{sec:training}).

Travel is accompanied by a shared directional preference despite the environment's left--right symmetry. Across the $172$ runs from the main conditions and a coarser $5\times5$ rate grid, signed velocity correlates with cooperator action bias $\chi_C$ at $r=0.984$, rising to $r=0.991$ within the $50$ traveling runs. Their signs agree in every traveling run (Figure~\ref{fig:rates}c). Averaged across traveling runs, $96\%$ of cooperators and $86\%$ of defectors favor the band direction, and these preferences persist even in mirror-symmetric observations. We call this shared, spontaneously selected preference a directional convention. Because bias and velocity are measured on the same rollout, their correlation alone does not establish causality. We therefore test the role of this preference through policy interventions.

Unlike the pooled rule of Section~\ref{sec:emerge}, the directional rule retains distinct left and right actions, with $27$ contexts defined by the resource level of the current patch and the resource differences from its two neighbors. We mirror-symmetrize this rule by averaging it with its reflection, exchanging both the neighboring observations and the left and right actions. This removes absolute directional preference while retaining a response to relative resource availability. Among the $50$ runs whose original learned policies produce coherent travel, replay of the directional rule preserves travel in $49$, whereas its symmetrized version produces none. Mean speed falls from $0.84$ under the original learned policies to $0.12$ under the symmetrized rule, while substantial aggregation remains, with $I_U=15.6$ compared with $16.3$. On $100$ fresh rollouts across the two ratio-$4$ conditions, the original policies, directional rules, and symmetrized rules produce coherent travel in $99$, $100$, and $0$ rollouts, respectively (see Appendix~\ref{sec:rules}).

We also remove directional preferences directly from the original policies, without distillation. For each agent, we use $\pi_j^{\mathrm{sym}}(a\mid s)=\tfrac12[\pi_j(a\mid s)+\pi_j(\mathcal M a\mid\mathcal M s)]$, where $\mathcal M$ swaps the two neighboring observations and the left and right actions. This retains all six observation components and a separate policy for each agent. Symmetrizing both populations reduces coherent travel from $49/50$ to $0/50$ fresh rollouts at $\alpha_C=0.2$ and $\alpha_D=0.05$, while clustering falls from $I_U=14.8$ to $7.3$. At $\alpha_C=0.05$ and $\alpha_D=0.0125$, travel falls from $50/50$ to $2/50$, while clustering remains strong at $I_U=20.9$ compared with $22.6$.

To test whether added randomness alone explains this result, we add uniform exploration matched to the mean action-disagreement rate. It also suppresses travel, leaving $4/50$ coherent rollouts in each condition, but reduces clustering much further to $I_U=3.0$ and $2.5$, respectively. Weaker exploration already reduces clustering below the symmetrized levels while most rollouts still travel (see Appendix~\ref{sec:mirror}). This contrast supports a specific role for directional preference, although the control matches only the average rate of action changes, not their dependence on the observation.

Symmetrizing only defectors keeps travel in all $100$ rollouts. Symmetrizing only cooperators leaves $38/50$ traveling rollouts in the first condition and $6/50$ in the second, with slower bands in the first. The populations' contributions therefore depend on the learning rates. Together, the interventions distinguish resource seeking that supports aggregation in the tested distilled rules from directional preferences that support coherent travel in the original policies.

\subsection{Traveling bands under continued training}
\label{sec:training}

\begin{figure}[t]
\centering
\includegraphics[width=0.7\columnwidth]{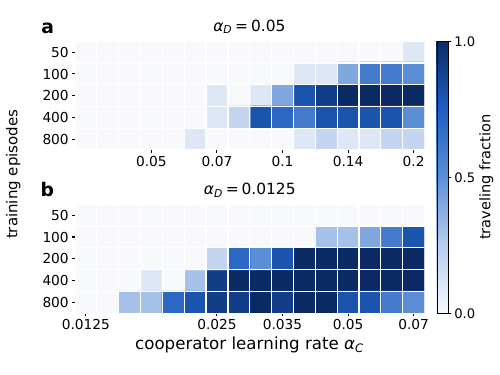}
\caption{Training duration reshapes coherent travel. 
Each heatmap shows the fraction of ten training seeds whose frozen policies produce coherent travel. Panel (a) uses \(\alpha_D=0.05\), and panel (b) uses \(\alpha_D=0.0125\). Travel appears, disappears, and shifts across learning rates as training continues, with different behavior at the two defector learning rates. These finite-training results do not define an asymptotic phase boundary.}
\label{fig:trainlength}
\end{figure}

The preceding results use policies frozen after $200$ episodes. To examine longer training, we train ten seeds at each of sixteen $\alpha_C$ values for each of $\alpha_D=0.05$ and $\alpha_D=0.0125$. We evaluate frozen policies after $50$, $100$, $200$, $400$, and $800$ episodes (Figure~\ref{fig:trainlength}). After $50$ episodes, at most one seed per condition produces coherent travel. Longer training allows travel at smaller $\alpha_C$. At $\alpha_C=0.025$ and $\alpha_D=0.0125$, the traveling counts at the five checkpoints are $0$, $0$, $2$, $9$, and $9$ out of ten. Travel can also disappear with further training. At $\alpha_C=0.141$ and $\alpha_D=0.05$, the corresponding counts are $0$, $4$, $10$, $8$, and $1$. After $800$ episodes, at most two seeds per condition travel along $\alpha_D=0.05$, even though mean clustering across this sweep increases from $I_U=10.5$ at episode $200$ to $16.5$.

We then extend the two conditions with $\alpha_C/\alpha_D=4$ to $1600$ episodes using five seeds each, alongside three seeds each for cooperator-only and equal-rate learning. We evaluate frozen policies every $20$ episodes (see Appendix~\ref{sec:seeds}). At $\alpha_C=0.05$ and $\alpha_D=0.0125$, most later checkpoints produce coherent travel, with one direction reversal per seed. At $\alpha_C=0.2$ and $\alpha_D=0.05$, travel appears intermittently as directional preferences repeatedly weaken and reverse. These changes compare policies learned at different checkpoints, not the motion of one continuously adapting band. Equal-rate learning produces travel at only $1$ of $213$ seed-checkpoints from episode $200$ onward, but aggregation grows with further training. Over episodes $1400$ to $1600$, mean clustering reaches $I_U=20.3$ compared with $24.9$ under cooperator-only learning. Mean welfare falls to $1587$ compared with $1731$, reversing their welfare ranking at episode $200$. Nevertheless, welfare stays below random movement at every checkpoint from episode $20$ onward in every long run and at all $1600$ seed-checkpoints of the training-length sweep. Thus, travel and welfare rankings change with training duration, while welfare losses persist across these evaluations. These finite runs establish neither a fixed threshold for travel nor an asymptotic outcome of learning.

\subsection{Crowding costs and the effect of a toll}
\label{sec:welfare}

\begin{figure*}[t]
\centering
\includegraphics[width=\textwidth]{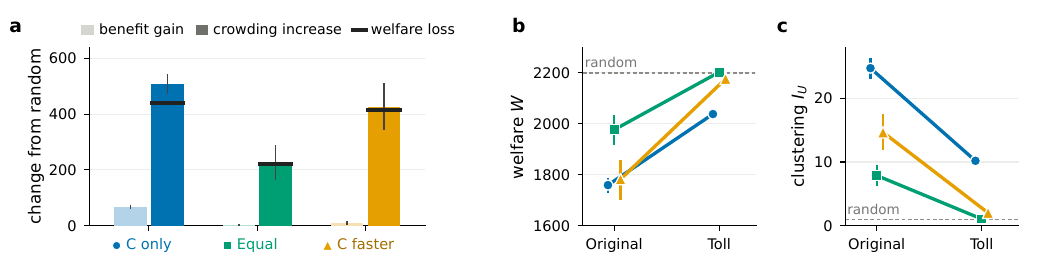}
\caption{Crowding drives the welfare loss, and a toll largely reverses it. (a)~Gain in resource benefit and increase in crowding cost relative to random movement. Black markers show the resulting welfare loss. (b)~Welfare and (c)~clustering after training with the original reward and with the crowding toll, which raises welfare and reduces clustering in all three conditions. Colors and marker shapes mark cooperator-only learning, equal-rate learning, and $(\alpha_C,\alpha_D)=(0.2,0.05)$, as labeled in panel a.}
\label{fig:toll}
\end{figure*}

After $200$ training episodes, every seed of the untolled conditions with cooperator learning in Table~\ref{tab:main} has lower welfare than both the random- and no-movement baselines, whose mean welfare is approximately $2199$. Mean welfare is also below the random-movement baseline in every cell of the learning-rate map.

The welfare decomposition shows that increased crowding accounts for most of the welfare loss (Table~\ref{tab:main} and Figure~\ref{fig:toll}a). Random movement gives resource benefit $B=2998$ and crowding cost $\mathcal K=352$. Across the clustered conditions, benefit ranges from $3001$ to $3066$, while crowding cost reaches $967$ and production costs remain fixed. Under cooperator-only learning, the mean resource level experienced by cooperators rises from $0.70$ to $0.85$, but their mean payoff falls from $3.38$ to $3.23$ relative to random movement.

Resource accounting explains why aggregation changes aggregate benefit only modestly. Total production is always $cN_C$. Defining $\Phi(t)=\sum_i\varphi_i(t)$, summing~\eqref{eq:phi} over the ring and averaging over an evaluation window $[t_0,t_1]$ of duration $T$ yields
\begin{equation}
\label{eq:resource_balance}
    \kappa\left\langle\sum_i n_i\varphi_i\right\rangle=cN_C-\delta\Omega\langle\Phi\rangle-\frac{\Omega}{T}\bigl[\Phi(t_1)-\Phi(t_0)\bigr].
\end{equation}
Decay and net changes in storage are small relative to production, leaving total uptake nearly unchanged across the evaluated conditions. The measured mean of $\sum_i n_i\varphi_i$ ranges from $446.7$ to $448.1$ across all seeds of the main and tolled conditions (see Appendix~\ref{sec:measure}).

Writing $B=r_v\sum_i n_i\varphi_i+(r_u-r_v)N_C\bar\varphi_C$, where $\bar\varphi_C$ is the mean resource level experienced by cooperators, movement can then change aggregate benefit mainly through the resource access of cooperators. Crowding cost instead grows with the spatial variance of occupancy $\mathcal K=\gamma N^2/(M\Omega)+(\gamma/\Omega)\sum_i(n_i-N/M)^2$, so aggregation can raise it substantially without a comparable gain in benefit.

Individual rewards account for an agent's own crowding cost but not the crowding it imposes on others. An agent's presence adds $\gamma/\Omega$ to the cost of each of the other $n_i-1$ occupants of its patch. To test whether accounting for this externality changes the learned organization, we subtract a toll $\gamma(n_i-1)/\Omega$ per unit time from the learning reward. This toll prices the direct crowding imposed on others, but not the effects of movement on resource production and consumption. At the reward-function level, this corresponds to doubling the crowding coefficient and adding an action-independent constant; finite-training trajectories are not exactly equivalent because the initial Q-values are not shifted accordingly (see Appendix~\ref{sec:rules}). Evaluation continues to use the original payoff in~\eqref{eq:payoff}, so the reported welfare changes are not changes in the evaluation objective.

With equal-rate co-learning, the toll reduces $I_U$ from $7.9$ to $1.06$ and raises welfare from $1977$ to $2203$, close to the random-movement baseline. The paired welfare increase is $226$, with a $95\%$ interval of $[170,285]$. At $(0.2,0.05)$, the toll reduces clustering from $14.7$ to $2.0$ and raises welfare from $1783$ to $2176$, approximately $1\%$ below the baseline. Neither tolled condition produces coherent travel, and welfare improves in every training seed (Figure~\ref{fig:toll}b, c).

When only cooperators learn, the toll improves welfare in all ten seeds, raising the mean from $1759$ to $2038$ and recovering approximately $63\%$ of the loss relative to random movement. Mean cooperator payoff reaches $3.59$ and exceeds the random-movement baseline in every seed. However, substantial clustering remains with $I_U=10.2$, and defectors still experience lower resource availability than under random movement.

At five further rate pairs with large untolled losses, the toll raises welfare in all ten seeds, but does not always restore the random-movement baseline. Recovery is nearly complete at $(0.05,0.0125)$, from $1599$ to $2193$ with no traveling seed, and partial at $(0.2,0.00625)$, from $781$ to $1617$ with eight seeds still traveling (see Appendix~\ref{sec:sweep}).

\subsection{Robustness}
\label{sec:robust}

At the same $200$-episode budget, the three core conditions retain their clustering and welfare orderings across separately tested variants with three training seeds each. These use rings of $32$ or $128$ patches at fixed densities, doubled resource diffusion, or defector valuation $r_v=6.5$. The ordering also persists with a $225$-state observation based on neighbor differences. Adding own-patch occupancy gives $675$ states and preserves the ordering of the two conditions tested. Coherent travel is more sensitive. At $\alpha_C=0.2$ and $\alpha_D=0.05$, it occurs in all three seeds at each alternative ring size, two with doubled diffusion, and none with $r_v=6.5$ (see Appendices~\ref{sec:env-app} and~\ref{sec:robust-app}).

\section{Discussion and Future Work}
\label{sec:discussion}

Our results show how learned movement connects individual adaptation to collective spatial organization. Previous spatial public goods models prescribe responses to payoffs or resource gradients~\cite{Wakano2009,Funk2019,yao2026pattern,Zhao2026pattern}, while classical chemotaxis models generate traveling bands through prescribed responses to a substrate that organisms also modify~\cite{KellerSegel1971Bands}. Here, resource seeking emerges through individual learning, and removing this preference eliminates aggregation in the tested distilled policies. Policy interventions also identify a shared directional preference as an important contributor to coherent travel. Despite the environment's left--right symmetry, learners can acquire a common directional preference whose removal strongly suppresses travel while retaining substantial aggregation.

The resulting patterns depend on how long and how quickly the agents learn. Traveling bands are reproducible across training seeds and fresh rollouts at the tested checkpoints, but their occurrence depends on both learning rates and training duration. Successive frozen policies can sustain travel, lose it, or generate movement in the opposite direction. These observations characterize finite-training outcomes rather than establish convergence to an equilibrium. The welfare loss is more consistent across the tested conditions. With production independent of location and total uptake nearly unchanged, aggregation increases crowding costs much more than aggregate resource benefit. Mean welfare remains below the random-movement baseline in every tested untolled learning-rate condition with cooperator learning. Charging agents for the crowding they impose on others recovers much of this loss at the evaluated training budget.

Future work could connect frozen policies to continuum descriptions with effective diffusion or taxis terms~\cite{yao2026pattern}. Other graph topologies may alter aggregation and collective motion by changing resource transport and agent connectivity. Learning contribution and movement jointly, or using function approximation, would test whether the identified mechanisms extend beyond fixed social types and tabular policies.

\begin{quote}
\begin{center}
\textbf{Acknowledgments}
\end{center}
We thank Prof. Daniel B. Cooney at the University of Illinois Urbana-Champaign for helpful discussions, insightful suggestions on the role of learning rates, and for pointing us to relevant literature.
\end{quote}

\begin{quote}
\begin{center}
\textbf{Declaration on the Use of AI}
\end{center}
We used Claude Fable 5.1 (Anthropic) and GPT-6 Astra (OpenAI) as assistive tools for language editing and for drafting and optimizing code for the numerical experiments. The authors reviewed and checked the AI-assisted content and code, and take full responsibility for the manuscript and reported results.
\end{quote}

\bibliographystyle{plainnat}
\bibliography{refs}

\appendix

\section*{Summary of Appendix Sections}

Section~\ref{sec:numerics} gives the numerical scheme and the learning update in full, Section~\ref{sec:measure} the measurements, including the complete algorithm that classifies coherent travel, and Section~\ref{sec:protocols} the protocols of policy recombination, distillation, and replay. Section~\ref{sec:main} gives the complete main table and seed-level intervals for the main comparisons. Section~\ref{sec:figs} collects the figures referred to in the main text: training curves, policy recombination, the directional replay on one fresh rollout, the complete learning-rate and directional-bias analysis, the aligned band profiles, and the long-training trajectories. Section~\ref{sec:rules} gives the distilled rules, the replay of every trained run, the per-agent analysis of the directional convention, and the crowding-toll runs per seed. Section~\ref{sec:grid} gives the learning-rate grid with traveling counts. Sections~\ref{sec:eps}--\ref{sec:mirror} report three additional controls: the sensitivity of the frozen policies to evaluation exploration, hand-designed movement rules, and the mirror-symmetrization of the original learned tables. Section~\ref{sec:env-app} gives the environment variants. Section~\ref{sec:seeds} gives per-seed values, Section~\ref{sec:robust-app} the observation variants, Section~\ref{sec:sweep} the finer learning-rate sweep, the training-length runs, and the additional toll runs, and Section~\ref{sec:code} the code and what is needed to reproduce every number.

Notation follows the main text. Agents are indexed by $j$, patches by $i$, and actions by $a$. All runs use the parameters of Table~\ref{tab:params} of the main text unless stated otherwise, and rate pairs are written as $(\alpha_C,\alpha_D)$. Seeds $0$--$9$ index both the training stream and the held-out evaluation placement. For a given seed, every policy starts from the same placement and uses the same decision phases. This holds for no movement, random movement, learned tables, and replayed rules.

\paragraph{Evaluation sets.} Frozen policies are evaluated in three settings, and the same condition gives slightly different numbers in each. Set \emph{E} is the held-out evaluation placement of each training seed, with one rollout per seed. It is used for the main table, the grid, the toll runs, the recombination, and the broad-scan replay. Set \emph{F} consists of five fresh rollouts per training seed with new placements and phases. It is used for the fresh-rollout replay, the exploration test, the hand-designed rules, and the mirror-symmetrized tables. Set \emph{L} covers the long-training runs, which are evaluated on \emph{E} at every checkpoint with three or five seeds. For example, cooperator-only learning has $W=1759$ on \emph{E} and $1742$ on \emph{F}.

\section{Numerical scheme and learning update}
\label{sec:numerics}

One step $t_m\to t_{m+1}$ performs, in order:
\begin{enumerate}\itemsep0pt
\item \emph{Decisions.} Every agent $j$ with $m\equiv\theta_j\pmod K$ observes its state $s$ from the configuration $(\varphi(t_m),n(t_m))$ before any move of this step. If it has already decided in this episode, it applies the Q-update to its previous state--action pair with the accumulated reward $R$ and next state $s$. It then chooses an action $\varepsilon$-greedily. With probability $\varepsilon$ the action is drawn uniformly from the three actions. Otherwise the agent takes a maximizer of its own table, with ties broken uniformly. Its reward accumulator is then reset to zero.
\item \emph{Moves.} All chosen moves are executed simultaneously, and $U$, $V$, and $n$ are updated. Agents that do not decide in this step stay where they are. This waiting is not a stay action and is not updated.
\item \emph{Public good.} The public good equation is integrated at the new occupancies with two explicit Euler substeps of length $\Delta t/2$, and negative values are clipped to zero after each substep. The step-averaged resource is the trapezoidal average of the three stages, $\bar\varphi^{(m)}=\tfrac14\varphi^{(0)}+\tfrac12\varphi^{(1/2)}+\tfrac14\varphi^{(1)}$.
\item \emph{Rewards.} Every agent adds $\Delta t\,\Pi_{\tau_j}(\bar\varphi^{(m)}_{x_j},n_{x_j})$ to its accumulator, using its post-move position and occupancy. In the toll runs it also subtracts $\Delta t\,\gamma(n_{x_j}-1)/\Omega$.
\end{enumerate}

The diffusion number is $D_\varphi\Delta t/(2\ell^2)=0.048$, well inside the stability bound $1/2$ of the explicit scheme. Each Euler substep writes the new value as a combination of $\varphi_i$, $\varphi_{i\pm1}$, and the production term. The coefficients are nonnegative whenever $(\Delta t/2)(\kappa n_i/\Omega+\delta+2D_\varphi/\ell^2)<1$. This holds for every possible occupancy of the main model, since $n_i\le N=640$ gives $0.90$. In the environment variants the worst case is not always covered. On the $128$-patch ring, for example, all $1280$ agents on one patch would violate it. The occupancies observed in the frozen evaluations of the variants do satisfy the condition, although training trajectories were not checked. The largest observed occupancy is $72$ for $M=32$, $135$ for $M=128$, $83$ with doubled diffusion, and $86$ with $r_v=6.5$. The corresponding coefficients are $0.81$, $0.74$, $0.70$, and $0.80$, and the largest concentration is below $0.98$. These checks are in \texttt{checks.py}.

Two consequences follow. First, the concentration stays nonnegative, so the clipping never acts. Second, since $c=\kappa$ and $U_i\le n_i$, the bound $\varphi\le1$ is preserved along every trajectory that starts below $1$, and all runs start from $\varphi^*<1$. For the continuous equation this is the maximum principle. At a spatial maximum with $\varphi_i=1$, the diffusion term is nonpositive and $cU_i/\Omega-(\kappa n_i/\Omega+\delta)\le-\delta<0$. For the discrete scheme, if all $\varphi_i\le1$ before a substep, then $\varphi_i\le1-(\Delta t/2)\big((n_i-U_i)/\Omega+\delta\big)\le1$ after it. The largest value in any frozen evaluation of the $172$ trained runs is $0.98$. The same-location payoff gap $\Pi_D-\Pi_C=c-(r_u-r_v)\varphi_i=1-\varphi_i$ is nonnegative in every reachable state of the main model. In the $r_v=6.5$ variant the gap is $1-0.5\varphi_i$, which is also nonnegative.

\paragraph{Episode boundaries.} At the start of every episode the positions are redrawn uniformly, $\varphi_i$ is reset to $\varphi^*$, and the phases $\theta_j$ are redrawn. The reward accumulators and previous state--action pairs are cleared, while the Q-tables persist. The first decision of an agent in an episode has no predecessor and triggers no update. The payoff collected after an agent's last decision of the episode covers an unfinished interval of fewer than $K$ steps and is discarded. Each agent makes $T_{\mathrm{ep}}/\tau_{\mathrm{dec}}=400$ decisions and $399$ updates per episode. Exploration in training episode $e$ is $\varepsilon_e=0.3-0.29\min\{1,e/(0.8E)\}$.

\paragraph{Observation levels.} The resource levels are $\varphi_i\le0.65$, $0.65<\varphi_i\le0.75$, and $\varphi_i>0.75$. The occupancy levels are $n_i\le7$, $8$--$12$, and $n_i\ge13$. A resource difference $\varphi_{i\pm1}-\varphi_i$ is lower if it is at most $-0.01$, similar if it lies in $(-0.01,0.01]$, and higher if it exceeds $0.01$. An occupancy difference $n_{i\pm1}-n_i$ is lower if it is at most $-3$, similar from $-2$ to $2$, and higher if it is at least $3$. Under random movement, agents see poor, intermediate, and rich patches with frequencies of 16\%, 70\%, and 14\%. The three occupancy levels occur with frequencies of 22\%, 57\%, and 21\% over patches and 13\%, 57\%, and 30\% as seen by agents, who are more often found on crowded patches.

\paragraph{Reward and reported payoff.} The learning reward uses the step-averaged resource $\bar\varphi^{(m)}$. Reported payoffs are computed from the configuration $(U,V,\varphi)$ stored once per time unit. In the main table, the toll comparison, the policy recombination, the exploration test, the hand-designed rules, and the mirror-symmetrized tables, $W$, $B$, $\mathcal K$, $\Pi_\tau$, and $\varphi_\tau$ are averages over the $201$ snapshots at $t=200,201,\dots,400$ of the frozen-policy evaluation. In the grid, replay, long-training, environment-variant, and observation-variant results, $W$ is the average of the instantaneous welfare over all numerical steps of the same interval. The two averages differ by about one unit of $W$ on average and by up to three units in a single seed. For equal-rate learning, for example, they give $1977$ and $1976$. This is why the same condition can differ in the last digit between tables. The no-movement baseline differs from $W^*$ because the random placement is not exactly uniform.

\section{Measurements}
\label{sec:measure}

\paragraph{Clustering and alignment.} $I_U=\mathrm{Var}_i(U_i)/\overline U$ and the Pearson correlation $\rho_{U\varphi}$ between $U_i$ and $\varphi_i$ over patches are averaged over the three snapshots $t=200,300,400$.

\paragraph{Coherent travel.} Let $U(t)\in\mathbb{R}^M$ be the cooperator profile at integer time $t$ and $\tilde U(t)$ the profile minus its spatial mean. For each window start $t=200,201,\dots,395$ ($196$ windows) and lag $\Delta=5$ we compute, for every cyclic shift $k\in\{-M/2,\dots,M/2-1\}$,
\[
    c_k(t)=\frac{\sum_i \tilde U_{i-k}(t)\,\tilde U_i(t+\Delta)}{\|\tilde U(t)\|\,\|\tilde U(t+\Delta)\|},
\]
and take $k^*(t)=\arg\max_k c_k(t)$. If several shifts attain the maximum, the smallest $k$ is taken. The window velocity is $v(t)=k^*(t)/\Delta$ patches per time unit, positive toward increasing patch index, and the largest detectable speed is $M/(2\Delta)=6.4$. We report the speed $|v|=\langle|v(t)|\rangle_t$, the signed velocity $\bar v=\langle v(t)\rangle_t$, the best-shift correlation $r=\langle c_{k^*(t)}(t)\rangle_t$, and the main-direction share
\[
    d=\max\Big\{\tfrac{1}{196}\#\{t: v(t)>0\},\ \tfrac{1}{196}\#\{t: v(t)<0\}\Big\}.
\]
Windows with $k^*(t)=0$ count toward neither direction, so a stationary profile has $d$ near zero. A run \emph{travels coherently} if $r>0.7$ and $d>0.9$. Both conditions are needed. A fluctuating profile without structure has a large speed because the best shift is arbitrary, but its correlation is low and it has no dominant direction. Under random movement, for example, $|v|=1.9$, $r=0.31$, and $d=0.46$. Stationary clusters have a high correlation but $d$ near zero. The classification is not sensitive to the thresholds on $r$ and $d$. Among the $172$ trained runs, $50$ are coherent at $(r,d)>(0.7,0.9)$, $50$ at $(0.6,0.9)$, $48$ at $(0.8,0.9)$, $56$ at $(0.7,0.8)$, and $46$ at $(0.7,0.95)$. Proportions of traveling runs are computed over training seeds, never over the correlated time windows of one run.

\paragraph{Directional bias.} $\chi_\tau=(\#R-\#L)/\#\text{decisions}$ over all decisions of type $\tau$ during the $400$ time units of the evaluation, including stays in the denominator. The per-agent bias uses the decisions of one agent.

\paragraph{Crowding cost.} Because the total number of agents is fixed,
\[
    \mathcal K=\frac{\gamma}{\Omega}\sum_i n_i^2=\frac{\gamma N^2}{M\Omega}+\frac{\gamma}{\Omega}\sum_i\Big(n_i-\frac NM\Big)^2 .
\]
The first term is $320$ for the parameters used, and the second is the spatial variance of the total occupancy. Random movement gives $\mathcal K\approx352$ and the clustered conditions up to $967$. $\mathcal K$ depends on the occupancy of both types, whereas $I_U$ measures cooperators only, so the two need not move together.

\paragraph{Resource benefit.} Summing the public good equation over patches cancels the diffusion term and gives $\frac{d}{dt}\sum_i\varphi_i=cN_C/\Omega-\sum_i(\kappa n_i/\Omega+\delta)\varphi_i$. Over an evaluation window in which $\sum_i\varphi_i$ does not drift, total uptake balances production minus decay, $\kappa\langle\sum_in_i\varphi_i\rangle=cN_C-\delta\Omega\langle\sum_i\varphi_i\rangle\approx448-0.9$, whatever the spatial arrangement. Since $B=r_v\sum_in_i\varphi_i+(r_u-r_v)\sum_iU_i\varphi_i=r_v\sum_in_i\varphi_i+(r_u-r_v)N_C\bar\varphi_C$, under the approximation of fixed total uptake, the benefit can change only through the resource level $\bar\varphi_C$ experienced by cooperators. We recover $\sum_in_i\varphi_i=(B-(r_u-r_v)N_C\bar\varphi_C)/r_v$ from the stored $B$ and $\bar\varphi_C$. It lies between $446.7$ and $448.1$ in every seed of the eleven rows of Table~\ref{tab:s_main}, which cover the baselines and the learned and tolled conditions. The approximation $B\approx r_v\cdot447.1+(r_u-r_v)N_C\bar\varphi_C$ reproduces the condition means of $B$ to within four units. Under this approximation and because $\varphi\le1$ (Section~\ref{sec:numerics}), $B$ can exceed its random-movement value by at most $(r_u-r_v)N_C(1-0.705)\approx132$, and $\mathcal K$ can fall at most $32$ below it. Under the approximation of fixed total uptake, the estimated maximum welfare gain over random movement is about $165$.

\begin{table*}[p]
\centering
\small

\caption{Complete version of Table~\ref{tab:main} of the main text.
Values are means over ten training seeds, and $W$ also gives the
between-seed standard deviation. $I_U$ is clustering,
$\rho_{U\varphi}$ the spatial correlation of cooperator occupancy
and public good, $|v|$ the speed, trav.\ the number of coherently
traveling seeds, and $|\chi_C|$ the absolute left--right action
bias of the cooperators. Welfare is $W=B-N_Cc-\mathcal K$ with
resource benefit $B$ and crowding cost $\mathcal K$.
$\Pi_\tau$ and $\varphi_\tau$ are the mean payoff and the mean
public good experienced by type $\tau$.
Rate pairs are $(\alpha_C,\alpha_D)$.}
\label{tab:s_main}

\medskip
\setlength{\tabcolsep}{4pt}

\begin{adjustbox}{max width=\textwidth,max totalheight=0.72\textheight}
\rotatebox{90}{%
\begin{tabular}{lrrrrr|rrrr|rrrr}
\toprule
condition
& $I_U$
& $\rho_{U\varphi}$
& $|v|$
& trav.
& $|\chi_C|$
& $W$
& (sd)
& $B$
& $\mathcal K$
& $\Pi_C$
& $\Pi_D$
& $\varphi_C$
& $\varphi_D$ \\
\midrule
No movement
& 0.97 & 0.36 & 0.00 & 0/10 & --
& 2199 & 4 & 2999 & 352
& 3.39 & 3.55 & 0.71 & 0.68 \\

Random movement
& 0.94 & 0.32 & 1.85 & 0/10 & --
& 2199 & 1 & 2998 & 352
& 3.38 & 3.56 & 0.70 & 0.68 \\
\midrule
Only D learns
& 0.94 & 0.61 & 1.85 & 0/10 & 0.00
& 2205 & 1 & 2997 & 344
& 3.36 & 3.65 & 0.70 & 0.69 \\

Only C learns
& 24.7 & 0.83 & 0.05 & 0/10 & 0.01
& 1759 & 49 & 3066 & 858
& 3.23 & 1.63 & 0.85 & 0.36 \\

Both learn, $(0.05,0.05)$
& 7.9 & 0.88 & 0.16 & 0/10 & 0.01
& 1977 & 94 & 3001 & 576
& 3.00 & 3.30 & 0.71 & 0.67 \\

\quad C co-trained, D random
& 11.0 & 0.84 & 0.12 & 0/10 & 0.01
& 2022 & 25 & 3039 & 570
& 3.49 & 2.39 & 0.79 & 0.48 \\

Both learn, $(0.2,0.05)$
& 14.7 & 0.91 & 1.06 & 10/10 & 0.33
& 1783 & 129 & 3008 & 777
& 2.73 & 2.92 & 0.73 & 0.63 \\

Both learn, $(0.05,0.0125)$
& 22.6 & 0.88 & 0.97 & 10/10 & 0.32
& 1599 & 169 & 3014 & 967
& 2.44 & 2.63 & 0.74 & 0.60 \\
\midrule
\multicolumn{14}{l}{\emph{Learning with a crowding toll}} \\

Only C learns
& 10.2 & 0.88 & 0.05 & 0/10 & 0.01
& 2038 & 22 & 3044 & 558
& 3.59 & 2.24 & 0.80 & 0.46 \\

Both learn, $(0.05,0.05)$
& 1.06 & 0.66 & 0.93 & 0/10 & 0.00
& 2203 & 3 & 2997 & 347
& 3.37 & 3.61 & 0.70 & 0.69 \\

Both learn, $(0.2,0.05)$
& 2.0 & 0.81 & 0.25 & 0/10 & 0.01
& 2176 & 9 & 3000 & 376
& 3.35 & 3.52 & 0.71 & 0.68 \\
\bottomrule
\end{tabular}%
}
\end{adjustbox}

\end{table*}

\section{Policy interventions}
\label{sec:protocols}

\paragraph{Recombination.} Each agent owns its table, so cooperator tables from one run can be evaluated with defector tables from another run of the same seed, or with untrained tables that move at random. Row $^\dagger$ of the main table evaluates the cooperator tables of the equal-rate run with untrained defector tables. The reverse pairing combines the defector tables of the equal-rate run with the cooperator tables of the cooperator-only run of the same seed.

\paragraph{Distillation.} During the frozen evaluation of a trained population, on the placement and phases of its seed, every decision is recorded with its context. The action frequencies per context and type define a stochastic rule shared by all agents of that type. The \emph{absolute rule} uses $27$ contexts, formed by the level of $\varphi_i$ and the levels of $\varphi_{i-1}-\varphi_i$ and $\varphi_{i+1}-\varphi_i$. Its actions are stay, left, and right. The \emph{pooled rule} merges mirror-image contexts into six, formed by the level of $\varphi_i$ and whether exactly one neighbor is in the higher level. If so, the actions are up, stay, and down relative to that neighbor. Otherwise the agent stays or moves, with left and right equally likely. The \emph{symmetrized rule} averages the absolute rule with its reflection, which exchanges the two neighbor differences and the labels left and right. It is applied to both types. Contexts that were never visited receive the uniform rule. Visited contexts are not smoothed, and every decision has equal weight. The recorded decisions include the $1\%$ exploratory actions of the evaluation, so a rule is replayed without additional exploration. None of the rules uses the occupancy components of the observation. A distilled rule is a different, smaller model that reproduces the collective behavior. It is not an equivalent representation of the tables, and interventions on it show what suffices and what is necessary within that model. Section~\ref{sec:mirror} gives an intervention on the original tables.

\paragraph{Ablations of the pooled rule.} \emph{Equal up/down}: in the three contexts with exactly one richer neighbor, the cooperators' up and down probabilities are both set to their mean, and the stay probability is unchanged. \emph{Stay $1/3$}: in every context, the cooperators' stay probability is set to $1/3$ and the remaining $2/3$ keeps the original split between up and down. In every condition, including equal-rate learning, both ablations modify only the cooperators' rule. Defectors keep their own distilled rule.

\paragraph{Replay.} A rule is replayed by letting every agent draw its actions from the rule of its type, and nothing is learned. The \emph{fresh-rollout} tests cover cooperator-only learning, equal-rate learning, and the two ratio-$4$ conditions. The rules estimated from the evaluation rollout of training seed $s$ are replayed on five new rollouts $j=0,\dots,4$ with new placements and new decision phases, using rollout seed $5000+10s+j$. These rollouts are never used for estimation, and the learned tables are run on the same five rollouts. In the \emph{broad scan} over all trained runs, a rule is replayed from the placement and phases on which it was estimated, with new action draws. This scan tests whether a rule reproduces the observed behavior, not whether it transfers.

\section{Complete main table and seed-level uncertainty}
\label{sec:main}

Table~\ref{tab:s_main} is the complete version of Table~\ref{tab:main} of the main text, and Table~\ref{tab:s_ci} gives seed-level intervals for the comparisons made there. The unit of replication is the training seed. The $50$ fresh rollouts of a condition come from ten trained populations, so the five rollouts of each seed are averaged first and every interval is computed over ten values. Intervals are percentile bootstrap intervals. The ten seed-level values or per-seed differences are resampled with replacement $10{,}000$ times, and we report the $2.5$ and $97.5$ percentiles of the resampled means. They need not coincide with $t$ intervals computed from rounded table entries. Differences are paired by seed, and the pairing holds by construction rather than by label, as described under Seeds in Section~\ref{sec:code}.

\begin{table*}[ht]\centering\small
\caption{Seed-level uncertainty of the main comparisons. The unit of replication is the training seed ($n=10$), and the five fresh rollouts of a seed are averaged before any statistic is computed. Entries are means with $95\%$ percentile bootstrap intervals over seeds from $10{,}000$ resamples. Differences are taken per seed between conditions that share the seed and therefore the evaluation placement and decision phases. For the toll, the paired runs also share the training placements and phases. The last column counts the seeds in which the difference has the sign of its mean.}\label{tab:s_ci}
{\tiny
\begin{tabular}{llrrr}\toprule 
& comparison & mean [95\% interval] & paired difference [95\% interval] & seeds\\\midrule
$I_U$ & random movement & 0.9 [0.9, 1.0] & -- & --\\
$I_U$ & only D learns & 0.9 [0.9, 1.0] & identical & --\\
$I_U$ & only C learns & 24.7 [23.0, 26.3] & $+23.8$ [$+22.0$, $+25.4$] & 10/10\\
$I_U$ & both learn $(0.05,0.05)$ & 7.9 [6.3, 9.6] & $+6.9$ [$+5.4$, $+8.6$] & 10/10\\
$I_U$ & C co-trained, D random & 11.0 [10.5, 11.6] & $+10.1$ [$+9.5$, $+10.7$] & 10/10\\
$I_U$ & both learn $(0.2,0.05)$ & 14.7 [11.9, 17.5] & $+13.7$ [$+10.9$, $+16.5$] & 10/10\\
$I_U$ & both learn $(0.05,0.0125)$ & 22.6 [19.1, 26.3] & $+21.7$ [$+18.2$, $+25.3$] & 10/10\\
\midrule
$W$ & random movement & 2199 [2198, 2199] & -- & --\\
$W$ & only D learns & 2205 [2205, 2206] & $+6$ [$+6$, $+7$] & 10/10\\
$W$ & only C learns & 1759 [1729, 1789] & $-439$ [$-470$, $-410$] & 10/10\\
$W$ & both learn $(0.05,0.05)$ & 1977 [1917, 2033] & $-222$ [$-282$, $-165$] & 10/10\\
$W$ & C co-trained, D random & 2022 [2006, 2036] & $-177$ [$-192$, $-163$] & 10/10\\
$W$ & both learn $(0.2,0.05)$ & 1783 [1702, 1860] & $-416$ [$-496$, $-339$] & 10/10\\
$W$ & both learn $(0.05,0.0125)$ & 1599 [1494, 1701] & $-599$ [$-705$, $-497$] & 10/10\\
\midrule
$I_U$ & both learn $(0.05,0.05)$ minus only C learns & 7.9 [6.3, 9.6] & $-16.8$ [$-19.4$, $-14.3$] & 10/10\\
$I_U$ & C co-trained, D random minus only C learns & 11.0 [10.5, 11.6] & $-13.7$ [$-15.2$, $-11.9$] & 10/10\\
$I_U$ & C co-trained: D learned minus D random & 7.9 [6.3, 9.6] & $-3.1$ [$-4.6$, $-1.7$] & 10/10\\
\midrule
$W$ & both learn $(0.05,0.05)$ minus only C learns & 1977 [1917, 2033] & $+217$ [$+152$, $+288$] & 10/10\\
$W$ & C co-trained, D random minus only C learns & 2022 [2006, 2036] & $+262$ [$+230$, $+294$] & 10/10\\
$W$ & C co-trained: D learned minus D random & 1977 [1917, 2033] & $-45$ [$-96$, $+4$] & 7/10\\
\midrule
$I_U$ & toll minus original: only C learns & 10.2 [9.4, 10.9] & $-14.5$ [$-16.3$, $-12.5$] & 10/10\\
$I_U$ & toll minus original: both learn $(0.05,0.05)$ & 1.1 [1.0, 1.2] & $-6.8$ [$-8.5$, $-5.2$] & 10/10\\
$I_U$ & toll minus original: both learn $(0.2,0.05)$ & 2.0 [1.8, 2.3] & $-12.6$ [$-15.4$, $-9.9$] & 10/10\\
\midrule
$W$ & toll minus original: only C learns & 2038 [2026, 2053] & $+279$ [$+249$, $+307$] & 10/10\\
$W$ & toll minus original: both learn $(0.05,0.05)$ & 2203 [2200, 2204] & $+226$ [$+170$, $+285$] & 10/10\\
$W$ & toll minus original: both learn $(0.2,0.05)$ & 2176 [2169, 2181] & $+393$ [$+316$, $+472$] & 10/10\\
\midrule
$I_U$ & only C learns: learned tables & 24.6 [23.9, 25.2] & -- & --\\
$I_U$ & only C learns: pooled replay minus learned & 27.3 [25.8, 28.7] & $+2.7$ [$+1.4$, $+3.9$] & 8/10\\
$I_U$ & only C learns: no preference minus pooled & 1.0 [1.0, 1.0] & $-26.3$ [$-27.7$, $-24.8$] & 10/10\\
$I_U$ & only C learns: stay 1/3 minus pooled & 28.0 [27.3, 28.7] & $+0.7$ [$-0.6$, $+1.9$] & 6/10\\
\midrule
$W$ & only C learns: pooled replay minus learned & 1689 [1661, 1717] & $-53$ [$-80$, $-25$] & 9/10\\
\midrule
$I_U$ & both learn $(0.05,0.05)$: learned tables & 7.4 [6.2, 8.9] & -- & --\\
$I_U$ & both learn $(0.05,0.05)$: pooled replay minus learned & 5.5 [4.6, 6.3] & $-2.0$ [$-3.1$, $-0.7$] & 9/10\\
$I_U$ & both learn $(0.05,0.05)$: no preference minus pooled & 1.0 [1.0, 1.0] & $-4.5$ [$-5.3$, $-3.6$] & 10/10\\
$I_U$ & both learn $(0.05,0.05)$: stay 1/3 minus pooled & 5.4 [4.6, 6.3] & $-0.0$ [$-0.5$, $+0.4$] & 4/10\\
\midrule
$W$ & both learn $(0.05,0.05)$: pooled replay minus learned & 2046 [2019, 2074] & $+42$ [$+6$, $+73$] & 8/10\\
\midrule
$|v|$ & both learn $(0.2,0.05)$: absolute replay & 1.03 [0.88, 1.16] & -- & --\\
$|v|$ & both learn $(0.2,0.05)$: symmetrized minus absolute & 0.15 [0.13, 0.17] & $-0.88$ [$-1.00$, $-0.73$] & 10/10\\
\midrule
$I_U$ & both learn $(0.2,0.05)$: symmetrized minus absolute & 9.1 [6.9, 11.5] & $-11.9$ [$-15.8$, $-7.7$] & 9/10\\
\midrule
$|v|$ & both learn $(0.05,0.0125)$: absolute replay & 0.94 [0.89, 0.99] & -- & --\\
$|v|$ & both learn $(0.05,0.0125)$: symmetrized minus absolute & 0.13 [0.11, 0.14] & $-0.82$ [$-0.86$, $-0.77$] & 10/10\\
\midrule
$I_U$ & both learn $(0.05,0.0125)$: symmetrized minus absolute & 20.3 [15.5, 25.8] & $-5.4$ [$-10.4$, $-0.2$] & 8/10\\
\bottomrule\end{tabular}
}
\end{table*}

In every untolled condition in which cooperators learn, the clustering and welfare differences from random movement have the same sign in all ten seeds. The same holds for every toll effect, computed as toll minus original reward and paired by seed, and for the effect of removing the preference for richer neighbors. The tolled populations are close to random-movement welfare, and their difference from it does not have one sign. With the toll, equal-rate learning is above random movement in $9/10$ seeds and the other two conditions in $0/10$.

With only defectors learning, cooperators move at random exactly as in the baseline, so $I_U$ is identical and only welfare differs. The defectors themselves do not aggregate. Their clustering $I_V=\mathrm{Var}_i(V_i)/\overline V$, measured at the same snapshots as $I_U$, is $0.45$ against $0.97$ under random movement and ranges from $0.37$ to $0.50$ over seeds. The learned defectors spread more evenly than under independent placement. Their spatial correlation with the public good is weaker, at $-0.27$ against $-0.53$. With consumption more uniform, $\rho_{U\varphi}$ rises from $0.32$ to $0.61$. Under co-learning the defectors cluster with the cooperators. For $(0.05,0.05)$, $(0.2,0.05)$, and $(0.05,0.0125)$, $I_V$ is $1.46$, $2.61$, and $3.21$, and the correlation of $U_i$ and $V_i$ over patches is $0.53$, $0.60$, and $0.56$. These values are computed by \texttt{vstats.py}. Standard deviations in Table~\ref{tab:s_main} and in Table~\ref{tab:main} of the main text are population standard deviations over the ten seeds.

Two differences are not distinguishable from zero. The first is the welfare change from replacing random by learned defectors next to fixed co-trained cooperators. It is $-45$ with interval $[-96,4]$ and is negative in $7/10$ seeds, and the main text reports it with its interval. The second is the change in clustering from setting the stay probability to $1/3$. For cooperator-only learning it is $+0.7$ with interval $[-0.6,1.9]$, which supports the statement that a raised stay probability is not required.

\section{Additional figures}
\label{sec:figs}

\begin{figure*}[ht]
\centering
\includegraphics[width=\textwidth]{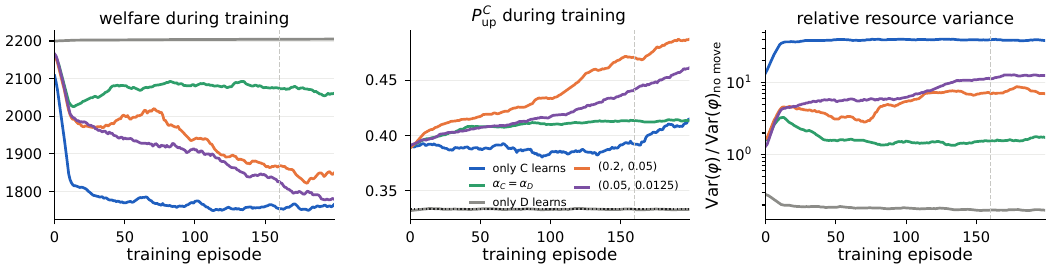}
\caption{Training dynamics, averaged over ten seeds with a 10-episode running average. Exploration is still on during training and reaches $\varepsilon=0.01$ at episode 160, marked by the dashed line. Left: welfare per training episode. Middle: the cooperators' probability of moving to the richer neighbor when exactly one neighbor is richer, with the random value dotted. Right: relative resource variance, the spatial variance of the public good divided by its no-movement value. Clusters appear within the first twenty training episodes.}
\label{fig:s_training}
\end{figure*}

\begin{figure}[ht]
\centering
\includegraphics[width=0.75\columnwidth]{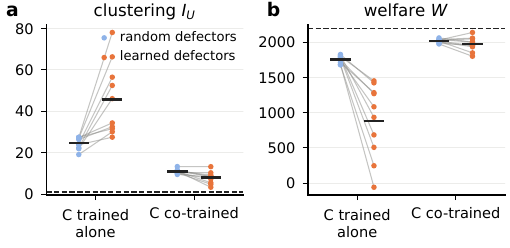}
\caption{Policy recombination. Cooperator tables were trained with non-learning defectors or co-trained at equal rates. They are evaluated with untrained, randomly moving defectors or with the defector tables of the equal-rate run of the same seed. (a)~Clustering. (b)~Welfare. Points are training seeds, lines join the two defector policies of a seed, bars mark means, and dashed lines show random movement.}
\label{fig:s_recombine}
\end{figure}

\begin{figure}[ht]
\centering
\includegraphics[width=0.8\columnwidth]{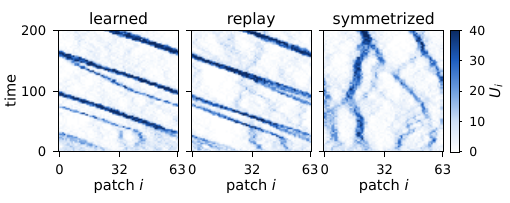}
\caption{Cooperator occupancy on one fresh rollout of the $(0.05,0.0125)$ run with training seed $0$. The panels show the learned tables, the distilled directional rule, and the mirror-symmetrized version of that rule, all from the same placement and decision phases. Across the two ratio-$4$ conditions, with $10$ training seeds and $5$ fresh rollouts each, the three policies travel coherently in $99$, $100$, and $0$ of the $100$ rollouts. Figure~\ref{fig:states} of the main text shows all three fields for the six main conditions.}
\label{fig:s_direction}
\end{figure}

\begin{figure*}[ht]
\centering
\includegraphics[width=\textwidth]{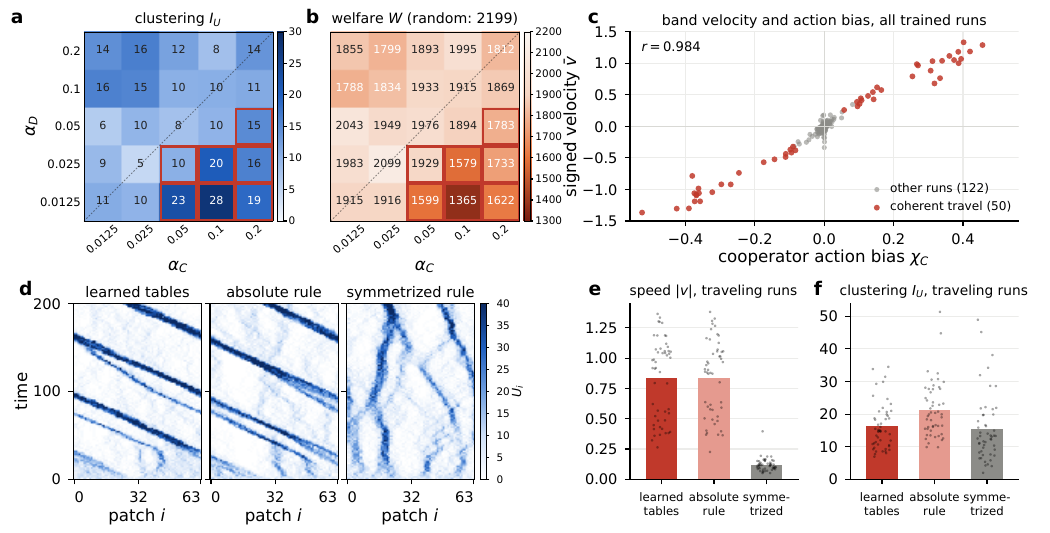}
\caption{Complete learning-rate and directional-bias analysis of the $172$ trained runs. (a,b)~Cell means of clustering and welfare on the $5\times5$ grid. Red boxes mark cells in which at least half of the seeds travel coherently, and the dashed line marks equal rates. (c)~Signed velocity $\bar v$ against the cooperators' action bias $\chi_C$ for every trained run. (d)~A fresh rollout of a ratio-$4$ run under the learned tables, the absolute rule, and the symmetrized rule. Figure~\ref{fig:s_direction} shows the same rollout enlarged. (e,f)~Speed and clustering of the $50$ coherently traveling runs under the three policies in the broad scan. Bars are means and dots are runs.}
\label{fig:s_morph}
\end{figure*}

\begin{figure}[ht]
\centering
\includegraphics[width=0.7\columnwidth]{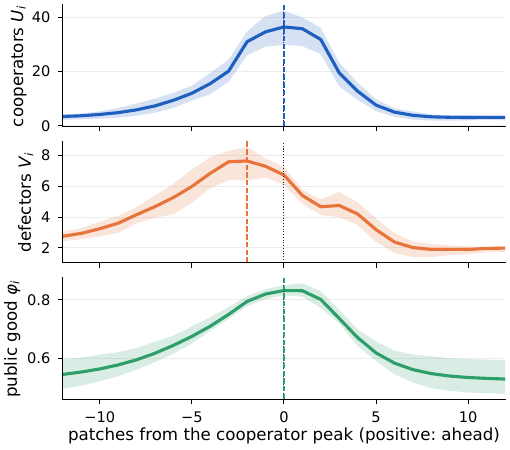}
\caption{Aligned profiles of the traveling bands. For each of the $50$ coherently traveling runs and each snapshot $t=200,\dots,400$, the profiles are shifted so that the maximum of the cooperator profile, smoothed over five patches, is at $0$. They are reflected so that the band travels to the right and are then averaged over snapshots. Lines show the mean over runs and shading the interquartile range over runs. Dashed lines mark the maximum of each mean profile, and the horizontal axis is in patches. The defector maximum lies two patches behind the cooperator maximum.}
\label{fig:s_anatomy}
\end{figure}

\begin{figure*}[ht]
\centering
\includegraphics[width=\textwidth]{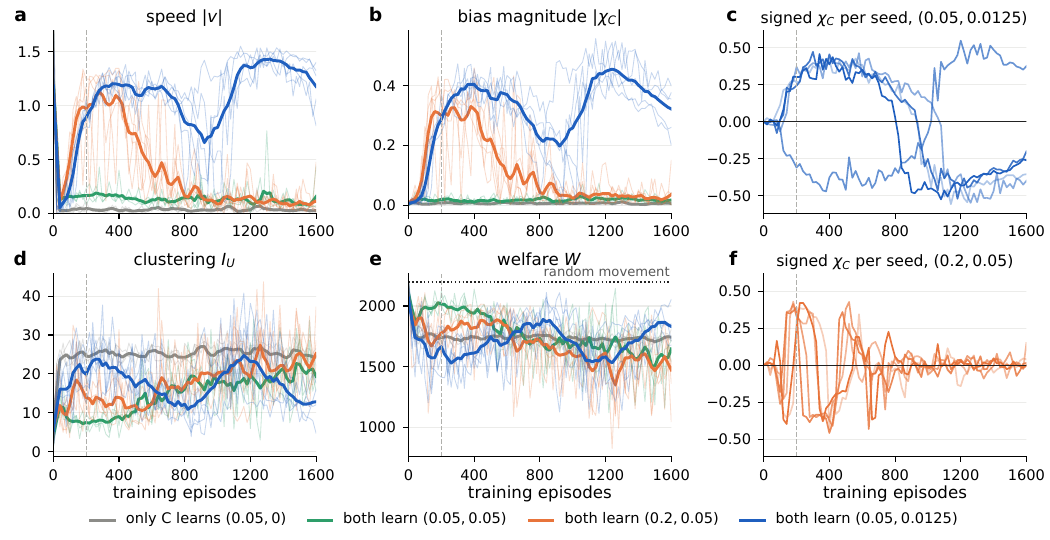}
\caption{Long training to $1600$ episodes, with frozen checkpoints every $20$ episodes and exploration $0.01$ after episode $160$. The dashed line marks the main budget of $200$ episodes. Thin lines are seeds and thick lines are means smoothed over three checkpoints. The ratio-$4$ conditions have five seeds and the two reference conditions three. (a)~Speed. (b)~Magnitude of the cooperators' action bias. The signed bias is not averaged because seeds travel in opposite directions. (c,f)~Signed bias of the individual seeds of the two ratio-$4$ conditions. At $(0.05,0.0125)$ every seed reverses once. At $(0.2,0.05)$ the bias forms, dissolves, and reverses repeatedly and is small at the end. (d)~Clustering. (e)~Welfare.}
\label{fig:s_long}
\end{figure*}

Figures~\ref{fig:s_training}--\ref{fig:s_long} are collected here and discussed in the sections that cite them. They show training dynamics (Section~\ref{sec:numerics}), policy recombination (Section~\ref{sec:env-app}), the directional replay on one fresh rollout, the complete learning-rate and convention analysis (Sections~\ref{sec:rules} and~\ref{sec:grid}), the aligned band profiles, and long training (Section~\ref{sec:seeds}).

\section{Distilled rules, replay, convention, and toll}
\label{sec:rules}

\paragraph{Distilled rules.} Table~\ref{tab:s_rules} gives the pooled rules of the core conditions for both types, by own-patch level and context. The shares are fractions of all decisions of that type, so the three contexts with exactly one richer neighbor do not sum to one. With only cooperators learning, cooperators on poor and intermediate patches move up strongly and almost never move down. In the traveling conditions they stay less on rich patches.

\begin{table*}[ht]\centering\small
\caption{Pooled rules, means over ten seeds. In contexts with exactly one richer neighbor, the table gives for each own-patch level the probabilities of moving up, staying, and moving down, and the share of all decisions made in these contexts. In all other contexts it gives the stay probability, with moves split equally between left and right, and the share of all decisions. Shares are fractions of all decisions of that type and sum to one over the six contexts.}\label{tab:s_rules}
\begin{tabular}{llcccc|cc}\toprule condition & own $\varphi_i$ & up & stay & down & share & stay (other) & share\\\midrule
only C learns, C & poor & 0.71 & 0.21 & 0.08 & 0.03 & 0.38 & 0.00\\
 & intermediate & 0.61 & 0.35 & 0.04 & 0.13 & 0.32 & 0.01\\
 & rich & 0.36 & 0.45 & 0.19 & 0.50 & 0.29 & 0.33\\
\midrule
both learn, $(0.05,0.05)$, C & poor & 0.44 & 0.32 & 0.24 & 0.11 & 0.34 & 0.12\\
 & intermediate & 0.42 & 0.35 & 0.23 & 0.23 & 0.35 & 0.21\\
 & rich & 0.37 & 0.41 & 0.22 & 0.14 & 0.35 & 0.18\\
\midrule
both learn, $(0.05,0.05)$, D & poor & 0.48 & 0.31 & 0.21 & 0.27 & 0.36 & 0.16\\
 & intermediate & 0.36 & 0.38 & 0.26 & 0.27 & 0.38 & 0.15\\
 & rich & 0.37 & 0.33 & 0.30 & 0.08 & 0.37 & 0.07\\
\midrule
both learn, $(0.2,0.05)$, C & poor & 0.52 & 0.29 & 0.19 & 0.12 & 0.33 & 0.14\\
 & intermediate & 0.54 & 0.28 & 0.18 & 0.17 & 0.30 & 0.08\\
 & rich & 0.46 & 0.27 & 0.27 & 0.24 & 0.25 & 0.24\\
\midrule
both learn, $(0.2,0.05)$, D & poor & 0.57 & 0.26 & 0.17 & 0.35 & 0.34 & 0.20\\
 & intermediate & 0.46 & 0.30 & 0.24 & 0.18 & 0.35 & 0.06\\
 & rich & 0.48 & 0.28 & 0.24 & 0.13 & 0.32 & 0.09\\
\midrule
both learn, $(0.05,0.0125)$, C & poor & 0.53 & 0.28 & 0.19 & 0.12 & 0.35 & 0.10\\
 & intermediate & 0.54 & 0.29 & 0.17 & 0.16 & 0.35 & 0.07\\
 & rich & 0.44 & 0.29 & 0.26 & 0.30 & 0.26 & 0.26\\
\midrule
both learn, $(0.05,0.0125)$, D & poor & 0.50 & 0.29 & 0.21 & 0.40 & 0.35 & 0.17\\
 & intermediate & 0.44 & 0.33 & 0.22 & 0.16 & 0.37 & 0.05\\
 & rich & 0.44 & 0.30 & 0.26 & 0.14 & 0.33 & 0.08\\
\midrule
only D learns, D & poor & 0.36 & 0.35 & 0.30 & 0.01 & 0.32 & 0.01\\
 & intermediate & 0.37 & 0.47 & 0.16 & 0.43 & 0.58 & 0.55\\
 & rich & 0.29 & 0.40 & 0.32 & 0.00 & 0.47 & 0.00\\
\bottomrule\end{tabular}\end{table*}

\paragraph{Replay in the broad scan.} Table~\ref{tab:s_replay} replays the pooled, absolute, and symmetrized rules of every trained run. Across all $172$ runs, clustering under the replayed absolute rule correlates with that of the learned population at $r=0.76$, and welfare correlates at $r=0.74$. For the pooled rule the clustering correlation is $0.69$. The pooled rule has no left--right bias and never travels coherently. The absolute rule travels coherently in $49$ of the $50$ runs whose learned population travels coherently and in $3$ of the other $122$. The symmetrized rule never does. In the traveling runs the mean speed is $0.84$ under the learned tables, $0.84$ under the absolute rule, and $0.12$ under the symmetrized rule. The corresponding clustering values are $16.3$, $21.3$, and $15.6$.

\begin{table*}[ht]\centering\small
\caption{Replay of distilled rules in the broad scan, given as cell means over all $n$ trained runs of a cell. Rules are replayed from the placement and decision phases of the estimation rollout, with new action draws. Learned denotes the frozen tables. Pooled uses six contexts with left and right merged, Absolute uses 27 contexts, and Symmetrized averages the absolute rule with its mirror image. The columns give clustering $I_U$, welfare $W$, speed $|v|$, and the share of runs traveling coherently (trav.). For the core conditions, the columns ``no pref.'' and ``stay 1/3'' give the clustering of the pooled rule with equal up and down probabilities and with the stay probability set to $1/3$. $\chi_C$ is the mean signed left--right bias and $|\chi_C|$ the mean absolute bias. Cells are labeled by $\alpha_C/\alpha_D$.}\label{tab:s_replay}
\setlength{\tabcolsep}{3pt}
{\tiny
\begin{tabular}{lr|rrrr|rrrr|rrrr|rrrr|rr|rr}\toprule
 & & \multicolumn{4}{c|}{learned} & \multicolumn{4}{c|}{pooled} & \multicolumn{4}{c|}{absolute} & \multicolumn{4}{c|}{symmetrized} & no & stay & & \\
$\alpha_C/\alpha_D$ & $n$ & $I_U$ & $W$ & $|v|$ & trav. & $I_U$ & $W$ & $|v|$ & trav. & $I_U$ & $W$ & $|v|$ & trav. & $I_U$ & $W$ & $|v|$ & trav. & pref. & 1/3 & $\chi_C$ & $|\chi_C|$\\\midrule
0/0.05 & 10 & 0.9 & 2205 & 1.85 & 0.0 & 1.0 & 2195 & 1.75 & 0.0 & 1.0 & 2194 & 2.05 & 0.0 & 1.0 & 2194 & 2.03 & 0.0 & 1.0 & 1.1 & +0.00 & 0.00\\
0.0125/0.0125 & 3 & 10.8 & 1915 & 0.07 & 0.0 & 6.5 & 2023 & 0.09 & 0.0 & 8.4 & 1958 & 0.19 & 0.0 & 9.3 & 1916 & 0.09 & 0.0 & -- & -- & -0.00 & 0.00\\
0.0125/0.025 & 3 & 9.0 & 1983 & 0.12 & 0.0 & 4.3 & 2090 & 0.11 & 0.0 & 6.2 & 2056 & 0.15 & 0.0 & 5.6 & 2039 & 0.14 & 0.0 & -- & -- & -0.01 & 0.01\\
0.0125/0.05 & 10 & 6.4 & 2043 & 0.18 & 0.0 & 4.7 & 2080 & 0.16 & 0.0 & 6.2 & 2037 & 0.15 & 0.0 & 5.6 & 2046 & 0.13 & 0.0 & -- & -- & -0.00 & 0.01\\
0.0125/0.1 & 3 & 15.9 & 1788 & 0.03 & 0.0 & 5.6 & 2057 & 0.09 & 0.0 & 8.9 & 1971 & 0.09 & 0.0 & 9.1 & 1987 & 0.07 & 0.0 & -- & -- & +0.00 & 0.00\\
0.0125/0.2 & 3 & 14.3 & 1855 & 0.03 & 0.0 & 4.5 & 2100 & 0.27 & 0.0 & 5.7 & 2079 & 0.31 & 0.0 & 5.0 & 2075 & 0.34 & 0.0 & -- & -- & +0.00 & 0.00\\
0.025/0.0125 & 10 & 9.8 & 1916 & 0.24 & 0.2 & 9.0 & 1928 & 0.08 & 0.0 & 13.5 & 1781 & 0.21 & 0.2 & 15.3 & 1712 & 0.06 & 0.0 & -- & -- & -0.03 & 0.04\\
0.025/0.025 & 3 & 4.6 & 2099 & 0.29 & 0.0 & 2.4 & 2138 & 0.27 & 0.0 & 4.1 & 2062 & 0.17 & 0.0 & 3.8 & 2100 & 0.22 & 0.0 & -- & -- & +0.01 & 0.01\\
0.025/0.05 & 10 & 9.6 & 1949 & 0.09 & 0.0 & 4.3 & 2094 & 0.21 & 0.0 & 5.7 & 2049 & 0.19 & 0.0 & 5.8 & 2037 & 0.17 & 0.0 & -- & -- & +0.00 & 0.01\\
0.025/0.1 & 3 & 15.4 & 1834 & 0.01 & 0.0 & 3.9 & 2116 & 0.22 & 0.0 & 5.6 & 2062 & 0.20 & 0.0 & 4.1 & 2087 & 0.14 & 0.0 & -- & -- & -0.00 & 0.00\\
0.025/0.2 & 3 & 16.1 & 1799 & 0.01 & 0.0 & 4.1 & 2099 & 0.21 & 0.0 & 4.7 & 2093 & 0.17 & 0.0 & 4.8 & 2085 & 0.20 & 0.0 & -- & -- & +0.00 & 0.00\\
0.05/0 & 10 & 24.7 & 1760 & 0.05 & 0.0 & 27.6 & 1674 & 0.00 & 0.0 & 30.3 & 1617 & 0.01 & 0.0 & 30.6 & 1620 & 0.00 & 0.0 & 1.0 & 26.7 & -0.00 & 0.01\\
0.05/0.0125 & 10 & 22.6 & 1599 & 0.97 & 1.0 & 16.4 & 1625 & 0.18 & 0.0 & 25.5 & 1513 & 0.96 & 1.0 & 23.1 & 1410 & 0.12 & 0.0 & 1.0 & 12.7 & +0.08 & 0.32\\
0.05/0.025 & 10 & 10.3 & 1929 & 0.47 & 1.0 & 8.9 & 1920 & 0.12 & 0.0 & 13.8 & 1824 & 0.50 & 1.0 & 14.5 & 1718 & 0.09 & 0.0 & -- & -- & +0.06 & 0.12\\
0.05/0.05 & 10 & 7.9 & 1976 & 0.16 & 0.0 & 4.9 & 2059 & 0.12 & 0.0 & 9.8 & 1905 & 0.13 & 0.0 & 8.4 & 1946 & 0.11 & 0.0 & 1.0 & 5.6 & -0.01 & 0.01\\
0.05/0.1 & 10 & 10.4 & 1933 & 0.08 & 0.0 & 3.7 & 2107 & 0.19 & 0.0 & 6.1 & 2048 & 0.16 & 0.0 & 5.6 & 2052 & 0.14 & 0.0 & -- & -- & -0.00 & 0.01\\
0.05/0.2 & 10 & 12.0 & 1893 & 0.18 & 0.0 & 3.9 & 2107 & 0.43 & 0.0 & 4.5 & 2084 & 0.38 & 0.0 & 4.9 & 2080 & 0.34 & 0.0 & -- & -- & -0.00 & 0.01\\
0.1/0.0125 & 3 & 27.7 & 1365 & 1.09 & 1.0 & 7.1 & 1931 & 0.17 & 0.0 & 34.5 & 1307 & 1.04 & 1.0 & 24.9 & 1169 & 0.13 & 0.0 & -- & -- & -0.37 & 0.37\\
0.1/0.025 & 3 & 20.5 & 1579 & 1.12 & 1.0 & 11.3 & 1817 & 0.18 & 0.0 & 27.0 & 1416 & 1.12 & 1.0 & 13.0 & 1815 & 0.14 & 0.0 & -- & -- & +0.37 & 0.37\\
0.1/0.05 & 10 & 9.9 & 1894 & 0.31 & 0.4 & 6.9 & 1989 & 0.10 & 0.0 & 13.9 & 1753 & 0.33 & 0.6 & 11.5 & 1804 & 0.10 & 0.0 & -- & -- & -0.02 & 0.07\\
0.1/0.1 & 3 & 10.5 & 1915 & 0.11 & 0.0 & 7.3 & 1975 & 0.06 & 0.0 & 14.8 & 1701 & 0.11 & 0.0 & 11.9 & 1835 & 0.08 & 0.0 & -- & -- & +0.00 & 0.01\\
0.1/0.2 & 3 & 8.2 & 1995 & 0.20 & 0.0 & 2.3 & 2153 & 0.35 & 0.0 & 4.5 & 2083 & 0.16 & 0.0 & 3.5 & 2096 & 0.21 & 0.0 & -- & -- & +0.00 & 0.01\\
0.2/0.0125 & 3 & 18.7 & 1622 & 1.13 & 1.0 & 9.0 & 1884 & 0.21 & 0.0 & 32.4 & 1289 & 1.11 & 1.0 & 17.3 & 1420 & 0.12 & 0.0 & -- & -- & -0.11 & 0.39\\
0.2/0.025 & 3 & 15.9 & 1733 & 1.16 & 1.0 & 8.2 & 1894 & 0.50 & 0.0 & 24.0 & 1477 & 1.16 & 1.0 & 9.1 & 1863 & 0.22 & 0.0 & -- & -- & -0.19 & 0.40\\
0.2/0.05 & 10 & 14.7 & 1783 & 1.06 & 1.0 & 8.9 & 1851 & 0.20 & 0.0 & 19.5 & 1637 & 1.03 & 1.0 & 8.0 & 1909 & 0.14 & 0.0 & 1.0 & 7.9 & +0.04 & 0.33\\
0.2/0.1 & 10 & 11.1 & 1869 & 0.20 & 0.2 & 11.8 & 1834 & 0.10 & 0.0 & 15.6 & 1717 & 0.19 & 0.2 & 19.5 & 1552 & 0.09 & 0.0 & -- & -- & +0.01 & 0.05\\
0.2/0.2 & 3 & 13.7 & 1812 & 0.09 & 0.0 & 8.5 & 1957 & 0.09 & 0.0 & 19.4 & 1598 & 0.11 & 0.0 & 12.9 & 1790 & 0.10 & 0.0 & -- & -- & -0.00 & 0.01\\
\bottomrule\end{tabular}
}
\end{table*}

\paragraph{Fresh-rollout replay.} Table~\ref{tab:s_oos} replays the rules of the core and ratio-$4$ conditions on five fresh rollouts per training seed. The pooled rule reproduces the clusters. Equalizing its up and down probabilities abolishes them in all $200$ rollouts, while setting the stay probability to $1/3$ does not. In the two ratio-$4$ conditions, the absolute rules travel coherently in $100/100$ fresh rollouts. The learned tables travel in $99/100$, and also in $99/100$ under greedy evaluation. The symmetrized rules never travel, and no training seed has a single traveling rollout under them.

\begin{table*}[ht]\centering\small
\caption{Fresh-rollout replay. Rules are estimated from the evaluation rollout of each training seed and replayed on five fresh rollouts per seed. These rollouts have new placements and decision phases and are never used for estimation. The learned tables run on the same fresh rollouts. Entries are means over the ten seed means. The column trav.\ gives the rollouts with coherent travel out of $50$, followed in parentheses by the number of training seeds with at least one such rollout. The column range gives the minimum and maximum $I_U$ over rollouts. Table~\ref{tab:s_ci} gives seed-level intervals.}\label{tab:s_oos}
\begin{tabular}{ll|rrrrc}\toprule condition & policy & $W$ & $I_U$ & $|v|$ & trav. & $I_U$ range\\\midrule
only C learns & learned tables & 1742 & 24.6 & 0.05 & 0/50 (0) & 20.3--28.5\\
 & pooled rule & 1689 & 27.3 & 0.01 & 0/50 (0) & 21.2--34.6\\
 & pooled, equal up/down & 2199 & 1.0 & 1.79 & 0/50 (0) & 0.8--1.2\\
 & pooled, stay 1/3 & 1664 & 28.0 & 0.01 & 0/50 (0) & 20.8--39.4\\
 & absolute rule & 1613 & 30.2 & 0.07 & 0/50 (0) & 21.7--39.7\\
 & symmetrized rule & 1635 & 29.5 & 0.03 & 0/50 (0) & 21.6--38.7\\
\midrule
both learn, $(0.05,0.05)$ & learned tables & 2004 & 7.4 & 0.17 & 0/50 (0) & 3.7--14.3\\
 & pooled rule & 2046 & 5.5 & 0.12 & 0/50 (0) & 2.3--10.8\\
 & pooled, equal up/down & 2196 & 1.0 & 1.76 & 0/50 (0) & 0.8--1.3\\
 & pooled, stay 1/3 & 2051 & 5.4 & 0.13 & 0/50 (0) & 2.5--11.6\\
 & absolute rule & 1910 & 9.1 & 0.13 & 0/50 (0) & 3.8--19.8\\
 & symmetrized rule & 1897 & 9.4 & 0.12 & 0/50 (0) & 3.8--22.6\\
\midrule
both learn, $(0.2,0.05)$ & learned tables & 1769 & 14.8 & 1.06 & 49/50 (10) & 6.0--32.3\\
 & pooled rule & 1880 & 9.4 & 0.20 & 0/50 (0) & 3.6--27.9\\
 & pooled, equal up/down & 2194 & 1.0 & 1.83 & 0/50 (0) & 0.8--1.3\\
 & pooled, stay 1/3 & 1915 & 8.5 & 0.19 & 0/50 (0) & 4.1--21.9\\
 & absolute rule & 1595 & 21.0 & 1.03 & 50/50 (10) & 8.3--33.8\\
 & symmetrized rule & 1886 & 9.1 & 0.15 & 0/50 (0) & 4.1--20.1\\
\midrule
both learn, $(0.05,0.0125)$ & learned tables & 1551 & 22.6 & 0.99 & 50/50 (10) & 12.3--35.5\\
 & pooled rule & 1577 & 17.2 & 0.17 & 0/50 (0) & 5.4--49.4\\
 & pooled, equal up/down & 2194 & 1.0 & 1.76 & 0/50 (0) & 0.7--1.4\\
 & pooled, stay 1/3 & 1681 & 15.1 & 0.16 & 0/50 (0) & 4.0--38.1\\
 & absolute rule & 1507 & 25.7 & 0.94 & 50/50 (10) & 16.1--35.7\\
 & symmetrized rule & 1422 & 20.3 & 0.13 & 0/50 (0) & 4.7--55.9\\
\bottomrule\end{tabular}\end{table*}

\paragraph{Directional convention.} Across all $172$ runs, the cooperators' bias $\chi_C$ and the signed velocity $\bar v$ correlate with $r=0.984$ (Figure~\ref{fig:s_morph}c). Most runs lie near the origin, so we also report the correlation within the $50$ traveling runs alone. It is $r=0.991$ for the signed values and $0.956$ between $|\chi_C|$ and $|\bar v|$. The fitted slope is $2.98$ patches per time unit per unit bias. An individual cooperator makes two decisions per time unit and drifts on average by $2\chi_C$, so the band moves faster than the net displacement of its members. In each of the $50$ traveling runs, the sign of $\chi_C$ matches the direction of travel. In these runs $|\chi_C|$ ranges from $0.057$ to $0.53$ with mean $0.26$. In the contexts of the absolute rule in which both neighbors fall in the same $\varphi$-difference level, it averages $0.22$, without conditioning on occupancy differences. In the $122$ other runs, $|\chi_C|\le0.087$ with mean $0.012$. The two ranges overlap slightly, so the bias magnitude alone does not classify a run.

Conditioning on the complete observation gives the same picture. In mirror-symmetric observations, $s=\mathcal Ms$, both the resource-difference levels and the occupancy-difference levels are equal on the two sides, so the observation offers no cue for either side. Such observations make up $11\%$ of the cooperators' decisions in the traveling runs and at least $9\%$ in every run. In these observations the cooperators' bias has the sign of the direction of travel in $50/50$ traveling runs. Its mean magnitude is $0.24$, with a minimum of $0.07$. The seed-level mean is $0.277$ with interval $[0.234,0.312]$ at $(0.2,0.05)$ and $0.284$ with interval $[0.264,0.305]$ at $(0.05,0.0125)$. In the $112$ other runs with learning cooperators, the mean magnitude is $0.02$ and the maximum $0.11$. Action bias and velocity are measured on the same rollout. Their agreement shows that the population moves the way most of its members step. It is not an independent test of what a deviating agent would earn. At the level of individual agents, on average $96\%$ of cooperators and $86\%$ of defectors in the traveling runs move more often in the band's direction than against it, with minima of $79\%$ and $64\%$. In the other runs with learning cooperators, on average $61\%$ of cooperators are on the majority side, with a maximum of $91\%$.

\paragraph{Crowding toll.} Table~\ref{tab:s_toll} gives the toll runs per seed. The original reward rate of an agent on patch $i$ contains $-\gamma n_i/\Omega$, and the toll subtracts $\gamma(n_i-1)/\Omega$. The tolled reward rate is therefore the original one with crowding coefficient $2\gamma$ plus the constant $\gamma/\Omega$. Because every decision interval has the same length, this constant shifts the fixed point of every action value by the same amount and leaves the greedy policy of the converged tables unchanged. Exact equivalence of the finite training trajectories would also require shifting the initial tables by $\tau_{\mathrm{dec}}\gamma/\Omega/(1-\gamma_{\mathrm{dec}})$. We did not apply this shift, so visited and unvisited entries absorb the constant at different rates. The toll is closely related to doubling the crowding coefficient in the learning reward and is not a separate control. It can be read as a price because $\gamma(n_i-1)/\Omega$ is the crowding cost the agent imposes on the other occupants of its patch. It does not price the effect of a move on production and consumption of the public good.

\begin{table*}[ht]\centering\small
\caption{Crowding toll, per seed. The table gives welfare $W$, clustering $I_U$, and the payoffs $\Pi_C$ and $\Pi_D$ of populations trained on the original reward (own) and on the reward minus the crowding toll (toll), with random movement for reference. The two runs of a seed share the training placements, the decision phases, and the evaluation placement. $W$ is always computed from the original payoff.}\label{tab:s_toll}
\setlength{\tabcolsep}{3pt}
{\tiny
\begin{tabular}{r|rr|rrrr|rrrr|rrrr|rrrr}\toprule
 & \multicolumn{2}{c|}{random} & \multicolumn{4}{c|}{only C, own / toll} & \multicolumn{4}{c|}{$(0.05,0.05)$, own / toll} & \multicolumn{4}{c|}{$(0.2,0.05)$, own / toll} & \multicolumn{4}{c}{$\Pi_C$ / $\Pi_D$, only C}\\
seed & $W$ & $\Pi_C$ & $W$ & $W$ & $I_U$ & $I_U$ & $W$ & $W$ & $I_U$ & $I_U$ & $W$ & $W$ & $I_U$ & $I_U$ & own & & toll & \\\midrule
0 & 2198 & 3.39 & 1813 & 2029 & 21.9 & 11.3 & 1954 & 2206 & 9.2 & 0.8 & 1799 & 2171 & 15.5 & 2.1 & 3.30 & 1.74 & 3.64 & 2.07\\
1 & 2199 & 3.38 & 1792 & 2020 & 22.4 & 10.7 & 2004 & 2205 & 7.3 & 1.0 & 1887 & 2179 & 9.7 & 1.7 & 3.17 & 1.93 & 3.55 & 2.23\\
2 & 2199 & 3.38 & 1688 & 2012 & 27.4 & 11.0 & 2141 & 2203 & 3.3 & 1.0 & 1964 & 2181 & 8.5 & 2.0 & 3.14 & 1.46 & 3.55 & 2.21\\
3 & 2197 & 3.37 & 1715 & 2029 & 27.5 & 10.6 & 1803 & 2205 & 13.3 & 1.0 & 1564 & 2185 & 20.8 & 2.2 & 3.18 & 1.51 & 3.56 & 2.26\\
4 & 2199 & 3.39 & 1801 & 2082 & 23.9 & 8.0 & 2005 & 2203 & 8.3 & 1.1 & 1738 & 2184 & 15.0 & 1.6 & 3.39 & 1.47 & 3.63 & 2.38\\
5 & 2200 & 3.39 & 1777 & 2076 & 24.3 & 7.9 & 1851 & 2194 & 10.3 & 1.4 & 1599 & 2152 & 21.9 & 2.9 & 3.32 & 1.51 & 3.60 & 2.41\\
6 & 2196 & 3.38 & 1752 & 2034 & 26.4 & 11.0 & 2051 & 2202 & 5.5 & 1.1 & 1731 & 2177 & 14.6 & 2.3 & 3.13 & 1.81 & 3.53 & 2.36\\
7 & 2200 & 3.38 & 1829 & 2030 & 19.1 & 11.4 & 1959 & 2202 & 8.5 & 1.2 & 1954 & 2169 & 8.5 & 1.9 & 3.26 & 1.91 & 3.62 & 2.14\\
8 & 2199 & 3.38 & 1680 & 2039 & 27.0 & 10.2 & 1938 & 2203 & 8.4 & 1.1 & 1851 & 2180 & 13.4 & 2.2 & 3.12 & 1.46 & 3.61 & 2.18\\
9 & 2200 & 3.39 & 1748 & 2028 & 27.3 & 9.9 & 2061 & 2201 & 4.8 & 0.9 & 1739 & 2180 & 18.7 & 1.5 & 3.28 & 1.44 & 3.59 & 2.20\\
\bottomrule\end{tabular}
}
\end{table*}

\section{Learning-rate grid and the ratio-2 boundary}
\label{sec:grid}

Table~\ref{tab:s_grid} gives every cell of the grid with the number of coherently traveling seeds and a $95\%$ Wilson interval for the probability that a trained run travels coherently. All four cells with $\alpha_C/\alpha_D=2$ have ten seeds, and Table~\ref{tab:s_ratio2} gives the per-seed values of the two cells that lie off the sweep lines. The cell $(0.05,0.025)$ travels in $10/10$ seeds with interval $[0.72,1]$, $(0.1,0.05)$ in $4/10$ with interval $[0.17,0.69]$, and $(0.025,0.0125)$ and $(0.2,0.1)$ in $2/10$ each with interval $[0.06,0.51]$. The outcome at ratio two depends on the absolute rates, and the intervals remain wide. Cells with three seeds give intervals of $[0.44,1]$ for $3/3$ and $[0,0.56]$ for $0/3$. We describe the grid as the empirical distribution of outcomes at a finite training budget and do not claim a threshold. Section~\ref{sec:sweep} extends the grid to a finer and wider scan of both rates.

\begin{table*}[ht]\centering\small
\caption{The $5\times5$ learning-rate grid with frozen policies after 200 episodes. For each cell the table gives the number of training seeds $n$ and the means over seeds of clustering, welfare, the cooperators' probabilities of moving up and of staying on a rich patch, speed, main-direction share, and best-shift correlation. It also gives the number of coherently traveling seeds with its $95\%$ Wilson interval for the traveling probability.}\label{tab:s_grid}
\setlength{\tabcolsep}{4pt}\begin{tabular}{rrrrrrrrrrrc}\toprule
$\alpha_C$ & $\alpha_D$ & $n$ & $I_U$ & $W$ & $P_{\text{up}}^C$ & $P_{\text{stay}|\text{rich}}^C$ & $|v|$ & dir. & corr & traveling & 95\% interval\\\midrule
0.0125 & 0.0125 & 3 & 10.8 & 1915 & 0.38 & 0.41 & 0.07 & 0.21 & 0.88 & 0/3 & [0.00, 0.56]\\
0.0125 & 0.025 & 3 & 9.0 & 1983 & 0.38 & 0.42 & 0.12 & 0.33 & 0.87 & 0/3 & [0.00, 0.56]\\
0.0125 & 0.05 & 10 & 6.4 & 2043 & 0.39 & 0.39 & 0.18 & 0.39 & 0.77 & 0/10 & [0.00, 0.28]\\
0.0125 & 0.1 & 3 & 15.9 & 1788 & 0.34 & 0.48 & 0.03 & 0.14 & 0.94 & 0/3 & [0.00, 0.56]\\
0.0125 & 0.2 & 3 & 14.3 & 1855 & 0.35 & 0.46 & 0.03 & 0.11 & 0.92 & 0/3 & [0.00, 0.56]\\
\addlinespace[2pt]
0.025 & 0.0125 & 10 & 9.8 & 1916 & 0.41 & 0.35 & 0.24 & 0.57 & 0.86 & 2/10 & [0.06, 0.51]\\
0.025 & 0.025 & 3 & 4.6 & 2099 & 0.41 & 0.38 & 0.29 & 0.44 & 0.71 & 0/3 & [0.00, 0.56]\\
0.025 & 0.05 & 10 & 9.6 & 1949 & 0.38 & 0.42 & 0.09 & 0.29 & 0.88 & 0/10 & [0.00, 0.28]\\
0.025 & 0.1 & 3 & 15.4 & 1834 & 0.36 & 0.45 & 0.01 & 0.05 & 0.94 & 0/3 & [0.00, 0.56]\\
0.025 & 0.2 & 3 & 16.1 & 1799 & 0.37 & 0.44 & 0.01 & 0.06 & 0.95 & 0/3 & [0.00, 0.56]\\
\addlinespace[2pt]
0.05 & 0.0125 & 10 & 22.6 & 1599 & 0.50 & 0.28 & 0.97 & 1.00 & 0.90 & 10/10 & [0.72, 1.00]\\
0.05 & 0.025 & 10 & 10.3 & 1929 & 0.43 & 0.34 & 0.47 & 0.97 & 0.84 & 10/10 & [0.72, 1.00]\\
0.05 & 0.05 & 10 & 7.9 & 1976 & 0.41 & 0.37 & 0.16 & 0.39 & 0.81 & 0/10 & [0.00, 0.28]\\
0.05 & 0.1 & 10 & 10.4 & 1933 & 0.39 & 0.40 & 0.08 & 0.28 & 0.88 & 0/10 & [0.00, 0.28]\\
0.05 & 0.2 & 10 & 12.0 & 1893 & 0.38 & 0.42 & 0.18 & 0.26 & 0.84 & 0/10 & [0.00, 0.28]\\
\addlinespace[2pt]
0.1 & 0.0125 & 3 & 27.7 & 1365 & 0.53 & 0.25 & 1.09 & 1.00 & 0.89 & 3/3 & [0.44, 1.00]\\
0.1 & 0.025 & 3 & 20.5 & 1579 & 0.52 & 0.25 & 1.12 & 1.00 & 0.92 & 3/3 & [0.44, 1.00]\\
0.1 & 0.05 & 10 & 9.9 & 1894 & 0.43 & 0.34 & 0.31 & 0.76 & 0.84 & 4/10 & [0.17, 0.69]\\
0.1 & 0.1 & 3 & 10.5 & 1915 & 0.41 & 0.36 & 0.11 & 0.35 & 0.86 & 0/3 & [0.00, 0.56]\\
0.1 & 0.2 & 3 & 8.2 & 1995 & 0.41 & 0.36 & 0.20 & 0.46 & 0.77 & 0/3 & [0.00, 0.56]\\
\addlinespace[2pt]
0.2 & 0.0125 & 3 & 18.7 & 1622 & 0.52 & 0.24 & 1.13 & 0.99 & 0.86 & 3/3 & [0.44, 1.00]\\
0.2 & 0.025 & 3 & 15.9 & 1733 & 0.54 & 0.24 & 1.16 & 1.00 & 0.86 & 3/3 & [0.44, 1.00]\\
0.2 & 0.05 & 10 & 14.7 & 1783 & 0.50 & 0.26 & 1.06 & 0.99 & 0.87 & 10/10 & [0.72, 1.00]\\
0.2 & 0.1 & 10 & 11.1 & 1869 & 0.43 & 0.34 & 0.20 & 0.62 & 0.86 & 2/10 & [0.06, 0.51]\\
0.2 & 0.2 & 3 & 13.7 & 1812 & 0.41 & 0.34 & 0.09 & 0.32 & 0.85 & 0/3 & [0.00, 0.56]\\
\bottomrule\end{tabular}\end{table*}

\begin{table}[ht]\centering
\caption{The two ten-seed cells of ratio $\alpha_C/\alpha_D=2$ that lie off the row and column of Tables~\ref{tab:s_sweep1}--\ref{tab:s_waves3}, per seed. The columns give clustering, welfare, the cooperators' probabilities of moving up and of staying on a rich patch, signed velocity $\bar v$, speed $|v|$, main-direction share $d$, best-shift correlation $r$, and coherent travel.}
\label{tab:s_ratio2}
\small\setlength{\tabcolsep}{2.6pt}
\resizebox{0.8\columnwidth}{!}{
\begin{tabular}{lrrrrrrrrrc}
\toprule
$\alpha_C$ / $\alpha_D$ & seed & $I_U$ & $W$ & $P_{\text{up}}^C$ & $P_{\text{stay}|\text{rich}}^C$ & $\bar v$ & $|v|$ & $d$ & $r$ & trav.\\
\midrule
0.025 / 0.0125 & 0 & 9.4 & 1944 & 0.41 & 0.35 & 0.11 & 0.13 & 0.546 & 0.87 & \\
0.025 / 0.0125 & 1 & 7.4 & 1962 & 0.41 & 0.36 & -0.02 & 0.10 & 0.265 & 0.86 & \\
0.025 / 0.0125 & 2 & 9.3 & 1940 & 0.41 & 0.37 & -0.43 & 0.56 & 0.969 & 0.83 & yes\\
0.025 / 0.0125 & 3 & 7.1 & 1989 & 0.43 & 0.34 & -0.05 & 0.14 & 0.383 & 0.82 & \\
0.025 / 0.0125 & 4 & 9.7 & 1923 & 0.42 & 0.34 & 0.00 & 0.16 & 0.352 & 0.85 & \\
0.025 / 0.0125 & 5 & 6.7 & 2026 & 0.41 & 0.36 & -0.25 & 0.25 & 0.816 & 0.79 & \\
0.025 / 0.0125 & 6 & 11.0 & 1926 & 0.40 & 0.36 & -0.21 & 0.24 & 0.776 & 0.86 & \\
0.025 / 0.0125 & 7 & 15.2 & 1751 & 0.41 & 0.35 & -0.48 & 0.48 & 1.000 & 0.91 & yes\\
0.025 / 0.0125 & 8 & 9.1 & 1929 & 0.41 & 0.35 & 0.03 & 0.20 & 0.255 & 0.86 & \\
0.025 / 0.0125 & 9 & 13.6 & 1772 & 0.40 & 0.37 & -0.04 & 0.11 & 0.332 & 0.91 & \\
\midrule
0.2 / 0.1 & 0 & 9.7 & 1956 & 0.42 & 0.34 & 0.00 & 0.10 & 0.240 & 0.83 & \\
0.2 / 0.1 & 1 & 10.4 & 1823 & 0.42 & 0.35 & 0.08 & 0.11 & 0.434 & 0.87 & \\
0.2 / 0.1 & 2 & 15.6 & 1733 & 0.41 & 0.34 & 0.00 & 0.09 & 0.224 & 0.91 & \\
0.2 / 0.1 & 3 & 10.7 & 1886 & 0.43 & 0.34 & 0.26 & 0.26 & 0.903 & 0.85 & yes\\
0.2 / 0.1 & 4 & 7.7 & 1939 & 0.44 & 0.33 & 0.36 & 0.36 & 0.980 & 0.82 & yes\\
0.2 / 0.1 & 5 & 12.8 & 1792 & 0.41 & 0.34 & -0.24 & 0.25 & 0.842 & 0.90 & \\
0.2 / 0.1 & 6 & 11.8 & 1779 & 0.42 & 0.33 & -0.04 & 0.10 & 0.316 & 0.90 & \\
0.2 / 0.1 & 7 & 10.0 & 1913 & 0.44 & 0.33 & -0.12 & 0.17 & 0.577 & 0.83 & \\
0.2 / 0.1 & 8 & 9.3 & 1974 & 0.43 & 0.34 & 0.23 & 0.23 & 0.827 & 0.81 & \\
0.2 / 0.1 & 9 & 12.8 & 1891 & 0.43 & 0.33 & -0.28 & 0.28 & 0.893 & 0.88 & \\
\bottomrule
\end{tabular}
}
\end{table}

\section{Exploration sensitivity of the frozen policies}
\label{sec:eps}

The main protocol evaluates frozen tables with $\varepsilon_{\mathrm{eval}}=0.01$. To test whether the patterns depend on this residual exploration, we evaluate the same saved tables, without retraining, with $\varepsilon_{\mathrm{eval}}=0$, $0.01$, and $0.05$ on the same five fresh rollouts per training seed (Table~\ref{tab:s_eps}). With $\varepsilon_{\mathrm{eval}}=0$ every decision is greedy, and ties between equal table entries are still broken uniformly, as in training. Clustering, welfare, the number of coherently traveling rollouts, and the directional bias are essentially unchanged between $\varepsilon_{\mathrm{eval}}=0$ and $0.01$ in all four conditions. Five percent exploration slightly weakens the clusters of the ratio-$4$ conditions and leaves travel intact in $49/50$ rollouts in both. The patterns come from the learned action preferences and not from exploration noise.

\begin{table}[ht]\centering\small
\caption{Exploration sensitivity of the frozen policies. The saved tables are evaluated with $\varepsilon_{\mathrm{eval}}\in\{0,0.01,0.05\}$ on the same five fresh rollouts per training seed. Entries are means over the ten seed means with $95\%$ bootstrap intervals over seeds. The column trav.\ counts coherently traveling rollouts out of $50$, and $|\chi_C|$ is the mean absolute bias.}\label{tab:s_eps}
\setlength{\tabcolsep}{3pt}\begin{tabular}{lrrrrr}\toprule condition & $\varepsilon_{\mathrm{eval}}$ & $I_U$ & $W$ & trav. & $|\chi_C|$\\\midrule
only C & 0 & 24.4 [23.5, 25.4] & 1755 [1737, 1772] & 0/50 & 0.01\\
 & 0.01 & 24.6 [23.9, 25.2] & 1742 [1727, 1756] & 0/50 & 0.01\\
 & 0.05 & 24.0 [23.0, 25.1] & 1762 [1739, 1781] & 0/50 & 0.01\\
\midrule
$(0.05,0.05)$ & 0 & 7.9 [6.5, 9.3] & 1995 [1955, 2033] & 0/50 & 0.01\\
 & 0.01 & 7.4 [6.2, 8.9] & 2004 [1964, 2042] & 0/50 & 0.01\\
 & 0.05 & 6.7 [5.7, 8.0] & 2023 [1988, 2054] & 0/50 & 0.01\\
\midrule
$(0.2,0.05)$ & 0 & 15.5 [12.5, 18.6] & 1761 [1662, 1851] & 50/50 & 0.35\\
 & 0.01 & 14.8 [11.7, 18.1] & 1769 [1675, 1856] & 49/50 & 0.34\\
 & 0.05 & 13.0 [11.1, 15.0] & 1833 [1774, 1892] & 49/50 & 0.30\\
\midrule
$(0.05,0.0125)$ & 0 & 22.9 [20.5, 25.3] & 1546 [1465, 1630] & 49/50 & 0.34\\
 & 0.01 & 22.6 [19.5, 25.8] & 1551 [1460, 1641] & 50/50 & 0.33\\
 & 0.05 & 19.7 [17.1, 22.3] & 1657 [1584, 1727] & 49/50 & 0.29\\
\bottomrule\end{tabular}\end{table}

\section{Hand-designed movement rules}
\label{sec:hand}

As a reference for the learned policies, we run rules that prescribe a response to the same neighbor comparisons that the agents observe. For a move to side $x\in\{\text{left},\text{right}\}$, let $z_\varphi(x)\in\{-1,0,1\}$ code whether $\varphi_x-\varphi_i$ falls in the lower, similar, or higher level. Similarly, $z_n(x)\in\{-1,0,1\}$ codes whether $n_x-n_i$ falls in the lower, similar, or higher level. For staying both are zero. The rule is
\[
\pi(a\mid s)\ \propto\ \exp\big(\lambda_\varphi z_\varphi(a)-\lambda_n z_n(a)\big).
\]
Three parameter settings were fixed before any run and were not tuned. They are weak resource seeking with $(\lambda_\varphi,\lambda_n)=(1,0)$, strong resource seeking with $(3,0)$, and resource seeking with a crowding penalty, $(3,1)$. Either the cooperators follow the rule while the defectors move at random, or both types follow it. Environment, decision phases, and measurements are those of the main model. The rollouts are the $50$ fresh placements used for the learned policies, and Table~\ref{tab:s_hand} gives the results.

\begin{table}[ht]\centering\small
\caption{Hand-designed movement rules on the same $10\times5$ fresh placements and decision phases as used for the learned policies. In rows marked `C', cooperators follow the rule and defectors move at random. In rows marked `C and D', both types follow it. The learned policies on the same rollouts are shown for comparison. Entries are means over rollouts, and trav.\ counts coherently traveling rollouts out of $50$.}\label{tab:s_hand}
\setlength{\tabcolsep}{3pt}\resizebox{0.8\columnwidth}{!}{\begin{tabular}{llrr|rrrrr}\toprule rule & followed by & $\lambda_\varphi$ & $\lambda_n$ & $I_U$ & $W$ & $B$ & $\mathcal K$ & trav.\\\midrule
random movement & -- & -- & -- & 1.0 & 2199 & 2998 & 352 & 0/50\\
weak resource seeking & C & 1 & 0 & 39.9 & 1378 & 3047 & 1221 & 0/50\\
 & C and D & 1 & 0 & 19.2 & 1452 & 2996 & 1096 & 0/50\\
strong resource seeking & C & 3 & 0 & 51.3 & 1105 & 3032 & 1478 & 0/50\\
 & C and D & 3 & 0 & 65.2 & -721 & 2994 & 3267 & 0/50\\
resource and crowding & C & 3 & 1 & 47.5 & 1194 & 3036 & 1394 & 0/50\\
 & C and D & 3 & 1 & 47.5 & 237 & 2995 & 2310 & 0/50\\
\midrule
learned, only C learns & \multicolumn{3}{l|}{Q-tables} & 24.6 & 1742 & 3066 & 876 & 0/50\\
learned, $(0.05,0.05)$ & \multicolumn{3}{l|}{Q-tables} & 7.4 & 2004 & 3001 & 549 & 0/50\\
learned, $(0.2,0.05)$ & \multicolumn{3}{l|}{Q-tables} & 14.8 & 1769 & 3009 & 792 & 49/50\\
learned, $(0.05,0.0125)$ & \multicolumn{3}{l|}{Q-tables} & 22.6 & 1551 & 3016 & 1017 & 50/50\\
\bottomrule\end{tabular}}\end{table}

Prescribed resource seeking by cooperators alone produces clusters and a welfare loss, as in the classical models that motivate this work. Both are larger than under the learned cooperator-only policies. Clustering reaches $I_U=39.9$ and $51.3$ against $24.6$, and welfare falls to $W=1378$ and $1105$ against $1742$. The benefit $B$ stays near its random-movement value and the loss is carried by crowding, the same decomposition as for the learned policies. The crowding penalty with $\lambda_n=1$ changes the outcome little when only cooperators follow the rule. When both types follow a strong resource-seeking rule, crowding becomes extreme. None of the six rule populations travels coherently. The rules are mirror-symmetric by construction and contain no directional preference. Aggregation requires only resource seeking, whereas the weaker clustering under co-learning and the traveling bands involve what the two populations learn from each other.

The three settings are a reference, not a search. They were not matched to the learned policies in movement rate or in the strength of the preference. A comparison with the best fixed rule would require such a search on separate validation rollouts, and the statement in the main text is limited to the tested rules. The hand-designed rules, the distilled rules of Section~\ref{sec:protocols}, and independent learning are three different objects. The first are prescribed, the second summarize learned tables for intervention, and only the third is the adaptation process studied in the paper.

\section{Mirror-symmetrized original tables}
\label{sec:mirror}

The symmetrized rule of Section~\ref{sec:protocols} removes the left--right preference from a $27$-context summary of the tables. To test the same intervention on the original policies, let $\mathcal M$ be the reflection of the ring. For an observation it exchanges the two neighbors, $\mathcal M s=(\varphi_i\text{ level},n_i\text{ level},\Delta\varphi_R,\Delta\varphi_L,\Delta n_R,\Delta n_L)$, and for an action it exchanges left and right. Every agent $j$ of the chosen type then acts on
\[
\pi_j^{\mathrm{sym}}(a\mid s)=\tfrac12\big[\pi_j(a\mid s)+\pi_j(\mathcal Ma\mid\mathcal Ms)\big],
\]
where $\pi_j$ is the $\varepsilon$-greedy policy of its own frozen table. The mixture is sampled exactly. At each decision, with probability $1/2$, the agent evaluates its table at $\mathcal Ms$ and takes the mirrored action. All six observation components and all differences between agents are kept. Symmetric individual policies do not by themselves imply that the population cannot move, so the outcome has to be measured.

Table~\ref{tab:s_mirror} gives the result on the fresh rollouts of the two ratio-$4$ conditions. With both types symmetrized, coherent travel disappears in $50/50$ rollouts at $(0.2,0.05)$ and in $48/50$ at $(0.05,0.0125)$, where the two remaining rollouts belong to one training seed. The speed falls from about $1$ to $0.33$ and $0.22$, and the cooperators' bias falls to $0.01$ and $0.03$. Clusters remain, with $I_U=7.3$ and $20.9$ against $14.8$ and $22.6$ under the learned tables. Symmetrizing only the defectors leaves coherent travel in all $100$ rollouts. Symmetrizing only the cooperators removes it in $44$ of $50$ rollouts at $(0.05,0.0125)$ and in $12$ of $50$ at $(0.2,0.05)$. In the latter condition the remaining bands are slower, with mean speed $0.46$ in the $38$ traveling rollouts against $1.06$ under the learned tables. At $(0.05,0.0125)$, symmetrizing only the cooperators also produces very strong stationary clusters, with $I_U=41.2$ and $W=899$. The directional preference that carries the bands is a property of the original tables and resides mainly in the cooperators. The defectors' preference alone can sustain a slower band in one of the two conditions. The defectors' bias has a similar size in both conditions, $|\chi_D|=0.15$ and $0.16$. What differs is the state of the symmetrized cooperators, which are strongly clustered and stationary at $(0.05,0.0125)$ and more weakly clustered at $(0.2,0.05)$.

\begin{table}[ht]\centering\small
\caption{Mirror-symmetrized original tables on the fresh rollouts of the two ratio-4 conditions, with ten training seeds and five rollouts each. Every agent of the indicated type acts on $\tfrac12[\pi_j(a\mid s)+\pi_j(\mathcal Ma\mid\mathcal Ms)]$ computed from its own frozen table. Entries are means over the ten seed means with $95\%$ bootstrap intervals over seeds. The column trav.\ counts coherently traveling rollouts out of $50$, followed in parentheses by the training seeds with at least one such rollout.}\label{tab:s_mirror}
\setlength{\tabcolsep}{3pt}\resizebox{0.9\columnwidth}{!}{\begin{tabular}{llrrrrr}\toprule condition & policy & $I_U$ & $W$ & $|v|$ & trav. & $|\chi_C|$\\\midrule
$(0.2,0.05)$ & learned tables & 14.8 [11.7, 18.1] & 1769 & 1.06 [0.93, 1.19] & 49/50 (10) & 0.34\\
 & C symmetrized & 13.4 [8.9, 20.6] & 1794 & 0.41 [0.34, 0.47] & 38/50 (9) & 0.10\\
 & D symmetrized & 19.9 [17.2, 22.9] & 1734 & 1.07 [1.00, 1.15] & 50/50 (10) & 0.42\\
 & both symmetrized & 7.3 [5.3, 9.7] & 1973 & 0.33 [0.26, 0.40] & 0/50 (0) & 0.01\\
\midrule
$(0.05,0.0125)$ & learned tables & 22.6 [19.5, 25.8] & 1551 & 0.99 [0.94, 1.04] & 50/50 (10) & 0.33\\
 & C symmetrized & 41.2 [37.1, 45.0] & 899 & 0.15 [0.10, 0.19] & 6/50 (2) & 0.06\\
 & D symmetrized & 21.9 [20.8, 23.3] & 1704 & 0.91 [0.87, 0.94] & 50/50 (10) & 0.36\\
 & both symmetrized & 20.9 [18.9, 23.1] & 1471 & 0.22 [0.20, 0.24] & 2/50 (1) & 0.03\\
\bottomrule\end{tabular}}\end{table}

\paragraph{Added randomness as an alternative explanation.} Symmetrization changes the action whenever the mirrored action at the mirrored observation differs from the original action. This happens at a third of the decisions of either type, with disagreement $0.33$ in Table~\ref{tab:s_randmatch}. The symmetrized policy is also more random, and a more random policy might stop the bands by itself. Two controls keep the original tables, and with them the directional preference, and add uniform exploration instead.

\begin{table}[ht]\centering\small
\caption{Symmetrization compared with added randomness on the fresh rollouts of the two ratio-4 conditions, with ten training seeds and five rollouts each. ``Disagreement'' is the fraction of a type's decisions at which the policy's action differs from the action the original policy would take at the same observation. The matched control keeps the original tables and adds uniform exploration at the rate that reproduces the disagreement of the symmetrized policy. The last three rows add uniform exploration $\varepsilon$ to both types. Entries are means over the ten seed means with $95\%$ bootstrap intervals for $I_U$, and trav.\ counts coherently traveling rollouts out of $50$.}\label{tab:s_randmatch}
\setlength{\tabcolsep}{3pt}\resizebox{0.9\columnwidth}{!}{\begin{tabular}{llrrrrrr}\toprule condition & policy & disagr. C & D & $I_U$ & $W$ & trav. & $|\chi_C|$\\\midrule
$(0.2,0.05)$ & learned tables & 0.00 & 0.00 & 14.8 [11.7, 18.1] & 1769 & 49/50 & 0.34\\
 & both symmetrized & 0.33 & 0.32 & 7.3 [5.3, 9.7] & 1973 & 0/50 & 0.01\\
 & control, both & 0.33 & 0.32 & 3.0 [2.7, 3.4] & 2136 & 4/50 & 0.07\\
 & C symmetrized & 0.33 & 0.00 & 13.4 [8.9, 20.6] & 1794 & 38/50 & 0.10\\
 & control, C & 0.33 & 0.00 & 1.6 [1.5, 1.7] & 2177 & 0/50 & 0.05\\
 & uniform eps 0.1 & 0.06 & 0.06 & 10.6 [9.2, 12.1] & 1904 & 49/50 & 0.26\\
 & uniform eps 0.2 & 0.13 & 0.13 & 8.0 [7.0, 9.0] & 1979 & 50/50 & 0.20\\
 & uniform eps 0.3 & 0.19 & 0.19 & 5.7 [5.1, 6.4] & 2052 & 40/50 & 0.15\\
\midrule
$(0.05,0.0125)$ & learned tables & 0.00 & 0.00 & 22.6 [19.5, 25.8] & 1551 & 50/50 & 0.33\\
 & both symmetrized & 0.33 & 0.32 & 20.9 [18.9, 23.1] & 1471 & 2/50 & 0.03\\
 & control, both & 0.33 & 0.32 & 2.5 [2.2, 2.9] & 2154 & 4/50 & 0.05\\
 & C symmetrized & 0.33 & 0.00 & 41.2 [37.1, 45.0] & 899 & 6/50 & 0.06\\
 & control, C & 0.33 & 0.00 & 1.3 [1.2, 1.4] & 2189 & 0/50 & 0.04\\
 & uniform eps 0.1 & 0.06 & 0.06 & 17.2 [15.3, 19.3] & 1746 & 50/50 & 0.25\\
 & uniform eps 0.2 & 0.13 & 0.13 & 11.5 [10.0, 13.1] & 1900 & 50/50 & 0.18\\
 & uniform eps 0.3 & 0.19 & 0.19 & 7.2 [6.2, 8.3] & 2018 & 44/50 & 0.12\\
\bottomrule\end{tabular}}\end{table}

The disagreement rate $d_\tau$ is measured on the fresh rollouts of the symmetrized policy itself. All rollouts, including the symmetrized policy and the matched control, execute actions exactly as in the main protocol, with uniform random tie-breaking among maximizing entries. At every decision of type $\tau$, the executed action is compared with a reference action of the original table at the same observation. The reference uses the same exploration draw, so exploration at the rate $0.01$ of the original policy counts as agreement. It breaks ties deterministically so that the comparison does not consume the random stream. Deterministic tie-breaking enters only this diagnostic and not the rollout policy. Where the original table has tied maxima, a randomly broken executed action can count as a disagreement. This effect is negligible here, because the learned tables themselves disagree with the reference at a rate of $0.000$ for both types.

The matched control evaluates the original tables on the same rollouts with $\varepsilon_\tau=0.01+1.5\,d_\tau\approx0.5$, where $d_\tau$ is measured per training seed on the symmetrized rollouts. A uniformly random action differs from the original one with probability $2/3$, so the additional exploration $1.5\,d_\tau$ replaces the original action at the rate $d_\tau$. For the cooperators, the measured disagreement of the control is $0.334$, against $0.333$ under symmetrization. The control also stops the bands and leaves $4/50$ coherent rollouts in each condition. Unlike symmetrization, it strongly weakens aggregation, to $I_U=3.0$ and $2.5$, which is still above the random-movement value of $1.0$. Symmetrization leaves clustering at $7.3$ and $20.9$.

The dose response shows the order in which uniform randomness removes the two features. At $\varepsilon=0.3$, which changes $19\%$ of the decisions, the bands still travel coherently in $40/50$ and $44/50$ rollouts, while clustering has fallen to $5.7$ and $7.2$. At $\varepsilon=0.1$ and $0.2$, all but one of the $200$ rollouts travel. Uniform randomness degrades aggregation before travel, whereas symmetrization removes travel while sparing aggregation. The control matches the mean rate of action changes but not the action entropy or the perturbation in each state, so it does not rule out every account based on randomness. The contrasting effects nevertheless support a specific role for the directional preference beyond a general increase in exploration. A matched control applied only to the cooperators gives the same answer.

\section{Policy recombination and environment variants}
\label{sec:env-app}

\paragraph{Recombination.} Table~\ref{tab:s_recombine} gives the four pairings of Figure~\ref{fig:s_recombine}. With random defectors, co-trained cooperators cluster less than cooperators trained alone. The paired difference in $I_U$ is $-13.7$ with interval $[-15.2,-11.9]$, negative in $10/10$ seeds. Next to co-trained cooperators, learned defectors reduce $I_U$ further, with a paired change of $-3.1$ and interval $[-4.6,-1.7]$ in $10/10$ seeds. They also reduce the resource benefit $B$ in every seed, with a change of $-38$ and interval $[-40,-36]$. The change in welfare is not distinguishable from zero. It is $-45$ with interval $[-96,4]$ and is negative in $7/10$ seeds. Next to cooperators trained alone, the same defector tables raise $I_U$ by $+20.9$ with interval $[11.5,31.1]$ in $10/10$ seeds, and they lower $W$ by $876$ with interval $[600,1180]$. The effects of the two populations are not additive, and we do not report them as shares of a total.

\begin{table}[ht]\centering\small
\caption{Policy recombination on the held-out evaluation placements. Entries are means over ten training seeds, with the between-seed standard deviation of $I_U$ and $W$. ``C alone'' denotes cooperator tables trained with non-learning defectors, and ``C co-trained'' the cooperator tables of the equal-rate run of the same seed. Defectors are either untrained and move at random or carry the tables of the equal-rate run.}\label{tab:s_recombine}
\setlength{\tabcolsep}{3pt}\resizebox{0.75\columnwidth}{!}{\begin{tabular}{llrrrrrrrr}\toprule cooperators & defectors & $I_U$ & (sd) & $W$ & (sd) & $B$ & $\mathcal K$ & $\varphi_C$ & $\varphi_D$\\\midrule
C alone & random & 24.7 & 2.7 & 1759 & 49 & 3066 & 858 & 0.85 & 0.36\\
C alone & learned & 45.6 & 16.5 & 883 & 495 & 3029 & 1698 & 0.76 & 0.55\\
C co-trained & random & 11.0 & 1.0 & 2022 & 25 & 3039 & 570 & 0.79 & 0.48\\
C co-trained & learned & 7.9 & 2.7 & 1977 & 94 & 3001 & 576 & 0.71 & 0.67\\
\bottomrule\end{tabular}}\end{table}

\paragraph{Environment variants.} Table~\ref{tab:s_envvar} repeats the three core conditions and the $(0.2,0.05)$ condition in four variants, with three seeds each. The ring size is changed at fixed densities and fixed patch geometry. For $M=32$ we use $L=4$, $N_C=224$, and $N_D=96$, and for $M=128$ we use $L=16$, $N_C=896$, and $N_D=384$. The spacing $\ell=L/M$, the patch volume $\Omega$, and the mean occupancies stay those of the main model. The other two variants change one parameter of the main model, setting either $D_\varphi=0.06$ or $r_v=6.5$. Directional consistency is the main-direction share $d$ of Section~\ref{sec:measure}. Coherent travel at $(0.2,0.05)$ persists in $3/3$ seeds on both ring sizes and in $2/3$ with doubled diffusion. With $r_v=6.5$ no seed travels coherently and $d$ lies between $0.63$ and $0.83$, so the bands are less robust than aggregation. We did not test sensitivity to the numerical step, for example halving $\Delta t$ at a fixed decision interval $K\Delta t$.

\begin{table}[ht]
\caption{Environment variants with three seeds each, compared with the main model with ten. Clustering and welfare are listed for cooperator-only, both-type, and defector-only learning. $W_{\mathrm{rnd}}$ is random-movement welfare. The last columns give the speed, directional consistency, and number of coherently traveling seeds at $(0.2,0.05)$.}
\label{tab:s_envvar}
\centering
\small
\setlength{\tabcolsep}{2.4pt}
\resizebox{0.75\columnwidth}{!}{%
\begin{tabular}{lccrccc}
\toprule
 & \multicolumn{3}{c}{only C / both / only D} & \multicolumn{3}{c}{ratio 4}\\
variant & $I_U$ & $W$ & $W_{\mathrm{rnd}}$ & $|v|$ & dir. & trav.\\
\midrule
$M=32$ & 26.6 / 4.4 / 1.0 & 847 / 1037 / 1102 & 1099 & 1.06 & 1.00 & 3/3\\
$M=128$ & 24.1 / 8.2 / 1.0 & 3554 / 3991 / 4410 & 4399 & 1.18 & 1.00 & 3/3\\
$D_\varphi=0.06$ & 21.5 / 5.5 / 0.8 & 1801 / 2023 / 2207 & 2198 & 0.79 & 0.81 & 2/3\\
$r_v=6.5$ & 23.9 / 2.0 / 0.8 & 1800 / 2238 / 2271 & 2264 & 0.37 & 0.74 & 0/3\\
\midrule
main (10 seeds) & 24.7 / 7.9 / 0.9 & 1760 / 1976 / 2205 & 2199 & 1.06 & 0.99 & 10/10\\
\bottomrule
\end{tabular}
}
\end{table}

\section{Per-seed results of the main model}
\label{sec:seeds}

\paragraph{Core conditions.}
Tables~\ref{tab:s_conly}--\ref{tab:s_donly} give, for each seed, the learned behavior and the clustering, the correlation between cooperators and the public good, and the welfare of the three policies run from the same placement.

\begin{table*}[ht]
\caption{Only cooperators learn ($\alpha_C=0.05$, $\alpha_D=0$), main model, per seed. $I_U$, $\rho_{U\varphi}$, and $W$ for the no-movement, random-movement, and learned policies from the same initial placement.}
\label{tab:s_conly}
\centering\small\setlength{\tabcolsep}{3pt}
\begin{tabular}{lrrrrrrrrrrr}
\toprule
seed & $P_{\text{up}}^C$ & $P_{\text{up}}^D$ & $I_U$ none & random & learned & $\rho$ none & random & learned & $W$ none & random & learned\\
\midrule
0 & 0.42 & 0.33 & 0.83 & 0.94 & 21.93 & 0.38 & 0.33 & 0.85 & 2202 & 2199 & 1811\\
1 & 0.43 & 0.33 & 1.06 & 0.79 & 22.40 & 0.43 & 0.42 & 0.82 & 2205 & 2199 & 1792\\
2 & 0.41 & 0.33 & 0.90 & 0.80 & 27.41 & 0.27 & 0.42 & 0.85 & 2198 & 2198 & 1691\\
3 & 0.35 & 0.33 & 0.92 & 0.95 & 27.49 & 0.35 & 0.28 & 0.81 & 2204 & 2197 & 1716\\
4 & 0.39 & 0.33 & 1.17 & 1.05 & 23.85 & 0.36 & 0.27 & 0.85 & 2193 & 2199 & 1800\\
5 & 0.42 & 0.33 & 0.85 & 1.04 & 24.31 & 0.28 & 0.30 & 0.82 & 2199 & 2201 & 1777\\
6 & 0.43 & 0.34 & 0.86 & 0.91 & 26.40 & 0.31 & 0.28 & 0.81 & 2198 & 2197 & 1753\\
7 & 0.39 & 0.33 & 1.32 & 1.08 & 19.12 & 0.49 & 0.34 & 0.82 & 2195 & 2200 & 1829\\
8 & 0.40 & 0.34 & 1.07 & 1.03 & 26.98 & 0.41 & 0.27 & 0.83 & 2194 & 2199 & 1680\\
9 & 0.40 & 0.33 & 0.75 & 0.85 & 27.33 & 0.29 & 0.29 & 0.83 & 2200 & 2200 & 1748\\
\bottomrule
\end{tabular}
\end{table*}

\begin{table*}[ht]
\caption{Both learn ($\alpha_C=\alpha_D=0.05$), main model, per seed. $I_U$, $\rho_{U\varphi}$, and $W$ for the no-movement, random-movement, and learned policies from the same initial placement.}
\label{tab:s_both}
\centering\small\setlength{\tabcolsep}{3pt}
\begin{tabular}{lrrrrrrrrrrr}
\toprule
seed & $P_{\text{up}}^C$ & $P_{\text{up}}^D$ & $I_U$ none & random & learned & $\rho$ none & random & learned & $W$ none & random & learned\\
\midrule
0 & 0.41 & 0.41 & 0.83 & 0.94 & 9.17 & 0.38 & 0.33 & 0.88 & 2202 & 2199 & 1953\\
1 & 0.42 & 0.41 & 1.06 & 0.79 & 7.31 & 0.43 & 0.42 & 0.87 & 2205 & 2199 & 2004\\
2 & 0.41 & 0.40 & 0.90 & 0.80 & 3.31 & 0.27 & 0.42 & 0.83 & 2198 & 2198 & 2141\\
3 & 0.39 & 0.40 & 0.92 & 0.95 & 13.25 & 0.35 & 0.28 & 0.93 & 2204 & 2197 & 1803\\
4 & 0.41 & 0.41 & 1.17 & 1.05 & 8.29 & 0.36 & 0.27 & 0.91 & 2193 & 2199 & 2004\\
5 & 0.40 & 0.42 & 0.85 & 1.04 & 10.30 & 0.28 & 0.30 & 0.93 & 2199 & 2201 & 1850\\
6 & 0.41 & 0.41 & 0.86 & 0.91 & 5.48 & 0.31 & 0.28 & 0.85 & 2198 & 2197 & 2050\\
7 & 0.41 & 0.42 & 1.32 & 1.08 & 8.53 & 0.49 & 0.34 & 0.88 & 2195 & 2200 & 1959\\
8 & 0.41 & 0.43 & 1.07 & 1.03 & 8.40 & 0.41 & 0.27 & 0.89 & 2194 & 2199 & 1936\\
9 & 0.41 & 0.41 & 0.75 & 0.85 & 4.76 & 0.29 & 0.29 & 0.87 & 2200 & 2200 & 2061\\
\bottomrule
\end{tabular}
\end{table*}

\begin{table*}[ht]
\caption{Only defectors learn ($\alpha_C=0$, $\alpha_D=0.05$), main model, per seed. $I_U$, $\rho_{U\varphi}$, and $W$ for the no-movement, random-movement, and learned policies from the same initial placement.}
\label{tab:s_donly}
\centering\small\setlength{\tabcolsep}{3pt}
\begin{tabular}{lrrrrrrrrrrr}
\toprule
seed & $P_{\text{up}}^C$ & $P_{\text{up}}^D$ & $I_U$ none & random & learned & $\rho$ none & random & learned & $W$ none & random & learned\\
\midrule
0 & 0.34 & 0.37 & 0.83 & 0.94 & 0.94 & 0.38 & 0.33 & 0.66 & 2202 & 2199 & 2205\\
1 & 0.33 & 0.38 & 1.06 & 0.79 & 0.79 & 0.43 & 0.42 & 0.64 & 2205 & 2199 & 2206\\
2 & 0.34 & 0.37 & 0.90 & 0.80 & 0.80 & 0.27 & 0.42 & 0.59 & 2198 & 2198 & 2204\\
3 & 0.34 & 0.38 & 0.92 & 0.95 & 0.95 & 0.35 & 0.28 & 0.62 & 2204 & 2197 & 2205\\
4 & 0.33 & 0.37 & 1.17 & 1.05 & 1.05 & 0.36 & 0.27 & 0.62 & 2193 & 2199 & 2205\\
5 & 0.33 & 0.36 & 0.85 & 1.04 & 1.04 & 0.28 & 0.30 & 0.65 & 2199 & 2201 & 2207\\
6 & 0.33 & 0.38 & 0.86 & 0.91 & 0.91 & 0.31 & 0.28 & 0.58 & 2198 & 2197 & 2203\\
7 & 0.33 & 0.37 & 1.32 & 1.08 & 1.08 & 0.49 & 0.34 & 0.59 & 2195 & 2200 & 2206\\
8 & 0.33 & 0.36 & 1.07 & 1.03 & 1.03 & 0.41 & 0.27 & 0.60 & 2194 & 2199 & 2206\\
9 & 0.33 & 0.37 & 0.75 & 0.85 & 0.85 & 0.29 & 0.29 & 0.60 & 2200 & 2200 & 2206\\
\bottomrule
\end{tabular}
\end{table*}

\paragraph{Learning-rate sweep.}
Tables~\ref{tab:s_sweep1}--\ref{tab:s_sweep3} give the frozen-policy statistics of every run on the row and column of the grid through $(0.05,0.05)$, and Tables~\ref{tab:s_waves1}--\ref{tab:s_waves3} the motion of the cooperator profile. The sign of $\bar v$ is the direction of travel ($+$ toward increasing patch index).

\begin{table*}[ht]\centering
\caption{Learning-rate sweep, per seed, part 1 of 3. Clustering, welfare, and learned behavior of the frozen policy.}
\label{tab:s_sweep1}
\centering\small\setlength{\tabcolsep}{3pt}
\begin{tabular}{lrrrrrrrr}
\toprule
$\alpha_C$ & $\alpha_D$ & seed & $I_U$ & $W$ & $P_{\text{up}}^C$ & $P_{\text{stay}|\text{rich}}^C$ & $P_{\text{leave}|\text{poor}}^C$ & $P_{\text{up}}^D$\\
\midrule
0.05 & 0 & 0 & 21.9 & 1811 & 0.42 & 0.38 & 0.85 & 0.33\\
0.05 & 0 & 1 & 22.4 & 1792 & 0.43 & 0.38 & 0.74 & 0.33\\
0.05 & 0 & 2 & 27.4 & 1691 & 0.41 & 0.39 & 0.83 & 0.33\\
0.05 & 0 & 3 & 27.5 & 1716 & 0.35 & 0.47 & 0.76 & 0.33\\
0.05 & 0 & 4 & 23.9 & 1800 & 0.39 & 0.40 & 0.70 & 0.33\\
0.05 & 0 & 5 & 24.3 & 1777 & 0.42 & 0.37 & 0.84 & 0.33\\
0.05 & 0 & 6 & 26.4 & 1753 & 0.43 & 0.37 & 0.83 & 0.34\\
0.05 & 0 & 7 & 19.1 & 1829 & 0.39 & 0.42 & 0.82 & 0.33\\
0.05 & 0 & 8 & 27.0 & 1680 & 0.40 & 0.39 & 0.73 & 0.34\\
0.05 & 0 & 9 & 27.3 & 1748 & 0.40 & 0.40 & 0.83 & 0.33\\
0.05 & 0.0125 & 0 & 25.6 & 1461 & 0.49 & 0.27 & 0.68 & 0.48\\
0.05 & 0.0125 & 1 & 31.5 & 1338 & 0.53 & 0.26 & 0.69 & 0.51\\
0.05 & 0.0125 & 2 & 16.1 & 1793 & 0.48 & 0.29 & 0.69 & 0.46\\
0.05 & 0.0125 & 3 & 29.9 & 1555 & 0.51 & 0.27 & 0.70 & 0.50\\
0.05 & 0.0125 & 4 & 18.3 & 1747 & 0.47 & 0.28 & 0.68 & 0.46\\
0.05 & 0.0125 & 5 & 29.3 & 1340 & 0.51 & 0.26 & 0.70 & 0.51\\
0.05 & 0.0125 & 6 & 19.5 & 1768 & 0.47 & 0.31 & 0.66 & 0.45\\
0.05 & 0.0125 & 7 & 23.1 & 1631 & 0.53 & 0.26 & 0.72 & 0.52\\
0.05 & 0.0125 & 8 & 14.8 & 1800 & 0.48 & 0.30 & 0.70 & 0.49\\
0.05 & 0.0125 & 9 & 18.4 & 1553 & 0.52 & 0.26 & 0.69 & 0.51\\
0.05 & 0.025 & 0 & 10.1 & 1898 & 0.44 & 0.34 & 0.67 & 0.43\\
0.05 & 0.025 & 1 & 9.0 & 1974 & 0.44 & 0.33 & 0.68 & 0.44\\
0.05 & 0.025 & 2 & 11.0 & 1891 & 0.42 & 0.34 & 0.66 & 0.40\\
0.05 & 0.025 & 3 & 9.2 & 1896 & 0.44 & 0.33 & 0.67 & 0.46\\
0.05 & 0.025 & 4 & 11.4 & 1929 & 0.42 & 0.34 & 0.67 & 0.41\\
0.05 & 0.025 & 5 & 10.5 & 1976 & 0.43 & 0.34 & 0.67 & 0.44\\
0.05 & 0.025 & 6 & 6.9 & 2004 & 0.44 & 0.35 & 0.69 & 0.45\\
0.05 & 0.025 & 7 & 14.8 & 1819 & 0.40 & 0.34 & 0.64 & 0.39\\
0.05 & 0.025 & 8 & 11.9 & 1913 & 0.42 & 0.36 & 0.67 & 0.43\\
0.05 & 0.025 & 9 & 8.5 & 1991 & 0.43 & 0.33 & 0.67 & 0.42\\
0.05 & 0.05 & 0 & 9.2 & 1953 & 0.41 & 0.41 & 0.67 & 0.41\\
0.05 & 0.05 & 1 & 7.3 & 2004 & 0.42 & 0.37 & 0.67 & 0.41\\
0.05 & 0.05 & 2 & 3.3 & 2141 & 0.41 & 0.35 & 0.67 & 0.40\\
0.05 & 0.05 & 3 & 13.3 & 1803 & 0.39 & 0.40 & 0.66 & 0.40\\
0.05 & 0.05 & 4 & 8.3 & 2004 & 0.41 & 0.35 & 0.67 & 0.41\\
0.05 & 0.05 & 5 & 10.3 & 1850 & 0.40 & 0.39 & 0.66 & 0.42\\
0.05 & 0.05 & 6 & 5.5 & 2050 & 0.41 & 0.36 & 0.67 & 0.41\\
\bottomrule
\end{tabular}
\end{table*}

\begin{table*}[ht]\centering
\caption{Learning-rate sweep, per seed, part 2 of 3. Clustering, welfare, and learned behavior of the frozen policy.}
\label{tab:s_sweep2}
\centering\small\setlength{\tabcolsep}{3pt}
\begin{tabular}{lrrrrrrrr}
\toprule
$\alpha_C$ & $\alpha_D$ & seed & $I_U$ & $W$ & $P_{\text{up}}^C$ & $P_{\text{stay}|\text{rich}}^C$ & $P_{\text{leave}|\text{poor}}^C$ & $P_{\text{up}}^D$\\
\midrule
0.05 & 0.05 & 7 & 8.5 & 1959 & 0.41 & 0.37 & 0.68 & 0.42\\
0.05 & 0.05 & 8 & 8.4 & 1936 & 0.41 & 0.37 & 0.67 & 0.43\\
0.05 & 0.05 & 9 & 4.8 & 2061 & 0.41 & 0.37 & 0.68 & 0.41\\
0.05 & 0.1 & 0 & 10.0 & 1984 & 0.40 & 0.42 & 0.67 & 0.41\\
0.05 & 0.1 & 1 & 17.9 & 1673 & 0.39 & 0.39 & 0.66 & 0.37\\
0.05 & 0.1 & 2 & 8.2 & 2018 & 0.41 & 0.38 & 0.67 & 0.41\\
0.05 & 0.1 & 3 & 6.1 & 2011 & 0.39 & 0.39 & 0.67 & 0.40\\
0.05 & 0.1 & 4 & 18.7 & 1729 & 0.37 & 0.43 & 0.66 & 0.40\\
0.05 & 0.1 & 5 & 8.4 & 1985 & 0.39 & 0.42 & 0.67 & 0.41\\
0.05 & 0.1 & 6 & 9.1 & 1970 & 0.41 & 0.37 & 0.67 & 0.38\\
0.05 & 0.1 & 7 & 6.1 & 2035 & 0.40 & 0.39 & 0.67 & 0.41\\
0.05 & 0.1 & 8 & 8.6 & 2001 & 0.40 & 0.39 & 0.67 & 0.40\\
0.05 & 0.1 & 9 & 10.4 & 1925 & 0.39 & 0.41 & 0.66 & 0.38\\
0.05 & 0.2 & 0 & 19.0 & 1688 & 0.38 & 0.42 & 0.65 & 0.38\\
0.05 & 0.2 & 1 & 3.4 & 2148 & 0.39 & 0.41 & 0.66 & 0.42\\
0.05 & 0.2 & 2 & 12.7 & 1885 & 0.36 & 0.46 & 0.65 & 0.38\\
0.05 & 0.2 & 3 & 9.5 & 1939 & 0.39 & 0.42 & 0.68 & 0.39\\
0.05 & 0.2 & 4 & 16.8 & 1750 & 0.37 & 0.44 & 0.65 & 0.38\\
0.05 & 0.2 & 5 & 6.5 & 2060 & 0.38 & 0.43 & 0.67 & 0.41\\
0.05 & 0.2 & 6 & 1.8 & 2191 & 0.39 & 0.35 & 0.67 & 0.43\\
0.05 & 0.2 & 7 & 23.8 & 1613 & 0.38 & 0.41 & 0.65 & 0.36\\
0.05 & 0.2 & 8 & 14.4 & 1747 & 0.39 & 0.39 & 0.66 & 0.37\\
0.05 & 0.2 & 9 & 11.7 & 1911 & 0.39 & 0.42 & 0.67 & 0.40\\
0 & 0.05 & 0 & 0.9 & 2205 & 0.34 & 0.34 & 0.64 & 0.37\\
0 & 0.05 & 1 & 0.8 & 2206 & 0.33 & 0.34 & 0.69 & 0.38\\
0 & 0.05 & 2 & 0.8 & 2204 & 0.34 & 0.32 & 0.66 & 0.37\\
0 & 0.05 & 3 & 0.9 & 2205 & 0.34 & 0.32 & 0.66 & 0.38\\
0 & 0.05 & 4 & 1.0 & 2205 & 0.33 & 0.32 & 0.66 & 0.37\\
0 & 0.05 & 5 & 1.0 & 2207 & 0.33 & 0.36 & 0.65 & 0.36\\
0 & 0.05 & 6 & 0.9 & 2203 & 0.33 & 0.35 & 0.66 & 0.38\\
0 & 0.05 & 7 & 1.1 & 2206 & 0.33 & 0.34 & 0.62 & 0.37\\
0 & 0.05 & 8 & 1.0 & 2206 & 0.33 & 0.35 & 0.68 & 0.36\\
0 & 0.05 & 9 & 0.8 & 2206 & 0.33 & 0.34 & 0.61 & 0.37\\
0.0125 & 0.05 & 0 & 3.7 & 2142 & 0.40 & 0.38 & 0.67 & 0.40\\
0.0125 & 0.05 & 1 & 7.3 & 2016 & 0.38 & 0.41 & 0.66 & 0.42\\
0.0125 & 0.05 & 2 & 4.9 & 2084 & 0.39 & 0.40 & 0.66 & 0.41\\
0.0125 & 0.05 & 3 & 6.9 & 2016 & 0.40 & 0.38 & 0.66 & 0.41\\
\bottomrule
\end{tabular}
\end{table*}

\begin{table*}[ht]\centering
\caption{Learning-rate sweep, per seed, part 3 of 3. Clustering, welfare, and learned behavior of the frozen policy.}
\label{tab:s_sweep3}
\centering\small\setlength{\tabcolsep}{3pt}
\begin{tabular}{lrrrrrrrr}
\toprule
$\alpha_C$ & $\alpha_D$ & seed & $I_U$ & $W$ & $P_{\text{up}}^C$ & $P_{\text{stay}|\text{rich}}^C$ & $P_{\text{leave}|\text{poor}}^C$ & $P_{\text{up}}^D$\\
\midrule
0.0125 & 0.05 & 4 & 12.2 & 1887 & 0.37 & 0.39 & 0.65 & 0.41\\
0.0125 & 0.05 & 5 & 10.6 & 1936 & 0.38 & 0.41 & 0.64 & 0.38\\
0.0125 & 0.05 & 6 & 3.9 & 2107 & 0.40 & 0.38 & 0.66 & 0.41\\
0.0125 & 0.05 & 7 & 4.0 & 2122 & 0.41 & 0.38 & 0.67 & 0.40\\
0.0125 & 0.05 & 8 & 5.7 & 2034 & 0.39 & 0.38 & 0.65 & 0.42\\
0.0125 & 0.05 & 9 & 4.2 & 2084 & 0.39 & 0.39 & 0.66 & 0.41\\
0.025 & 0.05 & 0 & 11.4 & 1903 & 0.38 & 0.42 & 0.65 & 0.37\\
0.025 & 0.05 & 1 & 14.3 & 1805 & 0.38 & 0.41 & 0.61 & 0.35\\
0.025 & 0.05 & 2 & 10.9 & 1918 & 0.38 & 0.43 & 0.66 & 0.41\\
0.025 & 0.05 & 3 & 8.1 & 2009 & 0.39 & 0.43 & 0.66 & 0.42\\
0.025 & 0.05 & 4 & 7.7 & 2021 & 0.38 & 0.43 & 0.67 & 0.44\\
0.025 & 0.05 & 5 & 8.0 & 1973 & 0.37 & 0.43 & 0.67 & 0.42\\
0.025 & 0.05 & 6 & 7.5 & 2010 & 0.39 & 0.40 & 0.66 & 0.42\\
0.025 & 0.05 & 7 & 10.0 & 1953 & 0.37 & 0.44 & 0.66 & 0.40\\
0.025 & 0.05 & 8 & 10.4 & 1878 & 0.37 & 0.41 & 0.64 & 0.35\\
0.025 & 0.05 & 9 & 7.8 & 2022 & 0.39 & 0.40 & 0.66 & 0.41\\
0.1 & 0.05 & 0 & 12.2 & 1838 & 0.41 & 0.36 & 0.68 & 0.41\\
0.1 & 0.05 & 1 & 12.0 & 1868 & 0.43 & 0.33 & 0.68 & 0.43\\
0.1 & 0.05 & 2 & 8.4 & 1925 & 0.43 & 0.34 & 0.68 & 0.44\\
0.1 & 0.05 & 3 & 6.6 & 2002 & 0.42 & 0.35 & 0.67 & 0.42\\
0.1 & 0.05 & 4 & 11.7 & 1862 & 0.44 & 0.33 & 0.69 & 0.45\\
0.1 & 0.05 & 5 & 5.9 & 2031 & 0.42 & 0.35 & 0.68 & 0.42\\
0.1 & 0.05 & 6 & 8.0 & 1880 & 0.43 & 0.34 & 0.68 & 0.46\\
0.1 & 0.05 & 7 & 10.8 & 1868 & 0.43 & 0.34 & 0.67 & 0.43\\
0.1 & 0.05 & 8 & 8.7 & 1896 & 0.43 & 0.34 & 0.69 & 0.43\\
0.1 & 0.05 & 9 & 14.6 & 1767 & 0.42 & 0.34 & 0.66 & 0.42\\
0.2 & 0.05 & 0 & 15.5 & 1800 & 0.47 & 0.28 & 0.68 & 0.46\\
0.2 & 0.05 & 1 & 9.7 & 1887 & 0.48 & 0.27 & 0.69 & 0.54\\
0.2 & 0.05 & 2 & 8.5 & 1965 & 0.48 & 0.28 & 0.71 & 0.52\\
0.2 & 0.05 & 3 & 20.8 & 1566 & 0.56 & 0.22 & 0.70 & 0.59\\
0.2 & 0.05 & 4 & 15.0 & 1739 & 0.52 & 0.23 & 0.69 & 0.58\\
0.2 & 0.05 & 5 & 21.9 & 1598 & 0.54 & 0.21 & 0.70 & 0.57\\
0.2 & 0.05 & 6 & 14.6 & 1730 & 0.53 & 0.23 & 0.68 & 0.58\\
0.2 & 0.05 & 7 & 8.5 & 1954 & 0.45 & 0.31 & 0.70 & 0.48\\
0.2 & 0.05 & 8 & 13.4 & 1851 & 0.46 & 0.28 & 0.69 & 0.51\\
0.2 & 0.05 & 9 & 18.7 & 1739 & 0.53 & 0.25 & 0.69 & 0.52\\
\bottomrule
\end{tabular}
\end{table*}

\begin{table*}[ht]\centering
\caption{Motion of the cooperator profile, per seed, part 1 of 3. The table gives the mean signed velocity $\bar v$ and the mean speed $|v|$ in patches per time unit at lag $\Delta=5$, the share $d$ of windows in the run's main direction to three decimals, the correlation $r$ at the best shift, and whether the run travels coherently, that is, whether $r>0.7$ and $d>0.9$.}
\label{tab:s_waves1}
\centering\small\setlength{\tabcolsep}{3pt}
\begin{tabular}{lrrrrrc}
\toprule
$\alpha_C$ / $\alpha_D$ & seed & $\bar v$ & $|v|$ & main dir. & corr & trav.\\
\midrule
0.05 / 0 & 0 & -0.01 & 0.03 & 0.097 & 0.93 & \\
0.05 / 0 & 1 & -0.01 & 0.03 & 0.112 & 0.91 & \\
0.05 / 0 & 2 & 0.04 & 0.05 & 0.245 & 0.94 & \\
0.05 / 0 & 3 & 0.01 & 0.21 & 0.107 & 0.95 & \\
0.05 / 0 & 4 & -0.00 & 0.01 & 0.031 & 0.95 & \\
0.05 / 0 & 5 & 0.02 & 0.03 & 0.112 & 0.93 & \\
0.05 / 0 & 6 & -0.01 & 0.02 & 0.066 & 0.92 & \\
0.05 / 0 & 7 & 0.01 & 0.02 & 0.061 & 0.94 & \\
0.05 / 0 & 8 & -0.03 & 0.05 & 0.204 & 0.94 & \\
0.05 / 0 & 9 & -0.01 & 0.03 & 0.107 & 0.93 & \\
0.0125 / 0.05 & 0 & -0.04 & 0.33 & 0.378 & 0.58 & \\
0.0125 / 0.05 & 1 & -0.03 & 0.33 & 0.423 & 0.84 & \\
0.0125 / 0.05 & 2 & 0.02 & 0.16 & 0.291 & 0.76 & \\
0.0125 / 0.05 & 3 & -0.13 & 0.17 & 0.597 & 0.79 & \\
0.0125 / 0.05 & 4 & 0.10 & 0.14 & 0.464 & 0.90 & \\
0.0125 / 0.05 & 5 & 0.04 & 0.06 & 0.235 & 0.88 & \\
0.0125 / 0.05 & 6 & -0.09 & 0.22 & 0.347 & 0.70 & \\
0.0125 / 0.05 & 7 & -0.06 & 0.12 & 0.342 & 0.68 & \\
0.0125 / 0.05 & 8 & 0.08 & 0.15 & 0.444 & 0.83 & \\
0.0125 / 0.05 & 9 & -0.06 & 0.14 & 0.403 & 0.76 & \\
0.025 / 0.05 & 0 & 0.10 & 0.16 & 0.163 & 0.90 & \\
0.025 / 0.05 & 1 & 0.01 & 0.02 & 0.071 & 0.93 & \\
0.025 / 0.05 & 2 & 0.08 & 0.10 & 0.434 & 0.90 & \\
0.025 / 0.05 & 3 & -0.01 & 0.07 & 0.194 & 0.85 & \\
0.025 / 0.05 & 4 & 0.07 & 0.13 & 0.464 & 0.85 & \\
0.025 / 0.05 & 5 & -0.09 & 0.12 & 0.474 & 0.89 & \\
0.025 / 0.05 & 6 & 0.03 & 0.10 & 0.301 & 0.85 & \\
0.025 / 0.05 & 7 & -0.07 & 0.10 & 0.408 & 0.90 & \\
0.025 / 0.05 & 8 & -0.00 & 0.02 & 0.056 & 0.91 & \\
0.025 / 0.05 & 9 & 0.03 & 0.12 & 0.296 & 0.84 & \\
0.05 / 0.2 & 0 & 0.03 & 0.04 & 0.194 & 0.95 & \\
0.05 / 0.2 & 1 & -0.11 & 0.23 & 0.505 & 0.60 & \\
0.05 / 0.2 & 2 & -0.01 & 0.03 & 0.102 & 0.93 & \\
0.05 / 0.2 & 3 & -0.07 & 0.08 & 0.342 & 0.91 & \\
0.05 / 0.2 & 4 & -0.01 & 0.03 & 0.092 & 0.95 & \\
0.05 / 0.2 & 5 & 0.06 & 0.09 & 0.367 & 0.85 & \\
0.05 / 0.2 & 6 & 0.14 & 1.15 & 0.536 & 0.39 & \\
\bottomrule
\end{tabular}
\end{table*}

\begin{table*}[ht]\centering
\caption{Motion of the cooperator profile, per seed, part 2 of 3. The table gives the mean signed velocity $\bar v$ and the mean speed $|v|$ in patches per time unit at lag $\Delta=5$, the share $d$ of windows in the run's main direction to three decimals, the correlation $r$ at the best shift, and whether the run travels coherently, that is, whether $r>0.7$ and $d>0.9$.}
\label{tab:s_waves2}
\centering\small\setlength{\tabcolsep}{3pt}
\begin{tabular}{lrrrrrc}
\toprule
$\alpha_C$ / $\alpha_D$ & seed & $\bar v$ & $|v|$ & main dir. & corr & trav.\\
\midrule
0.05 / 0.2 & 7 & 0.04 & 0.04 & 0.189 & 0.96 & \\
0.05 / 0.2 & 8 & -0.03 & 0.03 & 0.163 & 0.94 & \\
0.05 / 0.2 & 9 & -0.01 & 0.04 & 0.128 & 0.91 & \\
0.05 / 0.1 & 0 & -0.04 & 0.07 & 0.270 & 0.86 & \\
0.05 / 0.1 & 1 & 0.06 & 0.06 & 0.276 & 0.95 & \\
0.05 / 0.1 & 2 & -0.02 & 0.13 & 0.286 & 0.81 & \\
0.05 / 0.1 & 3 & -0.04 & 0.10 & 0.327 & 0.85 & \\
0.05 / 0.1 & 4 & -0.01 & 0.03 & 0.092 & 0.95 & \\
0.05 / 0.1 & 5 & 0.04 & 0.08 & 0.291 & 0.88 & \\
0.05 / 0.1 & 6 & -0.01 & 0.05 & 0.143 & 0.86 & \\
0.05 / 0.1 & 7 & -0.09 & 0.16 & 0.510 & 0.83 & \\
0.05 / 0.1 & 8 & -0.07 & 0.11 & 0.393 & 0.86 & \\
0.05 / 0.1 & 9 & -0.05 & 0.05 & 0.245 & 0.92 & \\
0.05 / 0.05 & 0 & -0.04 & 0.12 & 0.332 & 0.86 & \\
0.05 / 0.05 & 1 & -0.02 & 0.16 & 0.372 & 0.80 & \\
0.05 / 0.05 & 2 & -0.11 & 0.27 & 0.439 & 0.59 & \\
0.05 / 0.05 & 3 & 0.05 & 0.10 & 0.316 & 0.91 & \\
0.05 / 0.05 & 4 & -0.16 & 0.17 & 0.643 & 0.81 & \\
0.05 / 0.05 & 5 & 0.06 & 0.13 & 0.408 & 0.89 & \\
0.05 / 0.05 & 6 & 0.09 & 0.24 & 0.459 & 0.75 & \\
0.05 / 0.05 & 7 & -0.05 & 0.15 & 0.372 & 0.85 & \\
0.05 / 0.05 & 8 & 0.01 & 0.17 & 0.301 & 0.86 & \\
0.05 / 0.05 & 9 & -0.01 & 0.12 & 0.265 & 0.78 & \\
0.1 / 0.05 & 0 & -0.12 & 0.15 & 0.541 & 0.88 & \\
0.1 / 0.05 & 1 & -0.37 & 0.37 & 0.964 & 0.86 & yes\\
0.1 / 0.05 & 2 & 0.35 & 0.36 & 0.893 & 0.84 & \\
0.1 / 0.05 & 3 & 0.14 & 0.28 & 0.653 & 0.78 & \\
0.1 / 0.05 & 4 & -0.32 & 0.37 & 0.852 & 0.83 & \\
0.1 / 0.05 & 5 & -0.10 & 0.16 & 0.520 & 0.78 & \\
0.1 / 0.05 & 6 & -0.38 & 0.38 & 0.980 & 0.86 & yes\\
0.1 / 0.05 & 7 & 0.32 & 0.32 & 0.903 & 0.86 & yes\\
0.1 / 0.05 & 8 & 0.01 & 0.10 & 0.265 & 0.86 & \\
0.1 / 0.05 & 9 & -0.57 & 0.57 & 1.000 & 0.89 & yes\\
0.05 / 0.025 & 0 & -0.32 & 0.38 & 0.944 & 0.86 & yes\\
0.05 / 0.025 & 1 & 0.54 & 0.54 & 0.990 & 0.80 & yes\\
0.05 / 0.025 & 2 & 0.45 & 0.45 & 0.980 & 0.86 & yes\\
0.05 / 0.025 & 3 & 0.62 & 0.62 & 0.995 & 0.86 & yes\\
\bottomrule
\end{tabular}
\end{table*}

\begin{table*}[ht]\centering
\caption{Motion of the cooperator profile, per seed, part 3 of 3. The table gives the mean signed velocity $\bar v$ and the mean speed $|v|$ in patches per time unit at lag $\Delta=5$, the share $d$ of windows in the run's main direction to three decimals, the correlation $r$ at the best shift, and whether the run travels coherently, that is, whether $r>0.7$ and $d>0.9$.}
\label{tab:s_waves3}
\centering\small\setlength{\tabcolsep}{3pt}
\begin{tabular}{lrrrrrc}
\toprule
$\alpha_C$ / $\alpha_D$ & seed & $\bar v$ & $|v|$ & main dir. & corr & trav.\\
\midrule
0.05 / 0.025 & 4 & -0.52 & 0.52 & 0.985 & 0.83 & yes\\
0.05 / 0.025 & 5 & 0.58 & 0.58 & 0.995 & 0.82 & yes\\
0.05 / 0.025 & 6 & -0.41 & 0.41 & 0.903 & 0.78 & yes\\
0.05 / 0.025 & 7 & 0.43 & 0.43 & 0.990 & 0.90 & yes\\
0.05 / 0.025 & 8 & 0.39 & 0.39 & 0.954 & 0.86 & yes\\
0.05 / 0.025 & 9 & 0.42 & 0.42 & 0.985 & 0.81 & yes\\
0.2 / 0.05 & 0 & 0.68 & 0.91 & 0.969 & 0.80 & yes\\
0.2 / 0.05 & 1 & 0.99 & 0.99 & 1.000 & 0.87 & yes\\
0.2 / 0.05 & 2 & 0.97 & 0.97 & 1.000 & 0.84 & yes\\
0.2 / 0.05 & 3 & -1.30 & 1.30 & 1.000 & 0.90 & yes\\
0.2 / 0.05 & 4 & -1.30 & 1.30 & 1.000 & 0.89 & yes\\
0.2 / 0.05 & 5 & 1.29 & 1.29 & 1.000 & 0.92 & yes\\
0.2 / 0.05 & 6 & 1.34 & 1.34 & 1.000 & 0.89 & yes\\
0.2 / 0.05 & 7 & 0.48 & 0.49 & 0.974 & 0.81 & yes\\
0.2 / 0.05 & 8 & -0.92 & 0.92 & 1.000 & 0.87 & yes\\
0.2 / 0.05 & 9 & -1.06 & 1.06 & 1.000 & 0.86 & yes\\
0.05 / 0.0125 & 0 & -0.92 & 0.92 & 1.000 & 0.94 & yes\\
0.05 / 0.0125 & 1 & -1.06 & 1.06 & 1.000 & 0.94 & yes\\
0.05 / 0.0125 & 2 & 0.89 & 0.89 & 1.000 & 0.88 & yes\\
0.05 / 0.0125 & 3 & 1.04 & 1.04 & 1.000 & 0.91 & yes\\
0.05 / 0.0125 & 4 & 0.79 & 0.79 & 0.990 & 0.80 & yes\\
0.05 / 0.0125 & 5 & 1.07 & 1.07 & 1.000 & 0.94 & yes\\
0.05 / 0.0125 & 6 & -0.74 & 0.79 & 0.995 & 0.87 & yes\\
0.05 / 0.0125 & 7 & 1.00 & 1.00 & 1.000 & 0.90 & yes\\
0.05 / 0.0125 & 8 & 0.77 & 1.05 & 0.974 & 0.90 & yes\\
0.05 / 0.0125 & 9 & -1.04 & 1.04 & 1.000 & 0.92 & yes\\
random movement & 0 & -0.33 & 1.74 & 0.47 & 0.31 & \\
random movement & 1 & -0.04 & 2.18 & 0.47 & 0.31 & \\
random movement & 2 & -0.17 & 1.47 & 0.43 & 0.32 & \\
random movement & 3 & 0.04 & 1.65 & 0.48 & 0.31 & \\
random movement & 4 & -0.01 & 1.77 & 0.43 & 0.31 & \\
random movement & 5 & -0.07 & 2.22 & 0.48 & 0.31 & \\
random movement & 6 & -0.12 & 0.91 & 0.40 & 0.34 & \\
random movement & 7 & -0.18 & 2.13 & 0.50 & 0.31 & \\
random movement & 8 & 0.18 & 2.36 & 0.47 & 0.31 & \\
random movement & 9 & -0.24 & 2.10 & 0.45 & 0.31 & \\
\bottomrule
\end{tabular}
\end{table*}

\paragraph{Longer training.}
Table~\ref{tab:s_long} gives the checkpoint means of the long-training runs
shown in Figure 12 and lists every tenth checkpoint shown
in that figure. Table~\ref{tab:s_long_seeds} gives the per-seed summaries quoted in the main text, namely coherent checkpoints, median speed, reversals, final bias, and the change in clustering and welfare. Over episodes $1400$--$1600$, the seed means for cooperator-only learning, equal-rate learning, $(0.2,0.05)$, and $(0.05,0.0125)$ are $I_U=24.9$, $20.3$, $23.0$, and $13.5$ and $W=1731$, $1587$, $1537$, and $1817$. At episode $200$ of the same runs, welfare was $W=1722$, $2055$, $1713$, and $1664$, so the welfare ranking of the four conditions at episode $200$ is not preserved. In every long run, $W$ is below the random-movement value of $2197$--$2199$ at every checkpoint from episode $20$ onward, with a largest value of $2146$.

\begin{table*}[ht]\centering\small
\caption{Longer training, with means over seeds at each checkpoint of the frozen tables. Checkpoints are every 200 episodes of the runs of Figure~\ref{fig:s_long}. The columns give speed $|v|$, main-direction share, clustering $I_U$, welfare $W$, $P_{\text{up}}^C$, $P_{\text{stay}|\text{rich}}^C$, and the mean absolute cooperator bias.}\label{tab:s_long}
\setlength{\tabcolsep}{3pt}\begin{tabular}{lrrrrrrrr}\toprule condition & episode & $|v|$ & dir. & $I_U$ & $W$ & $P_{\text{up}}^C$ & $P_{\text{stay}|\text{rich}}^C$ & $|\chi_C|$\\\midrule
only C learns, $(0.05,0)$ (3 seeds) & 100 & 0.04 & 0.07 & 25.2 & 1739 & 0.35 & 0.47 & 0.00\\
 & 300 & 0.03 & 0.08 & 26.6 & 1714 & 0.40 & 0.40 & 0.00\\
 & 500 & 0.02 & 0.06 & 24.7 & 1759 & 0.41 & 0.39 & 0.00\\
 & 700 & 0.04 & 0.13 & 23.2 & 1764 & 0.40 & 0.41 & 0.00\\
 & 900 & 0.02 & 0.09 & 27.7 & 1722 & 0.38 & 0.42 & 0.01\\
 & 1100 & 0.01 & 0.05 & 25.4 & 1757 & 0.43 & 0.38 & 0.01\\
 & 1300 & 0.05 & 0.19 & 25.0 & 1739 & 0.39 & 0.41 & 0.01\\
 & 1500 & 0.03 & 0.09 & 23.9 & 1723 & 0.38 & 0.42 & 0.01\\
\midrule
both learn, $(0.05,0.05)$ (3 seeds) & 100 & 0.17 & 0.53 & 5.5 & 2051 & 0.41 & 0.38 & 0.02\\
 & 300 & 0.12 & 0.38 & 8.8 & 1929 & 0.41 & 0.36 & 0.01\\
 & 500 & 0.12 & 0.43 & 12.0 & 1897 & 0.42 & 0.36 & 0.02\\
 & 700 & 0.13 & 0.44 & 15.4 & 1735 & 0.44 & 0.33 & 0.02\\
 & 900 & 0.26 & 0.31 & 17.2 & 1723 & 0.46 & 0.31 & 0.01\\
 & 1100 & 0.18 & 0.33 & 16.5 & 1703 & 0.45 & 0.32 & 0.01\\
 & 1300 & 0.35 & 0.45 & 18.1 & 1665 & 0.46 & 0.30 & 0.03\\
 & 1500 & 0.07 & 0.25 & 28.1 & 1323 & 0.46 & 0.30 & 0.02\\
\midrule
both learn, $(0.2,0.05)$ (5 seeds) & 100 & 0.49 & 0.85 & 10.1 & 1843 & 0.44 & 0.34 & 0.15\\
 & 300 & 1.16 & 1.00 & 11.0 & 1883 & 0.50 & 0.25 & 0.32\\
 & 500 & 0.80 & 0.99 & 11.1 & 1851 & 0.47 & 0.29 & 0.23\\
 & 700 & 0.19 & 0.56 & 16.4 & 1695 & 0.44 & 0.32 & 0.04\\
 & 900 & 0.17 & 0.61 & 16.3 & 1685 & 0.44 & 0.33 & 0.04\\
 & 1100 & 0.07 & 0.24 & 23.3 & 1481 & 0.44 & 0.33 & 0.01\\
 & 1300 & 0.09 & 0.40 & 19.8 & 1614 & 0.45 & 0.32 & 0.02\\
 & 1500 & 0.10 & 0.42 & 27.3 & 1452 & 0.45 & 0.33 & 0.03\\
\midrule
both learn, $(0.05,0.0125)$ (5 seeds) & 100 & 0.19 & 0.51 & 17.7 & 1666 & 0.40 & 0.38 & 0.05\\
 & 300 & 1.20 & 1.00 & 26.8 & 1519 & 0.54 & 0.23 & 0.39\\
 & 500 & 1.14 & 1.00 & 15.4 & 1724 & 0.53 & 0.24 & 0.38\\
 & 700 & 1.08 & 1.00 & 13.7 & 1845 & 0.50 & 0.25 & 0.30\\
 & 900 & 0.71 & 0.81 & 13.1 & 1828 & 0.49 & 0.28 & 0.21\\
 & 1100 & 1.21 & 1.00 & 24.8 & 1477 & 0.56 & 0.22 & 0.42\\
 & 1300 & 1.42 & 1.00 & 19.9 & 1613 & 0.58 & 0.20 & 0.44\\
 & 1500 & 1.31 & 1.00 & 12.3 & 1865 & 0.52 & 0.23 & 0.36\\
\bottomrule\end{tabular}\end{table*}

\begin{table*}[ht]\centering\small
\caption{Longer training, per seed, for the runs of Figure~\ref{fig:s_long} with $81$ frozen checkpoints every $20$ episodes. Coherent counts the checkpoints at or after episode $200$ that travel coherently, out of $71$. The table also gives the median $|v|$ over the same checkpoints and the number of reversals, which are sign changes of $\bar v$ along the sequence of coherent checkpoints at or after episode $200$. The last columns give $|\chi_C|$ at episode $1600$, and $I_U$ and $W$ at episode $200$ and averaged over episodes $1400$--$1600$.}\label{tab:s_long_seeds}
\setlength{\tabcolsep}{4pt}
{\tiny
\begin{tabular}{lrrrrrrrrr}\toprule condition & seed & coherent & median $|v|$ & reversals & $|\chi_C|_{1600}$ & $I_U^{200}$ & $I_U^{1400\text{--}1600}$ & $W^{200}$ & $W^{1400\text{--}1600}$\\\midrule
only C learns, $(0.05,0)$ & 0 & 0/71 & 0.03 & 0 & 0.00 & 24.0 & 24.4 & 1782 & 1757\\
 & 1 & 0/71 & 0.03 & 0 & 0.00 & 25.7 & 24.8 & 1678 & 1728\\
 & 2 & 0/71 & 0.03 & 0 & 0.01 & 23.3 & 25.5 & 1706 & 1707\\
\midrule
both learn, $(0.05,0.05)$ & 0 & 0/71 & 0.11 & 0 & 0.00 & 7.5 & 20.4 & 2042 & 1564\\
 & 1 & 1/71 & 0.12 & 0 & 0.01 & 4.7 & 21.1 & 2120 & 1588\\
 & 2 & 0/71 & 0.13 & 0 & 0.04 & 9.7 & 19.4 & 2004 & 1610\\
\midrule
both learn, $(0.2,0.05)$ & 0 & 23/71 & 0.16 & 5 & 0.04 & 17.8 & 23.4 & 1637 & 1554\\
 & 1 & 23/71 & 0.15 & 6 & 0.15 & 12.3 & 23.3 & 1921 & 1525\\
 & 2 & 21/71 & 0.17 & 4 & 0.00 & 18.7 & 25.6 & 1597 & 1467\\
 & 3 & 23/71 & 0.19 & 3 & 0.01 & 21.2 & 20.4 & 1592 & 1611\\
 & 4 & 26/71 & 0.19 & 4 & 0.00 & 14.8 & 22.3 & 1818 & 1527\\
\midrule
both learn, $(0.05,0.0125)$ & 0 & 65/71 & 1.17 & 1 & 0.35 & 17.1 & 12.8 & 1840 & 1854\\
 & 1 & 65/71 & 1.16 & 1 & 0.40 & 25.4 & 17.7 & 1412 & 1669\\
 & 2 & 64/71 & 1.20 & 1 & 0.38 & 21.5 & 15.9 & 1653 & 1757\\
 & 3 & 67/71 & 1.18 & 1 & 0.20 & 12.5 & 11.5 & 1820 & 1875\\
 & 4 & 68/71 & 1.21 & 1 & 0.26 & 19.2 & 9.7 & 1593 & 1931\\
\bottomrule\end{tabular}
}
\end{table*}

\paragraph{Welfare decomposition.}
Tables~\ref{tab:s_w_cr}--\ref{tab:s_w_conly} give the decomposition $W=B-N_Cc-\mathcal K$ per seed for the policy recombination of Section~\ref{sec:emerge} of the main text.

\begin{table*}[ht]\centering
\caption{Welfare decomposition per seed, cooperators co-trained, defectors random: $W=B-N_Cc-\mathcal K$.}
\label{tab:s_w_cr}
\centering\small\setlength{\tabcolsep}{3pt}
\begin{tabular}{lrrrrrrrrr}
\toprule
seed & $W$ & $B$ & $K$ & $\Pi_C$ & $\Pi_D$ & $\varphi_C$ & $\varphi_D$ & $n_C$ & $n_D$\\
\midrule
0 & 2019 & 3038 & 571 & 3.47 & 2.42 & 0.79 & 0.49 & 21.2 & 10.1\\
1 & 2040 & 3037 & 549 & 3.47 & 2.53 & 0.78 & 0.51 & 20.0 & 10.4\\
2 & 2033 & 3035 & 554 & 3.48 & 2.47 & 0.78 & 0.50 & 20.3 & 10.3\\
3 & 1978 & 3048 & 622 & 3.45 & 2.24 & 0.80 & 0.46 & 23.4 & 10.2\\
4 & 2063 & 3040 & 529 & 3.58 & 2.41 & 0.79 & 0.48 & 19.3 & 9.9\\
5 & 2018 & 3039 & 573 & 3.48 & 2.39 & 0.79 & 0.48 & 21.3 & 10.0\\
6 & 2028 & 3039 & 563 & 3.47 & 2.46 & 0.79 & 0.49 & 20.7 & 10.3\\
7 & 1986 & 3043 & 609 & 3.47 & 2.25 & 0.80 & 0.46 & 22.9 & 10.0\\
8 & 2009 & 3040 & 583 & 3.52 & 2.26 & 0.80 & 0.46 & 21.6 & 10.2\\
9 & 2041 & 3034 & 545 & 3.49 & 2.49 & 0.78 & 0.50 & 19.9 & 10.2\\
\bottomrule
\end{tabular}
\end{table*}

\begin{table*}[ht]\centering
\caption{Welfare decomposition per seed, cooperators co-trained, defectors learned: $W=B-N_Cc-\mathcal K$.}
\label{tab:s_w_cl}
\centering\small\setlength{\tabcolsep}{3pt}
\begin{tabular}{lrrrrrrrrr}
\toprule
seed & $W$ & $B$ & $K$ & $\Pi_C$ & $\Pi_D$ & $\varphi_C$ & $\varphi_D$ & $n_C$ & $n_D$\\
\midrule
0 & 1954 & 3002 & 600 & 2.95 & 3.29 & 0.71 & 0.67 & 20.6 & 14.5\\
1 & 2004 & 3000 & 548 & 3.04 & 3.34 & 0.71 & 0.68 & 18.4 & 14.2\\
2 & 2141 & 2998 & 410 & 3.27 & 3.53 & 0.70 & 0.69 & 13.3 & 11.7\\
3 & 1803 & 3005 & 754 & 2.71 & 3.08 & 0.72 & 0.66 & 26.3 & 17.2\\
4 & 2005 & 2999 & 547 & 3.05 & 3.33 & 0.71 & 0.67 & 18.3 & 14.2\\
5 & 1851 & 3003 & 703 & 2.79 & 3.13 & 0.71 & 0.66 & 24.3 & 16.6\\
6 & 2051 & 2999 & 500 & 3.12 & 3.41 & 0.71 & 0.68 & 16.6 & 13.4\\
7 & 1959 & 3002 & 594 & 2.98 & 3.26 & 0.71 & 0.67 & 20.2 & 14.8\\
8 & 1938 & 3001 & 616 & 2.94 & 3.23 & 0.71 & 0.67 & 20.8 & 15.5\\
9 & 2061 & 3002 & 493 & 3.14 & 3.40 & 0.71 & 0.67 & 16.5 & 12.8\\
\bottomrule
\end{tabular}
\end{table*}

\begin{table*}[ht]\centering
\caption{Welfare decomposition per seed, only cooperators learn: $W=B-N_Cc-\mathcal K$.}
\label{tab:s_w_conly}
\centering\small\setlength{\tabcolsep}{3pt}
\begin{tabular}{lrrrrrrrrr}
\toprule
seed & $W$ & $B$ & $K$ & $\Pi_C$ & $\Pi_D$ & $\varphi_C$ & $\varphi_D$ & $n_C$ & $n_D$\\
\midrule
0 & 1813 & 3062 & 800 & 3.30 & 1.74 & 0.84 & 0.37 & 31.6 & 9.7\\
1 & 1792 & 3054 & 814 & 3.17 & 1.93 & 0.82 & 0.41 & 31.8 & 10.5\\
2 & 1688 & 3070 & 933 & 3.14 & 1.46 & 0.86 & 0.33 & 37.2 & 10.4\\
3 & 1715 & 3072 & 909 & 3.18 & 1.51 & 0.86 & 0.34 & 36.2 & 10.3\\
4 & 1801 & 3069 & 821 & 3.39 & 1.47 & 0.86 & 0.33 & 32.4 & 9.8\\
5 & 1777 & 3072 & 847 & 3.32 & 1.51 & 0.86 & 0.33 & 33.7 & 9.6\\
6 & 1752 & 3060 & 861 & 3.13 & 1.81 & 0.83 & 0.40 & 33.6 & 11.2\\
7 & 1829 & 3055 & 778 & 3.26 & 1.91 & 0.83 & 0.40 & 30.4 & 10.2\\
8 & 1680 & 3071 & 943 & 3.12 & 1.46 & 0.86 & 0.33 & 37.6 & 10.5\\
9 & 1748 & 3072 & 876 & 3.28 & 1.44 & 0.86 & 0.32 & 34.9 & 9.8\\
\bottomrule
\end{tabular}
\end{table*}

\section{Robustness of the observation}
\label{sec:robust-app}

Two variants of the agents' observation were run with the same learning rule, decision clocks, and protocol, using \texttt{run.py --obs relative} and \texttt{--obs relative\_n}. They do not simply drop components of the main observation, because each has its own discretization.
\emph{Neighbor differences only.} The agent observes the four neighbor differences. The resource differences $\varphi_{i\pm1}-\varphi_i$ have three levels with thresholds $\pm0.01$, and the occupancy differences $n_{i\pm1}-n_i$ have five levels with thresholds $-5.5$, $-2.5$, $2.5$, and $5.5$. This gives $3\cdot3\cdot5\cdot5=225$ states. The agent does not observe its own $\varphi_i$ or $n_i$. Tables~\ref{tab:s_rel_conly}--\ref{tab:s_rel_donly} give the results for 10 seeds. The three core conditions give $I_U=30.4$, $6.0$, and $0.94$ and $W=1638$, $2047$, and $2205$.
\emph{Neighbor differences plus own occupancy.} This observation adds $n_i$ in three levels, $\le7$, $8$--$12$, and $\ge13$, which gives $225\cdot3=675$ states. Tables~\ref{tab:s_absn_both}--\ref{tab:s_absn_conly} give the results for 5 seeds of cooperator-only and equal-rate learning, with $I_U=27.0$ and $6.2$ and $W=1712$ and $2059$.
With neighbor differences only, the ordering of the three learning conditions is the same as in the main model in every seed. Clustering is highest when only cooperators learn, intermediate when both learn, and lowest when only defectors learn, and welfare follows the reverse order. With own occupancy added, only the two listed conditions were run. In all five seeds, cooperator-only learning clusters more strongly and earns less than equal-rate learning. Keeping only the four three-level difference components of the main observation would give $81$ states, and that variant was not run.

\begin{table*}[ht]
\caption{Robustness, relative-only observation (225 states), only cooperators learn, per seed.}
\label{tab:s_rel_conly}
\centering\small\setlength{\tabcolsep}{3pt}
\begin{tabular}{lrrrrrrrr}
\toprule
seed & $\Pup^C$ & $\Pup^D$ & $I_U$ random & learned & $\rho$ random & learned & $W$ random & learned\\
\midrule
0 & 0.34 & 0.33 & 0.94 & 28.89 & 0.33 & 0.81 & 2199 & 1682\\
1 & 0.36 & 0.34 & 0.79 & 23.54 & 0.42 & 0.83 & 2199 & 1761\\
2 & 0.32 & 0.33 & 0.80 & 33.50 & 0.42 & 0.81 & 2198 & 1622\\
3 & 0.38 & 0.33 & 0.95 & 25.03 & 0.28 & 0.83 & 2197 & 1742\\
4 & 0.33 & 0.33 & 1.05 & 31.50 & 0.27 & 0.79 & 2199 & 1647\\
5 & 0.35 & 0.33 & 1.04 & 35.26 & 0.30 & 0.80 & 2201 & 1560\\
6 & 0.36 & 0.34 & 0.91 & 33.93 & 0.28 & 0.78 & 2197 & 1505\\
7 & 0.36 & 0.33 & 1.08 & 34.94 & 0.34 & 0.79 & 2200 & 1542\\
8 & 0.35 & 0.33 & 1.03 & 28.93 & 0.27 & 0.82 & 2199 & 1651\\
9 & 0.36 & 0.33 & 0.85 & 28.62 & 0.29 & 0.80 & 2200 & 1668\\
\bottomrule
\end{tabular}
\end{table*}

\begin{table*}[ht]
\caption{Robustness, relative-only observation (225 states), both learn, per seed.}
\label{tab:s_rel_both}
\centering\small\setlength{\tabcolsep}{3pt}
\begin{tabular}{lrrrrrrrr}
\toprule
seed & $\Pup^C$ & $\Pup^D$ & $I_U$ random & learned & $\rho$ random & learned & $W$ random & learned\\
\midrule
0 & 0.42 & 0.42 & 0.94 & 4.08 & 0.33 & 0.85 & 2199 & 2107\\
1 & 0.42 & 0.41 & 0.79 & 3.91 & 0.42 & 0.81 & 2199 & 2125\\
2 & 0.42 & 0.41 & 0.80 & 4.91 & 0.42 & 0.91 & 2198 & 2084\\
3 & 0.42 & 0.43 & 0.95 & 4.33 & 0.28 & 0.85 & 2197 & 2088\\
4 & 0.41 & 0.41 & 1.05 & 7.19 & 0.27 & 0.90 & 2199 & 2030\\
5 & 0.41 & 0.42 & 1.04 & 5.72 & 0.30 & 0.90 & 2201 & 2024\\
6 & 0.42 & 0.40 & 0.91 & 6.08 & 0.28 & 0.88 & 2197 & 2046\\
7 & 0.42 & 0.42 & 1.08 & 5.45 & 0.34 & 0.90 & 2200 & 2033\\
8 & 0.41 & 0.42 & 1.03 & 8.24 & 0.27 & 0.92 & 2199 & 1994\\
9 & 0.41 & 0.42 & 0.85 & 9.67 & 0.29 & 0.93 & 2200 & 1938\\
\bottomrule
\end{tabular}
\end{table*}

\begin{table*}[ht]
\caption{Robustness, relative-only observation (225 states), only defectors learn, per seed.}
\label{tab:s_rel_donly}
\centering\small\setlength{\tabcolsep}{3pt}
\begin{tabular}{lrrrrrrrr}
\toprule
seed & $\Pup^C$ & $\Pup^D$ & $I_U$ random & learned & $\rho$ random & learned & $W$ random & learned\\
\midrule
0 & 0.34 & 0.37 & 0.94 & 0.94 & 0.33 & 0.66 & 2199 & 2205\\
1 & 0.33 & 0.38 & 0.79 & 0.79 & 0.42 & 0.65 & 2199 & 2206\\
2 & 0.33 & 0.38 & 0.80 & 0.80 & 0.42 & 0.62 & 2198 & 2204\\
3 & 0.33 & 0.38 & 0.95 & 0.95 & 0.28 & 0.60 & 2197 & 2205\\
4 & 0.33 & 0.37 & 1.05 & 1.05 & 0.27 & 0.65 & 2199 & 2205\\
5 & 0.33 & 0.39 & 1.04 & 1.04 & 0.30 & 0.62 & 2201 & 2206\\
6 & 0.33 & 0.38 & 0.91 & 0.91 & 0.28 & 0.67 & 2197 & 2203\\
7 & 0.34 & 0.38 & 1.08 & 1.08 & 0.34 & 0.67 & 2200 & 2206\\
8 & 0.33 & 0.38 & 1.03 & 1.03 & 0.27 & 0.67 & 2199 & 2206\\
9 & 0.33 & 0.37 & 0.85 & 0.85 & 0.29 & 0.55 & 2200 & 2206\\
\bottomrule
\end{tabular}
\end{table*}

\begin{table*}[ht]
\caption{Robustness, neighbor differences plus own occupancy (675 states), both learn, per seed.}
\label{tab:s_absn_both}
\centering\small\setlength{\tabcolsep}{3pt}
\begin{tabular}{lrrrrrrrr}
\toprule
seed & $\Pup^C$ & $\Pup^D$ & $I_U$ random & learned & $\rho$ random & learned & $W$ random & learned\\
\midrule
0 & 0.40 & 0.40 & 0.94 & 8.65 & 0.33 & 0.89 & 2199 & 1968\\
1 & 0.42 & 0.42 & 0.79 & 4.22 & 0.42 & 0.84 & 2199 & 2118\\
2 & 0.39 & 0.41 & 0.80 & 8.98 & 0.42 & 0.91 & 2198 & 2005\\
3 & 0.42 & 0.42 & 0.95 & 4.29 & 0.28 & 0.83 & 2197 & 2116\\
4 & 0.42 & 0.42 & 1.05 & 4.88 & 0.27 & 0.92 & 2199 & 2090\\
\bottomrule
\end{tabular}
\end{table*}

\begin{table*}[ht]\centering
\caption{Robustness, neighbor differences plus own occupancy (675 states), only cooperators learn, per seed.}
\label{tab:s_absn_conly}
\centering\small\setlength{\tabcolsep}{3pt}
\begin{tabular}{lrrrrrrrr}
\toprule
seed & $\Pup^C$ & $\Pup^D$ & $I_U$ random & learned & $\rho$ random & learned & $W$ random & learned\\
\midrule
0 & 0.37 & 0.33 & 0.94 & 31.31 & 0.33 & 0.81 & 2199 & 1621\\
1 & 0.37 & 0.33 & 0.79 & 28.54 & 0.42 & 0.79 & 2199 & 1664\\
2 & 0.37 & 0.33 & 0.80 & 23.80 & 0.42 & 0.85 & 2198 & 1781\\
3 & 0.38 & 0.33 & 0.95 & 27.79 & 0.28 & 0.80 & 2197 & 1719\\
4 & 0.37 & 0.33 & 1.05 & 23.79 & 0.27 & 0.84 & 2199 & 1775\\
\bottomrule
\end{tabular}
\end{table*}

\section{Finer learning-rate sweep, training length, and further toll conditions}
\label{sec:sweep}

\begin{figure*}[ht]
\centering
\includegraphics[width=\textwidth]{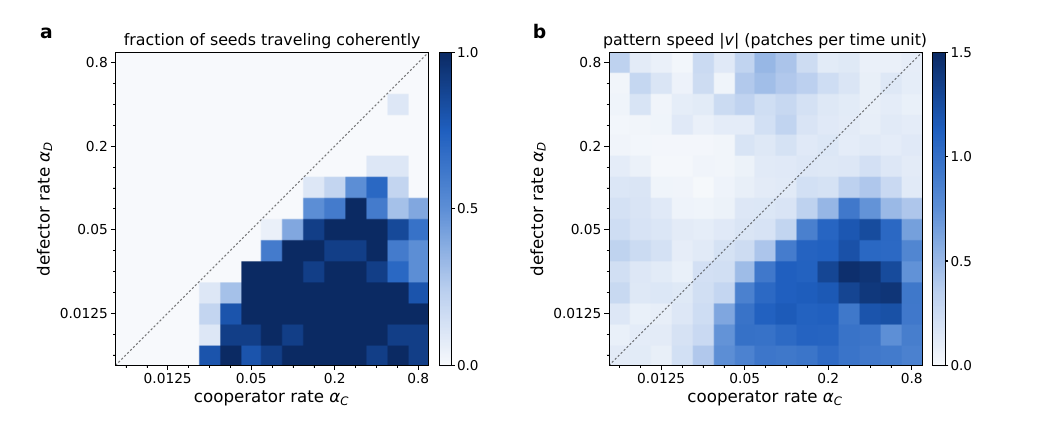}
\caption{The $15\times15$ learning-rate map after $200$ training episodes, with ten training seeds per cell, twenty for $\alpha_D=0.05$, and one frozen-policy evaluation per seed. (a)~Fraction of seeds that travel coherently. (b)~Mean pattern speed $|v|$. The dashed line is $\alpha_C=\alpha_D$. Figure~\ref{fig:rates} of the main text shows the welfare loss and clustering of the same cells.}
\label{fig:s_ratemap}
\end{figure*}

\begin{figure*}[ht]
\centering
\includegraphics[width=\textwidth]{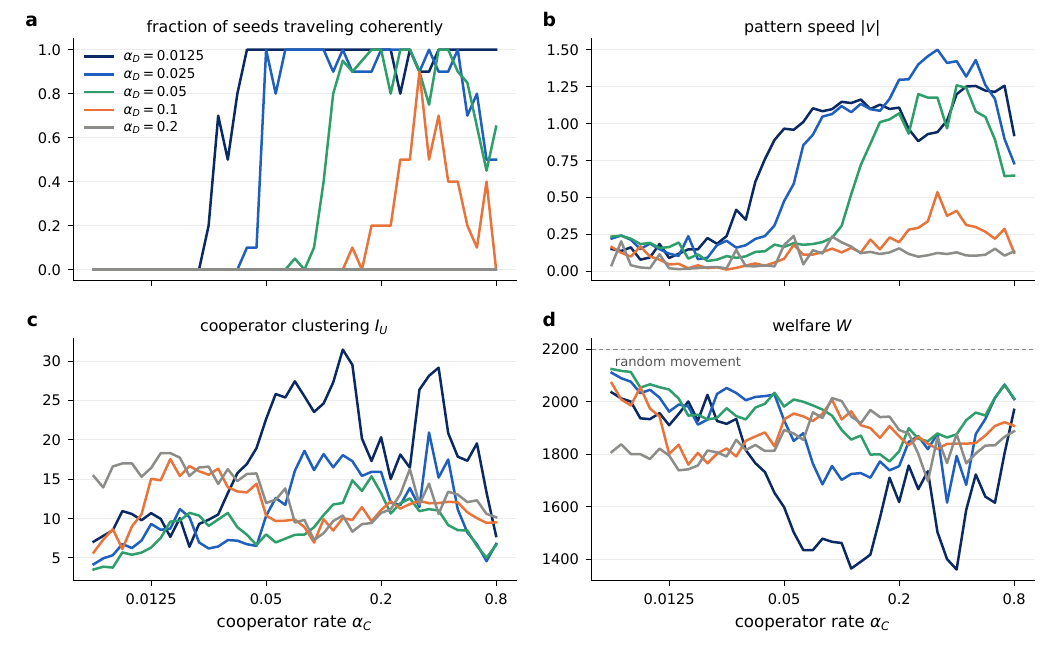}
\caption{Fine sweeps of $\alpha_C$ at five fixed values of $\alpha_D$ after $200$ training episodes, averaged over ten training seeds and over twenty for $\alpha_D=0.05$. The panels show the fraction of coherently traveling seeds, pattern speed, clustering, and welfare. Dashed lines mark random movement.}
\label{fig:s_ratelines}
\end{figure*}

\begin{figure*}[ht]
\centering
\includegraphics[width=\textwidth]{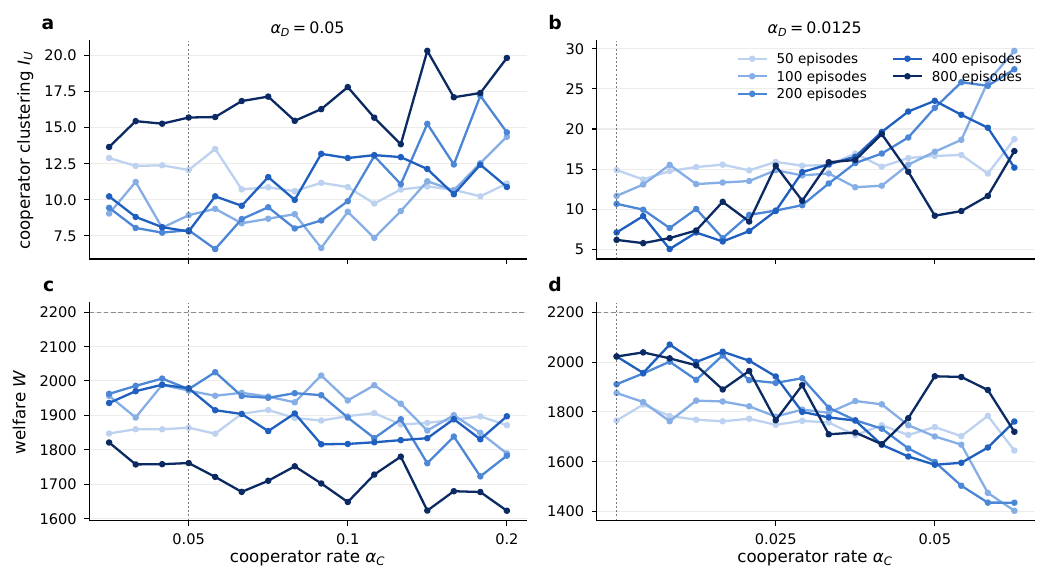}
\caption{Training length. Clustering (a,b) and welfare (c,d) of the frozen policies after $50$ to $800$ training episodes, averaged over ten training seeds. The dotted line marks $\alpha_C=\alpha_D$ and the dashed line random movement.}
\label{fig:s_trainlength}
\end{figure*}

\paragraph{Design.} Learning rates are taken from the lattice $\alpha=0.00625\cdot2^{k/6}$ with $k=0,\dots,42$, which runs from $0.00625$ to $0.8$ and contains the five rates of the $5\times5$ grid. The \emph{map} uses every third lattice point for both rates, giving $15\times15$ rate pairs with ten training seeds each. The \emph{fine lines} use all $43$ values of $\alpha_C$ at $\alpha_D=0.0125$, $0.025$, $0.05$, $0.1$, and $0.2$, with ten training seeds per cell and twenty at $\alpha_D=0.05$. In total these are $4080$ training runs with learning cooperators. Every run follows the main protocol exactly. It uses $200$ episodes, the same exploration schedule and stream seeds, and one frozen evaluation of $400$ time units from the held-out placement of its seed. The cells that coincide with the $5\times5$ grid reproduce the per-seed values of Table~\ref{tab:s_grid}. As in the grid, $W$ is the average of the instantaneous welfare over the second half of the evaluation. The random-movement reference is $W=2199$, the mean over the same ten evaluation placements.

\paragraph{Map.} Figure~\ref{fig:rates}a,b of the main text shows the welfare loss and the clustering of the map, and Figure~\ref{fig:s_ratemap} shows the fraction of coherently traveling seeds and the pattern speed for the same cells. Cell means of $W$ range from $782$ at $(0.2,0.00625)$ to $2125$ at $(0.00625,0.05)$, all below random movement. Cell means of $I_U$ range from $3.5$ to $49$. The correlation between the cell means of $W$ and $I_U$ is $-0.99$ across the $225$ cells and $-0.97$ across the $4080$ runs. Of the $4080$ runs, one has a higher $W$ than random movement on the evaluation placement of its seed. In $69$ cells at least half of the seeds travel coherently, and all of them have $\alpha_C\ge2\alpha_D$ and $\alpha_D\le0.1$. No seed travels in a cell with $\alpha_C\le\alpha_D$, and at most one seed in ten travels in any cell with $\alpha_D\ge0.14$. A second region of clustering without any traveling seed lies at large $\alpha_D$ and small $\alpha_C$, where cell means of $I_U$ reach $18$ for $\alpha_D\ge0.1$ and $\alpha_C\le0.035$.

\paragraph{Fine lines.} Figure~\ref{fig:s_ratelines} and Table~\ref{tab:s_fine} give the fine lines. At $\alpha_D=0.05$ the number of traveling seeds is $0$, $2$, $8$, $16$, and $19$ of $20$ at $\alpha_C=0.079$, $0.089$, $0.1$, $0.112$, and $0.126$. At $\alpha_D=0.0125$ and $0.025$ the corresponding rise occurs near $\alpha_C=0.03$ and $0.05$. At $\alpha_D=0.1$, at least half of the seeds travel only for $\alpha_C$ between $0.25$ and $0.4$, and never more than nine of ten. At $\alpha_D=0.2$ no seed travels. For the largest $\alpha_C$ the number of traveling seeds falls again at $\alpha_D=0.025$ and $0.05$. Welfare is not monotone in $\alpha_C$ along these lines. These are properties of frozen policies after $200$ episodes. The next paragraph shows that they change with the training budget, so we do not interpret them as thresholds.

\begin{table*}[ht]\centering\small
\caption{Finer learning-rate sweep with frozen policies after 200 episodes and one evaluation per training seed. The table gives the number of coherently traveling seeds, the mean clustering $I_U$, and the mean welfare $W$ for $43$ values of $\alpha_C$ at five fixed values of $\alpha_D$. Each cell has ten training seeds, and twenty for $\alpha_D=0.05$.}\label{tab:s_fine}
\setlength{\tabcolsep}{3.2pt}
{\tiny
\begin{tabular}{r|rrrrr|rrrrr|rrrrr}\toprule
 & \multicolumn{5}{c|}{traveling seeds, by $\alpha_D$} & \multicolumn{5}{c|}{$I_U$, by $\alpha_D$} & \multicolumn{5}{c}{$W$, by $\alpha_D$}\\
$\alpha_C$ & 0.0125 & 0.025 & 0.05 & 0.1 & 0.2 & 0.0125 & 0.025 & 0.05 & 0.1 & 0.2 & 0.0125 & 0.025 & 0.05 & 0.1 & 0.2\\\midrule
0.00625 & 0/10 & 0/10 & 0/20 & 0/10 & 0/10 & 7.1 & 4.2 & 3.5 & 5.7 & 15.4 & 2036 & 2111 & 2125 & 2072 & 1809\\
0.007015 & 0/10 & 0/10 & 0/20 & 0/10 & 0/10 & 7.8 & 4.9 & 3.9 & 7.3 & 13.9 & 2013 & 2090 & 2118 & 2009 & 1838\\
0.007875 & 0/10 & 0/10 & 0/20 & 0/10 & 0/10 & 8.5 & 5.3 & 3.8 & 8.6 & 16.6 & 2001 & 2077 & 2114 & 1986 & 1801\\
0.008839 & 0/10 & 0/10 & 0/20 & 0/10 & 0/10 & 10.9 & 6.8 & 5.7 & 6.1 & 17.0 & 1937 & 2033 & 2054 & 2055 & 1801\\
0.009921 & 0/10 & 0/10 & 0/20 & 0/10 & 0/10 & 10.6 & 6.2 & 5.4 & 9.0 & 17.0 & 1934 & 2046 & 2067 & 1974 & 1782\\
0.01114 & 0/10 & 0/10 & 0/20 & 0/10 & 0/10 & 9.8 & 7.2 & 5.7 & 10.5 & 15.3 & 1957 & 2016 & 2056 & 1944 & 1822\\
0.0125 & 0/10 & 0/10 & 0/20 & 0/10 & 0/10 & 10.7 & 9.3 & 6.3 & 15.0 & 16.4 & 1911 & 1963 & 2047 & 1801 & 1795\\
0.01403 & 0/10 & 0/10 & 0/20 & 0/10 & 0/10 & 10.0 & 8.6 & 7.5 & 14.9 & 18.3 & 1954 & 1990 & 2012 & 1837 & 1739\\
0.01575 & 0/10 & 0/10 & 0/20 & 0/10 & 0/10 & 7.7 & 8.7 & 9.6 & 17.5 & 18.3 & 2001 & 1982 & 1947 & 1761 & 1743\\
0.01768 & 0/10 & 0/10 & 0/20 & 0/10 & 0/10 & 10.0 & 11.2 & 9.7 & 15.4 & 17.7 & 1928 & 1913 & 1951 & 1806 & 1758\\
0.01984 & 0/10 & 0/10 & 0/20 & 0/10 & 0/10 & 6.4 & 10.2 & 10.7 & 16.7 & 15.4 & 2026 & 1932 & 1933 & 1766 & 1815\\
0.02227 & 0/10 & 0/10 & 0/20 & 0/10 & 0/10 & 9.3 & 7.0 & 10.4 & 16.0 & 16.5 & 1927 & 2030 & 1939 & 1802 & 1807\\
0.025 & 2/10 & 0/10 & 0/20 & 0/10 & 0/10 & 9.8 & 6.2 & 9.1 & 15.6 & 16.6 & 1916 & 2053 & 1976 & 1823 & 1792\\
0.02806 & 7/10 & 0/10 & 0/20 & 0/10 & 0/10 & 10.5 & 6.4 & 9.9 & 16.3 & 14.4 & 1935 & 2034 & 1946 & 1792 & 1855\\
0.0315 & 5/10 & 0/10 & 0/20 & 0/10 & 0/10 & 13.2 & 7.3 & 10.7 & 14.0 & 16.3 & 1816 & 2006 & 1933 & 1852 & 1816\\
0.03536 & 8/10 & 0/10 & 0/20 & 0/10 & 0/10 & 15.8 & 7.2 & 8.9 & 13.4 & 14.8 & 1766 & 2019 & 1978 & 1868 & 1836\\
0.03968 & 10/10 & 1/10 & 0/20 & 0/10 & 0/10 & 16.9 & 6.7 & 7.9 & 13.3 & 15.7 & 1732 & 2022 & 1993 & 1884 & 1813\\
0.04454 & 10/10 & 1/10 & 0/20 & 0/10 & 0/10 & 18.9 & 6.5 & 6.6 & 14.4 & 15.7 & 1653 & 2027 & 2035 & 1830 & 1814\\
0.05 & 10/10 & 10/10 & 0/20 & 0/10 & 0/10 & 22.6 & 10.3 & 8.0 & 10.4 & 12.0 & 1599 & 1929 & 1982 & 1933 & 1893\\
0.05612 & 10/10 & 8/10 & 0/20 & 0/10 & 0/10 & 25.8 & 12.6 & 7.0 & 9.7 & 12.4 & 1503 & 1851 & 2009 & 1955 & 1879\\
0.063 & 10/10 & 10/10 & 0/20 & 0/10 & 0/10 & 25.4 & 11.7 & 7.4 & 9.7 & 13.8 & 1435 & 1880 & 2000 & 1945 & 1855\\
0.07071 & 10/10 & 10/10 & 1/20 & 0/10 & 0/10 & 27.4 & 16.1 & 7.9 & 9.9 & 9.5 & 1435 & 1765 & 1986 & 1927 & 1960\\
0.07937 & 10/10 & 10/10 & 0/20 & 0/10 & 0/10 & 25.5 & 18.6 & 8.0 & 8.9 & 9.8 & 1479 & 1686 & 1971 & 1955 & 1938\\
0.08909 & 10/10 & 10/10 & 2/20 & 0/10 & 0/10 & 23.5 & 16.1 & 8.9 & 6.9 & 7.3 & 1467 & 1756 & 1947 & 2008 & 2014\\
0.1 & 10/10 & 10/10 & 8/20 & 0/10 & 0/10 & 24.6 & 18.2 & 10.4 & 9.9 & 8.1 & 1462 & 1703 & 1893 & 1933 & 2003\\
0.1122 & 10/10 & 9/10 & 16/20 & 0/10 & 0/10 & 27.3 & 16.5 & 11.8 & 8.5 & 9.7 & 1365 & 1725 & 1856 & 1964 & 1944\\
0.126 & 10/10 & 10/10 & 19/20 & 0/10 & 0/10 & 31.5 & 18.1 & 12.0 & 10.0 & 10.4 & 1391 & 1729 & 1872 & 1911 & 1919\\
0.1414 & 10/10 & 9/10 & 18/20 & 1/10 & 0/10 & 29.5 & 17.3 & 14.8 & 9.8 & 8.3 & 1418 & 1711 & 1799 & 1900 & 1969\\
0.1587 & 10/10 & 9/10 & 19/20 & 0/10 & 0/10 & 20.2 & 15.4 & 13.5 & 11.5 & 9.3 & 1558 & 1773 & 1800 & 1863 & 1942\\
0.1782 & 10/10 & 9/10 & 20/20 & 2/10 & 0/10 & 17.3 & 15.9 & 15.4 & 9.6 & 9.4 & 1711 & 1739 & 1773 & 1907 & 1944\\
0.2 & 10/10 & 10/10 & 20/20 & 2/10 & 0/10 & 20.3 & 15.9 & 13.2 & 11.1 & 10.7 & 1618 & 1755 & 1811 & 1869 & 1893\\
0.2245 & 10/10 & 10/10 & 16/20 & 2/10 & 0/10 & 15.0 & 12.0 & 10.6 & 12.2 & 11.2 & 1757 & 1848 & 1899 & 1834 & 1879\\
0.252 & 8/10 & 10/10 & 20/20 & 5/10 & 0/10 & 18.1 & 11.7 & 11.9 & 11.3 & 13.1 & 1667 & 1870 & 1859 & 1867 & 1802\\
0.2828 & 10/10 & 10/10 & 20/20 & 5/10 & 0/10 & 16.5 & 13.9 & 12.5 & 11.8 & 16.3 & 1736 & 1820 & 1850 & 1840 & 1697\\
0.3175 & 9/10 & 9/10 & 18/20 & 9/10 & 0/10 & 26.4 & 11.5 & 10.9 & 12.2 & 11.8 & 1504 & 1878 & 1879 & 1820 & 1864\\
0.3564 & 9/10 & 10/10 & 15/20 & 5/10 & 0/10 & 28.1 & 20.9 & 11.2 & 11.9 & 14.4 & 1401 & 1616 & 1864 & 1840 & 1767\\
0.4 & 10/10 & 9/10 & 20/20 & 7/10 & 0/10 & 29.2 & 15.2 & 11.0 & 12.0 & 10.6 & 1361 & 1794 & 1876 & 1841 & 1877\\
0.449 & 10/10 & 9/10 & 20/20 & 4/10 & 0/10 & 20.8 & 17.5 & 9.2 & 12.1 & 13.4 & 1588 & 1684 & 1930 & 1841 & 1766\\
0.504 & 10/10 & 10/10 & 18/20 & 4/10 & 0/10 & 17.9 & 11.1 & 8.5 & 12.1 & 13.0 & 1723 & 1878 & 1959 & 1843 & 1803\\
0.5657 & 10/10 & 7/10 & 17/20 & 2/10 & 0/10 & 17.3 & 8.2 & 8.5 & 10.8 & 12.1 & 1639 & 1935 & 1948 & 1872 & 1833\\
0.635 & 10/10 & 8/10 & 13/20 & 1/10 & 0/10 & 19.5 & 6.7 & 6.5 & 10.0 & 12.3 & 1616 & 2014 & 2016 & 1908 & 1835\\
0.7127 & 10/10 & 5/10 & 9/20 & 4/10 & 0/10 & 13.3 & 4.6 & 5.0 & 9.4 & 10.6 & 1804 & 2066 & 2064 & 1922 & 1867\\
0.8 & 10/10 & 5/10 & 13/20 & 0/10 & 0/10 & 7.8 & 6.8 & 6.6 & 9.5 & 10.2 & 1969 & 2010 & 2013 & 1909 & 1888\\
\bottomrule\end{tabular}
}
\end{table*}

\paragraph{Training length.} We use $16$ values of $\alpha_C$ from $0.035$ to $0.2$ at $\alpha_D=0.05$ and $16$ values from $0.0125$ to $0.071$ at $\alpha_D=0.0125$. For each rate pair we train ten seeds for $800$ episodes and evaluate the frozen tables after $50$, $100$, $200$, $400$, and $800$ episodes, each on the held-out placement of its seed. The exploration schedule is that of the main protocol, decreasing from $0.3$ to $0.01$ over the first $160$ episodes and then staying at $0.01$. The stream seeds are $1000s+e$, which are distinct for $e<1000$, so the checkpoint at episode $200$ is identical to the corresponding run of the sweep above. Table~\ref{tab:s_trainlength} gives the number of traveling seeds and the clustering at every checkpoint, and Figure~\ref{fig:trainlength} of the main text shows the corresponding fractions. Figure~\ref{fig:s_trainlength} shows the clustering and the welfare. After $50$ episodes at most one seed in ten travels. Between $100$ and $400$ episodes the range of $\alpha_C$ with traveling seeds extends to smaller values. At $\alpha_D=0.05$, at most two seeds in ten travel after $800$ episodes at any $\alpha_C$, and the mean clustering over the line rises from $10.5$ at episode $200$ to $16.5$. At $\alpha_D=0.0125$, most seeds with $\alpha_C\ge0.02$ travel after $800$ episodes, with fewer at the largest $\alpha_C$. In all $1600$ seed-checkpoints, $W$ is below random movement on the same placement. Together with the long runs of Figure~\ref{fig:s_long}, these runs show that traveling bands can appear, persist, weaken, disappear, or reverse under continued training. They do not identify an asymptotic state.

\begin{table*}[ht]\centering\small
\caption{Training length. The table gives the number of training seeds, out of ten, whose frozen policies travel coherently at the checkpoints after 50, 100, 200, 400, and 800 episodes of one 800-episode run per seed. It also gives the mean clustering $I_U$ at the same checkpoints.}\label{tab:s_trainlength}
\setlength{\tabcolsep}{3.5pt}
{\tiny
\begin{tabular}{r|rrrrr|rrrrr||r|rrrrr|rrrrr}\toprule
 & \multicolumn{10}{c||}{$\alpha_D=0.05$} & & \multicolumn{10}{c}{$\alpha_D=0.0125$}\\
 & \multicolumn{5}{c|}{traveling seeds} & \multicolumn{5}{c||}{$I_U$} & & \multicolumn{5}{c|}{traveling seeds} & \multicolumn{5}{c}{$I_U$}\\
$\alpha_C$ & 50 & 100 & 200 & 400 & 800 & 50 & 100 & 200 & 400 & 800 & $\alpha_C$ & 50 & 100 & 200 & 400 & 800 & 50 & 100 & 200 & 400 & 800\\\midrule
0.03536 & 0 & 0 & 0 & 0 & 0 & 12.9 & 9.0 & 9.4 & 10.2 & 13.6 & 0.0125 & 0 & 0 & 0 & 0 & 0 & 14.9 & 11.7 & 10.7 & 7.1 & 6.2\\
0.03968 & 0 & 0 & 0 & 0 & 0 & 12.3 & 11.2 & 8.0 & 8.8 & 15.4 & 0.01403 & 0 & 0 & 0 & 0 & 0 & 13.7 & 13.1 & 10.0 & 9.2 & 5.8\\
0.04454 & 0 & 0 & 0 & 0 & 0 & 12.4 & 8.0 & 7.7 & 8.1 & 15.3 & 0.01575 & 0 & 0 & 0 & 0 & 3 & 14.8 & 15.5 & 7.7 & 5.0 & 6.4\\
0.05 & 0 & 0 & 0 & 0 & 0 & 12.0 & 8.9 & 7.9 & 7.8 & 15.7 & 0.01768 & 0 & 0 & 0 & 1 & 3 & 15.3 & 13.1 & 10.0 & 7.1 & 7.4\\
0.05612 & 0 & 0 & 0 & 0 & 0 & 13.5 & 9.4 & 6.6 & 10.2 & 15.7 & 0.01984 & 0 & 0 & 0 & 0 & 7 & 15.6 & 13.3 & 6.4 & 6.0 & 10.9\\
0.063 & 0 & 0 & 0 & 0 & 1 & 10.7 & 8.4 & 8.6 & 9.6 & 16.8 & 0.02227 & 0 & 0 & 0 & 3 & 8 & 14.8 & 13.5 & 9.3 & 7.3 & 8.5\\
0.07071 & 0 & 0 & 1 & 1 & 0 & 10.8 & 8.7 & 9.5 & 11.6 & 17.1 & 0.025 & 0 & 0 & 2 & 9 & 9 & 15.9 & 14.9 & 9.8 & 9.8 & 15.4\\
0.07937 & 0 & 0 & 0 & 2 & 0 & 10.6 & 9.0 & 8.0 & 10.0 & 15.5 & 0.02806 & 0 & 0 & 7 & 10 & 9 & 15.4 & 14.2 & 10.5 & 14.6 & 11.1\\
0.08909 & 0 & 0 & 1 & 8 & 0 & 11.2 & 6.6 & 8.6 & 13.2 & 16.3 & 0.0315 & 0 & 0 & 5 & 10 & 10 & 15.5 & 14.5 & 13.2 & 15.6 & 15.9\\
0.1 & 0 & 0 & 4 & 7 & 0 & 10.9 & 9.2 & 9.9 & 12.9 & 17.8 & 0.03536 & 0 & 0 & 8 & 10 & 9 & 16.9 & 12.7 & 15.8 & 16.5 & 16.1\\
0.1122 & 0 & 1 & 8 & 6 & 1 & 9.7 & 7.3 & 13.0 & 13.1 & 15.7 & 0.03968 & 0 & 0 & 10 & 10 & 10 & 15.3 & 12.9 & 16.9 & 19.6 & 19.4\\
0.126 & 0 & 1 & 9 & 8 & 2 & 10.7 & 9.2 & 11.1 & 12.9 & 13.8 & 0.04454 & 0 & 3 & 10 & 10 & 10 & 16.4 & 15.6 & 18.9 & 22.2 & 14.7\\
0.1414 & 0 & 4 & 10 & 8 & 1 & 10.9 & 11.3 & 15.2 & 12.1 & 20.3 & 0.05 & 0 & 3 & 10 & 10 & 8 & 16.6 & 17.2 & 22.6 & 23.5 & 9.2\\
0.1587 & 0 & 6 & 10 & 8 & 1 & 10.7 & 10.6 & 12.4 & 10.4 & 17.1 & 0.05612 & 0 & 4 & 10 & 10 & 8 & 16.8 & 18.6 & 25.8 & 21.8 & 9.8\\
0.1782 & 0 & 6 & 10 & 8 & 2 & 10.2 & 12.5 & 17.2 & 12.4 & 17.4 & 0.063 & 0 & 6 & 10 & 10 & 6 & 14.4 & 25.9 & 25.4 & 20.2 & 11.7\\
0.2 & 1 & 5 & 10 & 5 & 2 & 11.1 & 14.4 & 14.7 & 10.9 & 19.8 & 0.07071 & 0 & 8 & 10 & 10 & 5 & 18.7 & 29.7 & 27.4 & 15.2 & 17.2\\
\bottomrule\end{tabular}
}
\end{table*}

\paragraph{Further toll conditions.} Table~\ref{tab:s_toll2} repeats the toll experiment of the main text at five rate pairs with large untolled losses. As before, the toll $\gamma(n_i-1)/\Omega$ enters the learning reward and evaluation uses the original payoff. The toll raises welfare in all ten seeds of every pair. Recovery relative to random movement is nearly complete at $(0.05,0.0125)$ and $(0.1,0.0125)$ and partial where defectors learn very slowly. At $(0.2,0.00625)$ the tolled policies still cluster strongly, and eight of ten seeds travel.

\begin{table*}[ht]\centering\small
\caption{Crowding toll at further rate pairs. All runs use 200 episodes and ten training seeds and are evaluated with the original payoff on the held-out placement of each seed, with $W$ averaged over snapshots as in Table~\ref{tab:s_toll}. The first row repeats the equal-rate condition of the main text. $\Delta W$ is the paired difference with its $95\%$ bootstrap interval over seeds. Recovered is $(W_{\mathrm{toll}}-W)/(W_{\mathrm{random}}-W)$ with $W_{\mathrm{random}}=2199$.}\label{tab:s_toll2}
\setlength{\tabcolsep}{5pt}\begin{tabular}{l|rrr|rrr|rcr}\toprule
 & \multicolumn{3}{c|}{original reward} & \multicolumn{3}{c|}{with toll} & & & \\
$(\alpha_C,\alpha_D)$ & $W$ & $I_U$ & traveling & $W$ & $I_U$ & traveling & $\Delta W$ & 95\% interval & recovered\\\midrule
$(0.05,0.05)$ & 1977 & 7.9 & 0/10 & 2203 & 1.06 & 0/10 & 226 & [170, 285] & 102\% \\
$(0.05,0.0125)$ & 1599 & 22.6 & 10/10 & 2193 & 1.45 & 0/10 & 594 & [491, 701] & 99\% \\
$(0.2,0.00625)$ & 781 & 49.2 & 10/10 & 1617 & 20.83 & 8/10 & 837 & [598, 1053] & 59\% \\
$(0.1,0.00625)$ & 893 & 45.1 & 10/10 & 2042 & 6.50 & 3/10 & 1150 & [960, 1330] & 88\% \\
$(0.1414,0.008839)$ & 1106 & 35.9 & 10/10 & 2020 & 7.34 & 4/10 & 914 & [731, 1114] & 84\% \\
$(0.1,0.0125)$ & 1463 & 24.6 & 10/10 & 2182 & 1.86 & 0/10 & 718 & [574, 854] & 98\% \\
\bottomrule\end{tabular}\end{table*}

\section{Code and reproducibility}
\label{sec:code}

{\raggedright \paragraph{Code.} The code will be made available upon publication. The simulation is a single Python module (\texttt{model.py}, NumPy and Numba) with the environment, the decision clocks, the state encoding, the Q-learning update (with the optional crowding toll), and the evaluation. \texttt{rules.py} distills frozen populations into stochastic rules and replays them, and \texttt{measures.py} holds the welfare decomposition, the coherent-travel classification, and the interval estimates. Scripts: \texttt{run.py} (one condition with baselines, training, and frozen evaluation, saving the Q-tables), \texttt{run\_all.sh} and \texttt{runs\_queue.sh} (grid and environment variants), \texttt{runs\_toll.sh}, \texttt{runs\_long2.sh}, and \texttt{longrun.py} (long training with periodic frozen evaluation), \texttt{maintable.py} (Table~\ref{tab:main} of the main text and Table~\ref{tab:s_main}), \texttt{grid.py}, \texttt{mechanism.py} (distillation, replay, and ablations of every run), \texttt{oos.py} (fresh-rollout replay), \texttt{convention.py} (per-agent bias), \texttt{anatomy.py}, \texttt{toll\_summary.py}, \texttt{eps\_sens.py} (Section~\ref{sec:eps}), \texttt{handrules.py} (Section~\ref{sec:hand}), \texttt{mirror.py} and \texttt{randmatch.py} (Section~\ref{sec:mirror}), \texttt{recombine.py} (Section~\ref{sec:env-app}), \texttt{seedstats.py} (Table~\ref{tab:s_ci}), \texttt{vstats.py} (defector clustering, Table~\ref{tab:main} of the main text), \texttt{supp\_tables.py} and \texttt{figs2.py} (Figures~\ref{fig:mechanism} and~\ref{fig:toll} of the main text and the figures of Section~\ref{sec:figs}), \texttt{phase.py}, \texttt{phase2.py}, \texttt{phase\_summary.py}, and \texttt{supp\_tables\_phase.py} (Section~\ref{sec:sweep}), and \texttt{figs3.py} (Figures~\ref{fig:states}, \ref{fig:rates}, and~\ref{fig:trainlength} of the main text and the figures of Section~\ref{sec:sweep}). \texttt{REPRODUCE.md} lists the commands in order.\par}

\paragraph{Seeds.} Training seed $s\in\{0,\dots,9\}$, or $s\in\{0,1,2\}$ for three-seed cells, fixes the placements of all training episodes through a NumPy generator seeded with $s$. It also fixes the random stream of episode $e$, which is seeded with $1000s+e$ and supplies the decision phases, exploration, and tie-breaking. The evaluation of seed $s$ uses the generator and stream seed $s+777777$. Fresh rollout $j$ of seed $s$ uses $5000+10s+j$ in place of $s$ for the evaluation. The placements and the decision phases are drawn before any action of the episode and from streams that depend only on $s$ and $e$. They are therefore identical for all runs with the same seed, whatever the learning rates, the reward, or the policies. Toll and no-toll runs of the same seed, and the pairings of the recombination, share the training placements, the decision phases, and the evaluation placement. The exploration and tie-breaking draws come later in the same stream and are consumed differently once the policies differ, so they are not shared. Training and evaluation use separate streams. The $200$-episode runs use the stream seed $1000s+e$, which is distinct for every pair $(s,e)$ with $e<1000$. The long-training runs over $1600$ episodes derive the stream seed of episode $e$ from NumPy's \texttt{SeedSequence} of $(s,e)$, so no stream is shared between seeds or episodes. Their placements are those of the main runs but their streams are not, so their checkpoint at episode $200$ differs from the main run of the same seed.

\paragraph{Determinism and cost.} A run is deterministic given its seed. All trained tables used in this paper were regenerated from the seeds on a second machine and reproduced the stored per-seed statistics exactly. One training run of $200$ episodes takes about 15 seconds on one CPU core, and the $172$ trained runs take about 45 minutes. Software: Python 3.13, NumPy 2.5, Numba 0.68, Matplotlib 3.11.

\paragraph{Data.} The per-seed values behind every table and figure are saved as arrays by the scripts above (\texttt{results/maintable.npz}, \texttt{grid\_seeds.npz}, \texttt{mechanism.npy}, \texttt{oos.npz}, \texttt{toll.npz}, \texttt{eps\_sens.npz}, \texttt{handrules.npz}, \texttt{mirror.npz}, \texttt{randmatch.npz}, \texttt{recombine.npz}, \texttt{seedstats.npz}, \texttt{robust\_seeds.npz}, \texttt{vstats.npz}, \texttt{long2/}) and will be released with the code.

\end{document}